\documentclass[%
 reprint,
superscriptaddress,
 amsmath,amssymb,
 aps,
pra,
]{revtex4-2}

\usepackage{graphicx}
\usepackage{dcolumn}
\usepackage{bm}
\usepackage{url}
\usepackage{verbatim}
\usepackage{float}
\usepackage{xcolor}
\usepackage{soul}
\usepackage[T1]{fontenc}
\usepackage{ragged2e}
\usepackage{textcomp}

\begin{document}

\preprint{APS/123-QED}

\title{Analytic solution to degenerate biphoton states \\generated in arrays of nonlinear waveguides}

\author{Jefferson Delgado-Quesada}

\affiliation{Centro de Investigaci{\'o}n en Ciencia e Ingenier{\'i}a de Materiales, Universidad de Costa Rica, Costa Rica}

\affiliation{Escuela de F{\'i}sica, Universidad de Costa Rica, Costa Rica}
\author{David Barral} 

\affiliation{Galicia Supercomputing Center (CESGA), Avda. de Vigo S/N, Santiago de Compostela, 15705, Spain}

\author{Kamel Bencheikh}

\affiliation{Centre de Nanosciences et de Nanotechnologies, CNRS, Universit{\'e} Paris-Saclay, 91120 Palaiseau, France}
 
\author{Edgar A. Rojas-Gonz{\'a}lez}%

\email{edgar.rojasgonzalez@ucr.ac.cr}
\affiliation{Centro de Investigaci{\'o}n en Ciencia e Ingenier{\'i}a de Materiales, Universidad de Costa Rica, Costa Rica}

\affiliation{Escuela de F{\'i}sica, Universidad de Costa Rica, Costa Rica}

\date{\today}

\begin{abstract}
The arrays of nonlinear waveguides are a powerful integrated photonics platform for studying and manipulating quantum states of light. Also, they are a valuable resource for various quantum technologies. In this work, we employed a supermode approach to obtain an analytic solution to the evolution of degenerate biphoton states in arrays of nonlinear waveguides. The solution accounts for arrays of an arbitrary number of waveguides and coupling profiles. In addition, it provides an explicit analytic expression without the need of computationally-expensive steps. Actually, it only relies on the calculation of the eigenvalues and eigenvectors of the coupling matrix. In general, this needs to be performed numerically, but there are relevant instances in which analytic expressions are available. Thus, in certain cases, the procedure proposed here does not require the use of any numerical method. We analyze the general properties of the solution and show some application examples. Particularly, simple solutions that can be obtained by special symmetric injection profiles, the results for small arrays, and the properties of the propagation when only the center waveguide is pumped in a symmetric odd array. In addition, we present a proof-of-principle example on how to use the analytic solution to tackle inversion problems. That is, obtaining the initial conditions required to achieve a desired quantum state---which is a valuable technological application. A relevant aspect of the method presented here is its scalability for large arrays due to the lack of resource-intensive steps---both for the direct and inverse problems.

\end{abstract}

\maketitle


\section{\label{sec:level1}Introduction\protect}

Integrated photonics platforms are crucial for advanced quantum technologies, offering significant advantages over bulk optics \cite{Wang2019,Pelucchi2021,Ramakrishnan2023,Baboux2023,Zhang2023,Shekhar2024,Labonte2024,Dutt2024}  in terms of scalability, enhanced stability, lower costs, and the ability to generate, process, and detect signals within compact optical chips.

Any discrete unitary operator $U(N)$ of size $N$ can be implemented in optics as a linear optical network by successive application of beam splitter and phase shifter building blocks performing $U(2)$ transformations \cite{Reck1994, Clements2016, Taballione2023}. In the context of quantum photonics, universal quantum processors based on this approach with few modes have reached groundbreaking milestones \cite{Carolan2015, Maring2024}. However, the number of sequential $U(2)$ transformations scales as $O(N^2)$ such that constraints of chip size, electronic control, channel crosstalk, losses, and imperfections set a limit on the experimentally attainable maximum number of modes $N$. In this regard, simpler schemes that parallelize specific multimode transformations can improve scalability. For example, a photonic lattice or array of linear waveguides, which performs continuous-time quantum walks or quantum discrete fractional Fourier transforms by suitable engineering of evanescent coupling \cite{Peruzzo2010,Weimann2016}. Notably, when nonlinear effects are added to an array of waveguides then this element can increase the degree of entanglement of an input state (active behavior), in contrast to its linear version, which in the best-case scenario can only preserve it (passive behavior) \cite{Solntsev2014}. Thus, an array of nonlinear waveguides (ANW) is able to manipulate quantum light in non-trivial ways due to the intertwined combination of evanescent coupling and nonlinear effects in a compact scheme \cite{Christodoulides2003}. Moreover, it holds an inherit versatility attributed to suitable shaping of the pump, coupling, and phase-matching which opens the possibility to engineer, for instance, cluster states \cite{Barral2020a}.

An ANW can be implemented by fabricating the waveguides with nonlinear materials, such as periodically poled lithium niobate (PPLN) \cite{Kruse2013} or semiconductor AlGaAs platforms \cite{Raymond2024}. The properties of ANW have been explored both in the continuous \cite{Herec2003,Rai2012,Barral2020} and discrete variable regimes. In this work, we focus on the latter---in which biphoton, and even three-photon states \cite{Bacaoco2024} have been investigated in connection with ANW. Regarding biphoton states in ANW, these have been studied in relation to different topics---for example, quantum walks \cite{Solntsev2014,Hamilton2014}, topological photonics \cite{Leykam2015,Doyle2022}, Anderson localization \cite{Bai2016}, engineering of spatio-spectral correlations \cite{Kruse2013,Yang2014,Raymond2024}, and the generation of nonclassical states \cite{Solntsev2014,Hamilton2022}. In addition, nonclassical biphoton states in multimode systems, such as those that can be produced in ANW, could be key resources for diverse integrated-photonics quantum technologies. For example, in quantum communication \cite{Gisin2007,Couteau2023,Luo2023}, quantum computing \cite{Kok2007,Couteau2023,Maring2024}, and distributed quantum sensing \cite{Kim2024}.

The output biphoton states in ANW can be readily obtained by means of numerical methods for a given set of input parameters \cite{Solntsev2012,Graefe2012,Solntsev2012a,Kruse2013,Solntsev2014,Hamilton2014,Yang2014,Raymond2024}. 
However, the inverse problem of determining the input conditions to engineer a particular desired output state is challenging.
Indeed, interesting and valuable steps have been taken in this direction. Particularly, Belsley et al. \cite{Belsley2020} made use of the Cayley-Hamilton theorem to obtain analytic expressions that describe the biphoton output states in ANW as well as an inversion method that expresses the input conditions in terms of the output state. In addition, He et al. \cite{He2024} proposed to build the scattering tensor which describes the propagation in ANW, and compute its pseudo-inverse to obtain the input parameters that approximate a desired output biphoton state. The above proposals may work efficiently in ANW with few waveguides. However, in general, they are computationally-expensive for large arrays due to the nature of certain steps present in these methods.
Indeed, the approach presented in \cite{Belsley2020} requires to calculate the inverse of a Vandermonde matrix, and in \cite{He2024} it is necessary to build the related scattering tensor.         

In this work, we propose to use a supermode approach to obtain analytic solutions for the degenerate biphoton states produced in ANW through spontaneous parametric down-conversion (SPDC). The solution is obtained for arrays of an arbitrary number of waveguides and a general coupling profile. Our method relies on finding the eigenvalues and eigenvectors describing the supermodes of the array. Though this task might require in general numerical calculations, for certain coupling profiles the eigenvalues and eigenvectors present explicit analytic expressions \cite{Meng2004,Bosse2017,PerezLeija2013,PerezLeija2010,RodriguezLara2011}. 
For instance, in comparison with the method presented in \cite{Belsley2020}, our approach avoids the step of calculating an inverse matrix---which can be computationally expensive for large arrays. Thus, in addition of obtaining analytic output expressions, our method is a valuable alternative especially for analyzing lattices with a large number of waveguides.

Our model of SPDC dynamics in ANW presents the following advantages: i) It gives \emph{physical insight}. As mentioned above, there are established methods to solve the direct propagation in ANW numerically. Nevertheless, an analytic solution is more convenient for exploring the characteristics of the system and the interplay of the different relevant parameters in ANW. Moreover, our method shares features with those used in the spectral domain \cite{Grice1997}. The parallelism between both approaches enables to acquire understanding of a complex problem. ii) Our method is a tool to \emph{tackle inverse problems}. This is, arguably, the most important point of this work. That is, an analytic solution gives the explicit relation between the output state and the input parameters. This particular aspect allows to implement optimization algorithms to obtain the input conditions required to generate a desired target state, or at least to get the best possible approximation. In fact, this is significant for the design and engineering of ANW. It should be pointed out that the inverse problem was analyzed in \cite{Belsley2020, He2024}. However, we claim that our method presents features that can complement or be advantageous with respect to these approaches in terms of scalability. Indeed, it does not require potentially
computationally expensive steps, such as calculating inverse matrices or building the scattering tensor. iii) Our method is \emph{simpler} than others: Admittedly, as mentioned above, there exist other methods that calculate the output state in ANW. Nevertheless, using an explicit analytic expression is more convenient and simpler than solving numerically a system of coupled differential equations---as, for example, in \cite{Solntsev2014}.

The structure of this article is the following. First, in Sec. \ref{sec:level2} we introduce the relevant parameters that describe the propagation in an ANW.
Second, in Sec. \ref{sec:level3} we obtain an analytic solution to the propagation problem.
Third, in Sec. \ref{sec:level4} we comment on the general properties of this solution. Next, in Sec. \ref{sec:level5}, we present some application examples. Then, in Sec. \ref{sec:level6}, we give a proof-of-principle demonstration of the use of the analytic solution in an inverse problem. Finally, we summarize the main results, and provide some relevant remarks.

\section{\label{sec:level2}ANW and input parameters\protect}

\begin{figure}
    \centering
    \includegraphics[width=\columnwidth]{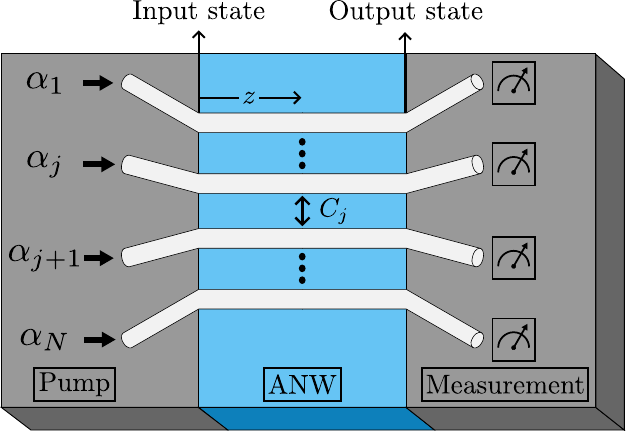}
    \caption{Schematic of the ANW with $N$ identical waveguides. The $j$th waveguide is injected with the pump field $\alpha_j$. The linear coupling and nonlinear effects take place only within the ANW region. The propagation occurs along the $z$ axis, and $z=0$ (related to the input state $|\Psi(0)\rangle$) corresponds to the beginning of the ANW region. The linear coupling between waveguides $j$ and $j+1$ is characterized by the parameter $C_j$, and the nonlinear effects are determined by the nonlinear constant $g$---which, in this case, is the same for all waveguides. The output state is taken at the end of the ANW region. The output of each individual waveguide could be measured, distributed, or sent as an input of a certain operation.}
    \label{fig:ANW_figure}
\end{figure}

A schematic of the ANW model is depicted in Fig. \ref{fig:ANW_figure}. It consists of $N$ identical waveguides with a quadratic nonlinearity described by the second-order susceptibility $\chi^{(2)} $ and where the $j$th waveguide can be independently excited by a strong coherent pump field with amplitude $\alpha_j=|\alpha_j|e^{i\phi_j}$. We define $\vec{\alpha}=(\alpha_1,\dots,\alpha_N)^T$. In these circumstances, a SPDC process takes place in the ANW. Here, we consider a situation in which the undepleted pump approximation is valid and focus on the degenerate case, where a pump photon with frequency $\omega_\mathrm{p}$ is down-converted into a pair of twin and undistinguishable signal photons with frequency $\omega_\mathrm{s}=\omega_\mathrm{p}/2$. This is the case in type-0 or type-I interactions, where both twin photons have the same polarization. Another process that takes place in the ANW is evanescent coupling between waveguides. Here, we consider only nearest-neighbor coupling of the signal photons. As the modal confinement increases with the frequency \cite{Noda1981}, we assume that the pump field is evanescently uncoupled to adjacent waveguides. Throughout this paper, we can thus reasonably assume that the pump photons remain in their initial waveguide.

The spatial evolution of the different modes propagating in the ANW is described by the momentum operator $\hat{M}$ \cite{Toren1994,Horoshko2022}, which can be expressed in the interaction picture as follows \cite{Linares2008}

\begin{eqnarray} \label{Eq:1_M}
\hat{M} &=&\hbar \sum_{j=1}^{N-1} C_j \hat{A}_{j+1}\hat{A}_j^\dagger e^{i[\beta_{j+1}(\omega_{\mathrm{s}})-\beta_{j}(\omega_{\mathrm{s}})]z}\nonumber\\
& &+\hbar \sum_{j=1}^N \gamma_j (\hat{A}_j^\dagger)^2 e^{i[\beta_{j}(\omega_{\mathrm{p}})-2\beta_{j}(\omega_{\mathrm{s}})]z}+\mathrm{h.c.},
\end{eqnarray}
with $\mathrm{h.c.}$ denoting the Hermitian conjugate, $\beta_j(\omega)$ the propagation constant at the $j$th waveguide and frequency $\omega$, and $\hat{A}_j$ ($\hat{A}_j^\dagger$) the monochromatic slowly-varying amplitude annihilation (creation) operator associated with the signal photons in the individual mode $j$ (the $j$th waveguide). These operators depend on the position, and frequency. In addition, they satisfy the commutation relation $[\hat{A}_j(z,\omega),\hat{A}_{j'}^\dagger(z',\omega')]=\delta(z,z')\delta(\omega-\omega')\delta_{j,j'}$ \cite{BenAryeh1991}, with $z$ the direction of propagation along the ANW. The first term in Eq. \eqref{Eq:1_M} describes the coupling between neighboring waveguides, with $C_j \equiv f_j C_0$ the linear coupling constant between waveguides $j$ and $j+1$, $C_0$ a characteristic coupling strength, and $f_j$ a real scaling factor. The coupling constants are usually fixed once the ANW is fabricated. However, it is possible to design and fabricate arrays with adjustable coupling profiles \cite{Yang2024}. Furthermore, a recent work has reported a 2D-programmable waveguide \cite{Onodera2024}. Here, the coupling is considered to be constant along the propagation direction. The second term in Eq. \eqref{Eq:1_M} represents the SPDC process, with $\gamma_j \equiv g \alpha_j$ the nonlinear coupling constant related to the $j$th waveguide, and $g$ a nonlinear constant which determines the strength of the nonlinear process---it is proportional to $\chi^{(2)}$ and the overlap between the pump and signal fields. We define $\vec{\gamma}=g \vec{\alpha}$, which can be conveniently expressed as $\vec{\gamma}=g ||\vec{\alpha}|| \hat{\eta}$, with $||\vec{\alpha}||$ the norm of $\vec{\alpha}$, $\hat{\eta}$ a unit vector, and $\eta_i$ the $i$th component of $\hat \eta$.
In general, $g$ and $C_0$ are frequency-dependent, and we take them as real (note that this assumption does not affect the underlying physics of the problem).

In this work, we focus on the case of identical waveguides ($\beta_j(\omega)\equiv \beta(\omega)$ for any $j$), and assume the phase-matching condition for the degenerate SPDC process is satisfied within the propagation region---that is, $\Delta \beta=\beta(\omega_\mathrm{p})-2 \beta(\omega_\mathrm{s})=0$. Hence, Eq. \eqref{Eq:1_M} simplifies to 
\begin{equation} \label{Eq:2_M_simple}
\hat{M} =\hbar \sum_{j=1}^{N-1} C_j \hat{A}_{j+1}\hat{A}_j^\dagger +\hbar \sum_{j=1}^N \gamma_j (\hat{A}_j^\dagger)^2 +\mathrm{h.c.}
\end{equation}

\section{\label{sec:level3}Analytic Solution\protect}

In the ANW, the output state $|\Psi(z)\rangle$ at a propagation distance $z$ can be obtained by
\begin{equation}
\label{Eq:3_evolution}
|\Psi(z)\rangle = \hat{O} \left[ e^{\frac{i}{\hbar} \int^z_0 \hat{M} \mathrm{d}z'} \right] |\Psi(0)\rangle,
\end{equation}
with $|\Psi(0)\rangle$ the input state at $z=0$, which in our case corresponds to the vacuum state $|0\rangle$. Here, $\hat{O}$ denotes a space-ordering operator, which is required because, in general, $\hat{M}$ does not commute with itself at different positions. However, the space-ordering effects can be neglected in the low-injection regime \cite{Quesada2014}, which is the case in this work. As a result, we can write
\begin{equation}
\label{Eq:4_evolution_b}
|\Psi(z)\rangle =  e^{\frac{i}{\hbar} \int^z_0 \hat{M} \mathrm{d}z'} |\Psi(0)\rangle.
\end{equation}
Next, the problem is transformed into a convenient basis. We begin by expressing the momentum operator in the following compact form
\begin{equation}
\label{Eq:5_M_op_matriz_a}
\hat{M} = \hbar(A_{\mathrm{D}}^T\Omega A + g ||\vec{\alpha}|| A^T_{\mathrm{D}}{P} A_{\mathrm{D}} +  g ||\vec{\alpha}|| A^T{P}^\dagger A),
\end{equation}
with $ A \equiv (\hat{A}_1,\dots,\hat{A}_N)^T$, $ A_{\mathrm{D}} \equiv (\hat{A}^\dagger_1,\dots,\hat{A}^{\dagger}_N)^T$, ${P}\equiv \mathrm{diag}(\eta_1,\dots,\eta_N)$, and $\Omega$ a real, symmetric, and tridiagonal matrix of the form

\begin{equation}
\label{Eq:6_Omega_form}
\Omega=\begin{pmatrix}
0   & C_1 &         &        &         \\
C_1 &  0  & C_2     &        &         \\
    & C_2 & \ddots  & \ddots &         \\ 
    &     & \ddots  & \ddots & C_{N-1} \\ 
    &     &         &  C_{N-1}   & 0   \\ 
\end{pmatrix}.
\end{equation}

Let us consider now the linear supermode basis $\{\hat{b}_j\}$ corresponding to the propagation eigenmodes of the ANW without the nonlinear effects \cite{Kapon1984}---for example, taking $g=0$. The transformation from the individual-mode $\{\hat{A}_j\}$ to the supermode basis $\{\hat{b}_j\}$ can be expressed as follows
\begin{equation}
\label{Eq:7_A_to_b}
 b = S  A,
\end{equation}
with $ b \equiv (\hat{b}_1,\dots,\hat{b}_N)^T$, and $S$ an orthogonal transformation matrix of the form
\begin{equation}
\label{Eq:8_S}
S = \begin{pmatrix}
\vec{v}_1^{\hspace{0.1cm}T}\\
\vdots\\
\vec{v}_N^{\hspace{0.1cm}T}
\end{pmatrix},
\end{equation}
with $\vec{v}_k$ the $k$th normalized eigenvector of $\Omega$ associated with the eigenvalue $\lambda_k$. Due to the nature of $\Omega$ stated above, its eigenvalues are real, and it possesses a set of $N$ orthogonal eigenvectors, which can be defined as real \cite{Efremidis2005}. In this work, the indexing of the eigenvalues was chosen in such a way that $\lambda_1>\dots>\lambda_k>\dots>\lambda_N$. 

Then, by means of the relation given in Eq. \eqref{Eq:7_A_to_b}, the momentum operator $\hat{M}$ defined in Eq. \eqref{Eq:1_M} can be written in the supermode basis as follows
\begin{equation}
\hat{M} = \hat{M}_{\mathrm{L}}+\hat{M}_{\mathrm{NL}},\label{Eq:9_MLS}
\end{equation}
with 
\begin{equation}
\hat{M}_{\mathrm{L}} = \hbar \sum^N_{j=1} \lambda_j \hat{b}^{\dagger}_j \hat{b}_j \label{Eq:10_ML_b},
\end{equation}
the term corresponding to the linear coupling effects (which is diagonal in the supermode basis), and
\begin{equation}
\hat{M}_{\mathrm{NL}} = \hbar \left\{\sum^N_{n,m=1} \sum^N_{j=1} \gamma_j S_{nj} S_{mj}  \hat{b}^{\dagger}_n \hat{b}^{\dagger}_m + \mathrm{h.c.}, \right\}\label{Eq:11_MNL_b}
\end{equation}
the term related to the nonlinear coupling. The Eqs. \eqref{Eq:9_MLS}, \eqref{Eq:10_ML_b}, and \eqref{Eq:11_MNL_b} illustrate the main characteristics of light propagation in ANW. That is, the first, and second terms on the right-hand side of Eq. \eqref{Eq:9_MLS} represent, respectively, the linear coupling between waveguides described as a superposition of supermodes, and the coupling between supermodes due to the nonlinear effects. 

It is possible to obtain the following solution for the output state in the supermode basis assuming the case of biphoton states (the detailed derivation can be found in Appendix \ref{Appendix_A}) 
\begin{equation}
|\Psi(z)\rangle_\mathrm{b}
= \frac{1}{\sqrt{\mathcal{N}}} \left[1+\sum^N_{n,m=1} \tilde{Q}_{nm}(z) \hat{b}^{\dagger}_n \hat{b}^{\dagger}_m  \right]  |0\rangle_\mathrm{b},
\label{Eq:12_sol_phi_z_1_order_an} 
\end{equation}
with the subscript $\mathrm{b}$ denoting that the eigenstates are in the supermode basis, and $\mathcal{N}$ an appropriate normalization constant. In matrix form, $\tilde{Q}(z)$ is given by the product of a scalar factor $\delta(z) = i z g ||\vec{\alpha}||$, and the pump $\tilde{P}$ and phase-matching $\tilde{T}(z)$  matrices as
\begin{equation}
\tilde{Q}(z)=\delta(z) \tilde{P}\odot \tilde{T}(z),
\label{Eq:13_Qtilde_Def} 
\end{equation}
with $\odot$ corresponding to a Hadamard product (element-wise) and entries given by
\begin{subequations}
\begin{align} 
\tilde{P}_{nm} &= \sum^N_{j=1} |\eta_j| e^{i\phi_j} S_{nj} S_{mj}, \label{Eq:14A_P_tilde} \\
\tilde{T}_{nm}(z) &=e^{i(\lambda_n+\lambda_m)z/2}\mathrm{sinc}\left[(\lambda_n+\lambda_m)z/2\right].\label{Eq:14B_T_tilde}
\end{align}
\end{subequations}
Due to the degeneracy of the photons, we can write
\begin{eqnarray}
|\Psi(z)\rangle_\mathrm{b}
&=&  \frac{1}{\sqrt{\mathcal{N}}} \left\{ |0\rangle_\mathrm{b} +  \sum^N_{n=1} \tilde{K}_{nn}(z) |\dots,2_n,\dots\rangle_{\mathrm{b}} \right. \nonumber \\
& &\left.+ \sum^{N-1}_{n=1} \sum^N_{m=n+1} \tilde{K}_{nm}(z) |\dots,1_n,\dots,1_m,\dots\rangle_{\mathrm{b}}\right\}, \nonumber\\ 
\label{Eq:15_sol_phi_z_1_order_sm} 
\end{eqnarray}
with $\tilde{K}(z)$ given by
\begin{equation}
\tilde{K}(z)=  D \odot\tilde{Q}(z) = \delta(z) D \odot \tilde{{P}}\odot \tilde{T}(z),\label{Eq:16_K_tilde_matrix}
\end{equation}
where $D$ is the degeneracy matrix with $nm$ entries ${D}_{nm} = 2^{1-\delta_{nm}/2}$. $\tilde{K}(z)$ is the supermode joint spatial amplitude (sJSA) matrix and its elements are the amplitudes of probability to measure specific spatial correlations in the supermode basis---in the same way as in the frequency domain we have a joint spectral amplitude \cite{Grice1997}.

Note that every matrix $\tilde{\mathcal{O}}$ in the supermode basis is related to an individual-mode matrix $\mathcal{O}$ through  $\tilde{\mathcal{O}}=S \mathcal{O} S^{T}$. Thus, once we have this simple solution in the supermode basis, it is possible to return to the individual-mode basis and obtain the corresponding output state with the form (see details in Appendix \ref{Appendix_A})
\begin{equation}
|\Psi(z)\rangle = \frac{1}{\sqrt{\mathcal{N}}} \left[1+\sum^N_{k,q=1} Q_{kq}(z) \hat{a}^{\dagger}_k \hat{a}^{\dagger}_q  \right]  |0\rangle, \label{Eq:17_sol_phi_z_1_A_im} 
\end{equation}
with 
\begin{equation}
Q(z)=S^T\tilde{Q}(z)S. \label{Eq:18_Q_transf} 
\end{equation}
Considering the degeneracy of the photons, we obtain
\begin{eqnarray}
|\Psi(z)\rangle
&=&\frac{1}{\sqrt{\mathcal{N}}}\left\{ |0\rangle +  \sum^N_{q=1} K_{qq}(z) |\dots,2_q,\dots\rangle \right. \nonumber \\
& &\left. + \sum^{N-1}_{k=1} \sum^N_{q=k+1} K_{kq}(z) |\dots,1_k,\dots,1_q,\dots\rangle \right\}, \nonumber\\ 
\label{Eq:19_sol_phi_z_1_A_im} 
\end{eqnarray}
with ${K}_{kq}(z)$ the $kq$ entry of the individual-mode joint spatial amplitude (iJSA) matrix of the created photons, given by
\begin{align}
{K}(z) &= D \odot Q(z).
\label{Eq:20_K_matrix}
\end{align}
Importantly, the above equation shows that calculating the supermodes of the system simplifies notably the calculation of the iJSA matrix, as it is equivalent to a change of basis of $\tilde{Q}(z)$ multiplied element-wise by the degeneracy matrix $D$. Interestingly, the phases in the sJSA matrix are important because they could create interferences in the entries of the iJSA matrix.

In summary, the biphoton amplitudes of probability in the supermode and individual-mode bases are given by the entries of the matrices $\tilde{K}(z)$ and $K(z)$, respectively. In the supermode basis, element-wise multiplication of the degeneracy, pump, and phase-matching matrices ensures scalability of numerical simulations at the cost of calculating the eigenvectors and eigenvalues of the tridiagonal matrix defined in Eq. \eqref{Eq:6_Omega_form}. Moreover, as the photons are collected from each single waveguide, it is suitable to solve the system at the level of the individual-mode basis, carried out through a simple change of basis by means of Eqs. \eqref{Eq:18_Q_transf} and \eqref{Eq:20_K_matrix}.

In general, the procedure for obtaining an analytic solution for an array of arbitrary size and coupling profile could be decomposed in the following steps. First, calculate the matrices describing the pump profile in the supermode basis ($\tilde{P}$) and the phase-matching ($\tilde{T}$). Second, multiply element-wise $\tilde{P}$ and $\tilde{T}$ according to Eq. \eqref{Eq:16_K_tilde_matrix}. Finally, return to the individual-mode space applying a change of basis. In fact, for certain cases, the eigenvectors $\vec{v}_n$ and eigenvalues $\lambda_n$ can be expressed analytically. For example, for the homogeneous, parabolic, and Glauber-Fock (square-root) coupling profiles the eigenvalues present an analytic form and the entries of the eigenvectors ($S_{nm}$) can be written in terms of orthonormal Chebyshev \cite{Meng2004}, Krawtchouk \cite{Bosse2017,PerezLeija2013}, and Hermite polynomials \cite{PerezLeija2010,RodriguezLara2011}, respectively. In these instances, the final solution could be obtained without the need of implementing any numerical method. Comments on the scalability of the calculation of the output state can be found in Appendix \ref{Appendix_scalability}, see Fig. \ref{fig:benchmarking_direct}.

\section{\label{sec:level4}General properties of the solution\protect}

\subsection{\label{sec:sublevel41} General properties of the supermodes}

Light propagation in an ANW is, to a large extent, governed by the characteristics of the eigenvalues and eigenvectors of the matrix $\Omega$. The nature of $\Omega$, as defined in Eq. \eqref{Eq:6_Omega_form}, imposes special properties on its eigenvalue spectrum $\{\lambda_k\}$, and the corresponding set of eigenvectors $\{\vec{v}_k\}$ \cite{Efremidis2005}. Some of these are stated below.

First, the set of eigenvectors is orthonormal, thus
\begin{eqnarray}
\sum_{k=1}^N S_{nk}S_{mk}&=&\delta_{nm}.
\label{Eq:21_orthonormal_S_A}
\end{eqnarray}
Second, the following relations are satisfied between eigenvalues, and eigenvectors corresponding to symmetric pairs of supermodes---that is, with indexes symmetric with respect to the center of the set $\{1,\dots,N\}$ (for example, $k$ and $N+1-k$),
\begin{eqnarray}
\lambda_{N+1-k} &=& -\lambda_k, \label{Eq:22_lambda_sym_pair}\\
S_{N+1-k,m}&=&(-1)^{m+1}S_{km}. \label{Eq:23_S_sym_pair}
\end{eqnarray}
In addition, Eqs. \eqref{Eq:21_orthonormal_S_A} and \eqref{Eq:23_S_sym_pair} can be combined to obtain the relations
\begin{eqnarray}
\sum_{k=1}^N(-1)^{k+1} S_{nk} S_{mk}  &=&  \delta_{n,N+1-m},\label{Eq:24_S_relation_A}\\
\sum_{k=1}^{\lceil N/2 \rceil} S_{n,2k-1} S_{m,2k-1}  &=&  \frac{1}{2}(\delta_{nm}+\delta_{n,N+1-m}), \label{Eq:25_S_relation_B}\\
\sum_{k=1}^{\lfloor N/2 \rfloor} S_{n,2k} S_{m,2k}  &=&  \frac{1}{2}(\delta_{nm}-\delta_{n,N+1-m}), \label{Eq:26_S_relation_C}
\end{eqnarray}
with ${\lceil x \rceil}$ (${\lfloor x \rfloor}$) denoting the ceiling (floor) function---that is, the smallest (largest) integer greater (smaller) than or equal to $x$. As we show in Sec. \ref{sec:sublevel51}, as a consequence of these relations, a suitable pump shaping can lead to a considerable simplification of the pump matrix $\tilde{P}$, and thus the solution of the system.

\subsection{\label{sec:sublevel42}Supermode phase-matching}
The entries of the phase-matching matrix $\tilde{T}_{nm}$ and hence the amplitudes of the output biphoton eigenstates in the supermode basis $\tilde{K}_{nm}(z)$ are proportional to $\mathrm{sinc}[(\lambda_n+\lambda_m)z/2]$---see Eqs. \eqref{Eq:14B_T_tilde} and \eqref{Eq:16_K_tilde_matrix}. This is a phase-matching factor which determines the eigenstates (in the supermode basis) that become dominant in the output state as the propagation distance $z$ increases. That is, those with $|\lambda_n+\lambda_m|$ as small as possible. Note that this is different from the phase-matching condition for the degenerate SPDC process ($\Delta \beta=\beta(\omega_\mathrm{p})-2 \beta(\omega_\mathrm{s})=0$), which was assumed in this work. Particularly, the limiting case is obtained when $\lambda_n+\lambda_m=0$, which is a condition satisfied by any symmetric pair of supermodes (with $m=N+1-n$), see Eq. \eqref{Eq:22_lambda_sym_pair}. Thus, the dominant eigenstates are those with form $|\dots,1_n,\dots,1_{N+1-n},\dots\rangle_{\mathrm{b}}$. In fact, this dominance increases as $C_0 z$---note that the eigenvalues $\lambda_j$ are proportional to $C_0$. As a result, the effect commented above would be observed at small distances for large couplings, and the other way around.

In addition, an ANW with odd size $N$ possesses a supermode with a vanishing eigenvalue, which we denote here as the zero supermode. According to the relation given in Eq. \eqref{Eq:22_lambda_sym_pair}, this corresponds to the central supermode associated with the index $(N+1)/2$. Evidently, the condition $\lambda_n+\lambda_m=0$ is satisfied for the central supermode when $n=m=(N+1)/2$, and this is related to an eigenstate of the form $|\dots,2_{(N+1)/2},\dots \rangle_{\mathrm{b}}$. In summary, as the propagation increases the output state in the supermode basis will be dominated by the kind of eigenstates which satisfy the phase-matching condition described above. That is, $|\dots,1_n,\dots,1_{N+1-n},\dots\rangle_{\mathrm{b}}$ (in general) and $|\dots,2_{(N+1)/2},\dots \rangle_{\mathrm{b}}$ (for odd arrays). Note that for certain injection profiles some of the pump matrix elements $\tilde{P}_{nm}$ may vanish due to destructive interferences---see Eq. \eqref{Eq:14A_P_tilde}. In that case, the corresponding amplitude $\tilde{K}_{nm}(z)$ becomes null for any $z$. Thus, the related eigenstate will not participate in the evolution---even if they satisfy the phase-matching condition described above.

\section{\label{sec:level5} Application examples\protect}

\subsection{\label{sec:sublevel51}Simple solutions: Symmetric injection profiles}

In the following, we show that a special symmetric injection profile leads to simple analytic solutions, which are valid for arrays of any number of waveguides and any coupling profile. As mentioned above, it is possible to choose certain symmetric injection profiles which, due to the relations in Eqs. \eqref{Eq:21_orthonormal_S_A}-\eqref{Eq:26_S_relation_C}, give rise to a significant simplification of the analytic expression for the output state \cite{Barral2020}. The most notable simplification is obtained when the injection is identical for all the odd (even) waveguides. That is, with nonlinear coupling constants given by
\begin{equation}
\eta_j=\begin{cases}
			|\eta_1| e^{i\phi_1}, & \text{if $j$ is odd,}\\
            |\eta_2| e^{i\phi_2}, & \text{if $j$ is even.}
		 \end{cases}
\label{Eq:27_eta_cases}
\end{equation}
In this case, we can use Eqs. \eqref{Eq:25_S_relation_B} and \eqref{Eq:26_S_relation_C} together with Eq. \eqref{Eq:16_K_tilde_matrix} to obtain
\begin{align}
    \tilde{K}_{n,m}(z) =& F_{\mathrm{A}} \delta(z) D_{n, n}  \tilde{T}_{n, n} \delta_{n,m} \nonumber \\
    &+ F_{\mathrm{B}} \delta(z) D_{n, (N+1-n)} \delta_{n,(N+1-m)} 
    \label{Eq:24_Knm_simple},
\end{align}
with $ F_\mathrm{A/B}  \equiv \frac{1}{2} (|\eta_1|e^{i\phi_1} \pm |\eta_2|e^{i\phi_2})$, and where we have employed $\tilde{T}_{n,(N+1-n)} = 1$. Thus, the sJSA matrix is composed by a diagonal and an antidiagonal term with amplitudes proportional to $F_\mathrm{A}$ and $F_\mathrm{B}$, respectively.

\subsubsection{\label{sec:subsublevel511}Simple solution in the supermode basis}

The result from Eq. \eqref{Eq:24_Knm_simple} can be plugged into Eq. \eqref{Eq:15_sol_phi_z_1_order_sm} to obtain the following simplified solution in the supermode basis
\begin{eqnarray}
|\Psi(z)\rangle_\mathrm{b}
 &=& \frac{1}{\sqrt{\mathcal{N}}} \Bigg\{ |0\rangle_\mathrm{b} \nonumber\\
 && + i z \sqrt{2} F_{\mathrm{A}} \sum^N_{n=1} e^{i \lambda_n z} \mathrm{sinc}(\lambda_n z) |\dots,2_n,\dots\rangle_{\mathrm{b}}  \nonumber \\
 &&+2 i z F_{\mathrm{B}} \sum_{n=1}^{\lfloor N/2 \rfloor} |\dots,1_n,\dots,1_{N+1-n},\dots\rangle_{\mathrm{b}} \Bigg\}. \nonumber \\ \label{Eq:29_sol_Psi_simple} 
\end{eqnarray}

The output state given in Eq. \eqref{Eq:29_sol_Psi_simple} presents interesting physical properties. First, the eigenstates with mixed supermodes of the form $|\dots,1_n,\dots,1_m,\dots\rangle_{\mathrm{b}}$ ($n \neq m$) show nonvanishing amplitudes only for pairs of symmetric supermodes---with $m=N+1-n$. Second, the antibunching condition in the supermode basis is obtained when $F_\mathrm{A}=0$, which is satisfied when $|\eta_1|=|\eta_2|$ and $\phi_1=\phi_2+2 \pi (m+1/2)$, with $m \in \mathbb{Z}$---note we are not considering the trivial solution $|\eta_1|=|\eta_2|=0$. Third, the bunching case in the supermode basis occurs when $F_\mathrm{B}=0$. That is, for $|\eta_1|=|\eta_2|$ and $\phi_1=\phi_2+2 \pi m$, with $m \in \mathbb{Z}$. In the resulting bunching output state, the amplitudes of the eigenstates $|\dots,2_n,\dots\rangle_{\mathrm{b}}$ are proportional to the factor $\mathrm{sinc}(\lambda_n z)$. Thus, in an odd array, all the probabilities associated with the eigenstates decrease with the propagation $z$ except that of $|\dots,2_{(N+1)/2},\dots\rangle_{\mathrm{b}}$---corresponding to the zero supermode---because it is the only one that satisfies the phase-matching condition described in section \ref{sec:sublevel42}. Hence, for a sufficiently large $z$, the bunching output state in an odd array would practically resemble $|\dots,2_{(N+1)/2},\dots\rangle_{\mathrm{b}}$---that is, two photons in the zero supermode.   

In principle, it is experimentally feasible to move from the individual-mode to the supermode basis by means of linear transformations \cite{Reck1994}---which can be implemented with off-the-shelf optical elements. As a result, the state presented in Eq. \eqref{Eq:29_sol_Psi_simple} could work as a $N$-mode state from which bunching, and antibunching could be selected only by adjusting the relative phase of the injected fields.  

\subsubsection{\label{sec:subsublevel512}Simple solution in the individual-mode basis}

The output state in the individual-mode basis equivalent to the expression given in Eq. \eqref{Eq:29_sol_Psi_simple} can be obtained by transforming the corresponding matrix $\tilde{Q}(z)$ according to Eq. \eqref{Eq:18_Q_transf} and applying Eq. \eqref{Eq:20_K_matrix}. This gives the related iJSA matrix, whose entries can be plugged into Eq. \eqref{Eq:19_sol_phi_z_1_A_im} to get
\begin{eqnarray}
|\Psi(z)\rangle &=& \frac{1}{\sqrt{\mathcal{N}}} \Bigg\{ |0\rangle \nonumber\\ & &
+ i \sqrt{2} \sum^N_{q=1} \left[ F_\mathrm{A} \left(\sum_{n=1}^N S_{nq}^2\tilde{T}_{nn} \right) + (-1)^{q+1} z F_\mathrm{B} \right] \nonumber \\
& &\,\,\,\times |\dots,2_q,\dots\rangle \nonumber \\
& &+ i 2 \sum^{N-1}_{k=1} \sum^N_{q=k+1} F_\mathrm{A} \sum_{n=1}^N S_{nk}S_{nq} \tilde{T}_{nn}  \nonumber \\
& &\,\,\,\times |\dots,1_k,\dots,1_q,\dots\rangle  \Bigg\}. 
\label{Eq:30_output_simple_physical} 
\end{eqnarray}
From Eq. \eqref{Eq:30_output_simple_physical}, we can see the condition for bunching is $F_{\mathrm{A}}=0$---that is, $|\eta_1|=|\eta_2|$, and $\phi_1=\phi_2+2\pi(m+1/2)$, with $m\in \mathbb{Z}$. Interestingly, this corresponds to the antibunching condition in the supermode basis. On the other hand, no simple general condition can be obtained for achieving a state in which the two photons can not be detected at the same waveguide---that is, without eigenstates of the form $|\dots,2_q,\dots\rangle$. In this regard, each particular case should be addressed individually.

It is important to remark that the properties stated here, in relation to the injection profile given in Eq. \eqref{Eq:27_eta_cases}, are general for ANW of any size $N$ and coupling profile $\{C_j\}$.

\subsection{\label{sec:sublevel52}Solution for two waveguides with a homogeneous coupling profile}

Here, we apply the solution presented in Sec. \ref{sec:level3} to the case of a two-waveguide array. In this instance, the coupling profile consists of only one element. That is, $C_1$, and we define $C_1=C_0$. Note that this problem can be considered as a special case of a homogeneous array of $N$ waveguides (taking $N=2$). The transformation matrix $S$ takes a simple form when the array is homogeneous---that is, when all the elements of the coupling profile $\{C_j\}$ are equal to $C_0$. In this case, the elements of the matrix $S$ are \cite{Meng2004}
\begin{equation}
    S_{nm}=S_{mn} = \frac{\displaystyle \sin\left(\frac{nm\pi}{N+1}\right)}{\displaystyle \sqrt{\sum_{k=1}^N\sin^2\left(\frac{kn\pi}{N+1}\right)}},
    \label{Eq:31_homogeneous_eigenvectors}
\end{equation}
while its eigenvalues are given by
\begin{equation}
    \lambda_n = 2C_0\cos\left(\frac{n\pi}{N+1}\right).
    \label{Eq:32_homogeneous_eigenvalues}
\end{equation}
In terms of the entries of the iJSA matrix, the output state for the case of a two-waveguide homogeneous array is given by
\begin{subequations}
\begin{align}
    K_{11} &= ig ||\vec{\alpha}||\frac{2zC_0(\eta_1 - \eta_2) + (\eta_1 + \eta_2)\sin(2C_0z)}{2\sqrt{2}C_0}, \label{eq:33_K_2waveguides_A}\\
    K_{12} &= K_{21} = g ||\vec{\alpha}||\frac{(\eta_1 + \eta_2)[\cos(2C_0z)-1]}{2C_0}, \label{eq:33_K_2waveguides_B}\\
    K_{22} &= ig ||\vec{\alpha}||\frac{2zC_0(\eta_2 - \eta_1) + (\eta_1 + \eta_2)\sin(2C_0z)}{2\sqrt{2}C_0}. \label{eq:33_K_2waveguides_C}
\end{align}
\end{subequations}
In fact, this solution could also be obtained directly from Eq. \eqref{Eq:30_output_simple_physical} for the particular case of a two-waveguide homogeneous array. It is worth mentioning that the expressions given in Eqs. \eqref{eq:33_K_2waveguides_A}-\eqref{eq:33_K_2waveguides_C} are consistent with the results found by Belsley et al. \cite{Belsley2020}. It is important to remark that the coefficients presented in \cite{Belsley2020} correspond to the entries of $Q(z)$, which is linked to $K(z)$ by means of Eq. \eqref{Eq:20_K_matrix}.

It is interesting to consider the conditions that must be satisfied to obtain a diagonal state $|B\rangle_\pm=\tfrac{1}{\sqrt{2}}(|2_1,0_2\rangle\pm|0_1,2_2\rangle)$ or an antidiagonal state $|A\rangle=|1_1,1_2\rangle$. In fact, the experimental generation of a diagonal state for the case of a two-waveguide array is described and implemented by Kruse et al. \cite{Kruse2015}.

The diagonal case is achieved when $K_{12}=0$. This gives rise to two possibilities. First, $|\eta_1|=|\eta_2|$, and $\phi_1=\phi_2+2\pi(m+1/2)$ with $m\in \mathbb{Z}$---which is identical to the condition found in Sec. \ref{sec:subsublevel512}. Second, $z = \pi m/C_0$, with $m\in \mathbb{Z}$. We can see from Eqs. \eqref{eq:33_K_2waveguides_A} and (\ref{eq:33_K_2waveguides_C}) that $K_{11}=-K_{22}$. This implies that regardless which of the previous conditions is satisfied, only the antisymmetric diagonal state $|B\rangle_-$ is achievable. Also, Fock states in individual modes are not possible, such as $|2_1,0_2\rangle$ or $|0_1,2_2\rangle$ . It is worth noting that the first condition offers the advantage of being position independent, while the second one is compatible with any injection profile. 

On the other hand, an antidiagonal state requires $K_{11}=K_{22}=0$. This is only possible when the conditions $|\eta_1|=|\eta_2|$, $\phi_1-\phi_2=2\pi m$, and $z = \pi(2m+1)/(2C_0)$ with $m\in \mathbb{Z}$ are simultaneously satisfied. Therefore, the realization of an antidiagonal state is subject to a position-dependent constraint.

\subsection{\label{sec:sublevel53}Solution for three waveguides with a homogeneous coupling profile}

Here, we provide the form taken by the analytic solution presented in Sec. \ref{sec:level3} for the case of a three-waveguide homogeneous array. In this situation, the coupling profile is given by $\{C_1,C_2\}$ with $C_1=C_2=C_0$, and the elements of the transformation matrix $S$ and the eigenvalues $\lambda_n$ satisfy Eqs. \eqref{Eq:31_homogeneous_eigenvectors} and \eqref{Eq:32_homogeneous_eigenvalues}, respectively. Accordingly, the relevant entries of the iJSA matrix are the following 

\begin{subequations}
\begin{eqnarray}
    K_{11}  &=&  \frac{ig ||\vec{\alpha}||}{16\sqrt{2}C_0} [4C_0z(3\eta_1- 2 \eta_2 + 3\eta_3) \nonumber\\
            &+& 8\sqrt{2}(\eta_1 - \eta_3)\sin(\sqrt{2}C_0 z) \nonumber\\
            &+&  \sqrt{2}(\eta_1 + 2\eta_2 + \eta_3) \sin( 2\sqrt{2} C_0 z)], \label{Eq:34_K_3w_A}\\
    K_{12}  &=&K_{21}=\frac{-g ||\vec{\alpha}||}{2C_0} \sin^2\left(\frac{C_0 z}{\sqrt{2}}\right)[3\eta_1 + 2\eta_2 -\eta_3\nonumber\\
            &+&(\eta_1 + 2\eta_2 +\eta_3)\cos(\sqrt{2}C_0 z) ], \label{Eq:34_K_3w_B}\\
    K_{13}  &=&K_{31}=ig ||\vec{\alpha}||\frac{\eta_1+2\eta_2+\eta_3}{16C_0}[-4C_0z\nonumber\\
            &+&\sqrt{2}\sin(2\sqrt{2}C_0z)], \label{Eq:34_K_3w_C}\\
    K_{22}  &=& \frac{ig ||\vec{\alpha}||}{8\sqrt{2}C_0}[-4C_0z(\eta_1-2\eta_2+\eta_3)\nonumber\\
            &+&\sqrt{2}(\eta_1+2\eta_2+\eta_3)\sin(2\sqrt{2}C_0z)], \label{Eq:34_K_3w_D}\\
    K_{23}  &=&K_{32}=\frac{-g ||\vec{\alpha}||}{2C_0} \sin^2\left(\frac{C_0 z}{\sqrt{2}}\right)[-\eta_1+2\eta_2+3\eta_3\nonumber\\
            &+&(\eta_1+2\eta_2+\eta_3)\cos(\sqrt{2}C_0z)], \label{Eq:34_K_3w_E}\\
    K_{33}  &=& \frac{ig ||\vec{\alpha}||}{16\sqrt{2}C_0} [4C_0z(3\eta_1- 2 \eta_2 + 3\eta_3) \nonumber\\
            &+& 8\sqrt{2}(\eta_3 - \eta_1)\sin(\sqrt{2}C_0 z) \nonumber\\
            &+&  \sqrt{2}(\eta_1 + 2\eta_2 + \eta_3) \sin( 2\sqrt{2} C_0 z)]. \label{Eq:34_K_3w_F}
\end{eqnarray}
\end{subequations}
The elements of $K(z)$ are also consistent with those found by Belsley et al. \cite{Belsley2020}. Furthermore, we point out that the solutions for injection profiles with $\eta_1=\eta_3$ and arbitrary $\eta_2$---which are of the kind described by Eq. \eqref{Eq:27_eta_cases}---are contained in Eq. \eqref{Eq:30_output_simple_physical}. We can study again the requirements for obtaining diagonal and antidiagonal states. Now, three equations must be simultaneously satisfied in order to obtain a diagonal state. That is, $|\eta_1|=|\eta_2|=|\eta_3|$, $\phi_1-\phi_2=2\pi \left(m+\frac{1}{2}\right)$, and $\phi_1-\phi_3 = 2\pi m$, with $m\in \mathbb{Z}$. This result is equivalent to the condition found in Sec. \ref{sec:subsublevel512}, and therefore produces a state with alternating phases---that is, $|B\rangle=\tfrac{1}{\sqrt{3}}\left(|2_1,0_2,0_3\rangle-|0_1,2_2,0_3\rangle+|0_1,0_2,2_3\rangle\right)$.

For an antidiagonal state, there are two possible systems of equations. One of them has no solution, whereas the other has an infinite number of solutions, subject only to the constraints $|\eta_1|=|\eta_3|$, $|\eta_2|=0$, $\phi_1-\phi_3=2\pi \left(m+\frac{1}{2}
\right)$, and $z = \frac{\pi}{\sqrt{2}C_0}m$, with $m\in \mathbb{Z}$. Although this condition is verified at certain positions, it also implies that $K_{13}=0$. Thus, for example, it is not possible to achieve a Dicke state such as $\frac{1}{\sqrt{3}}(|1_1,1_2,0_3\rangle+|0_1,1_2,1_3\rangle+|1_1,0_2,1_3\rangle)$. However, the state $|A\rangle=\frac{1}{\sqrt{2}}(|1_1,1_2,0_3\rangle+|0_1,1_2,1_3\rangle)$ is realizable.

The results presented so far, for homogeneous arrays consisting of two and three waveguides, are consistent with what has been reported elsewhere \cite{Belsley2020}. This is important for checking the validity of our method.

\begin{figure*}[t]
    \includegraphics[scale=1.20]{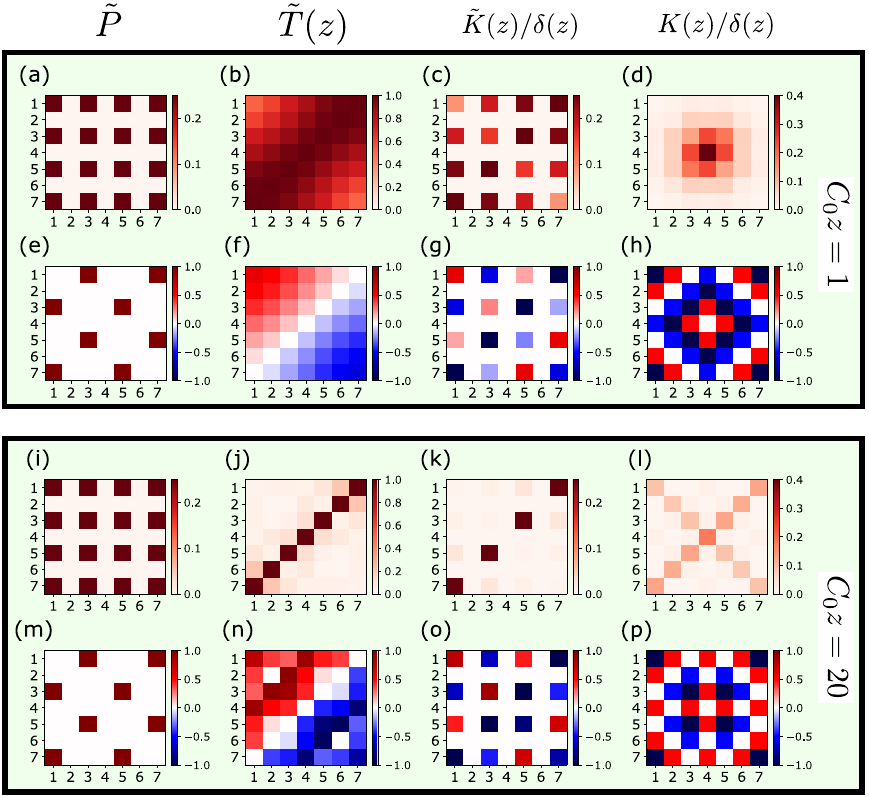}
    \caption{Absolute values---first and third rows---and phases---second and fourth rows---of the entries of the matrices $\tilde{P}$ (first column), $\tilde{T}(z)$ (second column), $\tilde{K}(z)/\delta(z)$ (third column), and $K(z)/\delta(z)$ (fourth column) for the case of an injection only in the center waveguide of a homogeneous ANW consisting of seven waveguides. The results are shown for the propagation distances given by $C_0z=1$ (first and second rows) and $C_0z=20$ (third and fourth rows). For convenience, the phase values are shown in units of $\pi$, and $\tilde{K}(z)$ and $K(z)$ are divided by the scalar factor $\delta(z)=izg||\vec{\alpha}||$. The matrix $\tilde{P}$ does not depend on $z$. Thus, its entries at $C_0z=20$ are the same as those at $C_0z=1$. The vertical, and horizontal axis in each plot correspond to the rows, and columns of the related matrix, respectively.
    }    \label{fig:matrices_inj_center}
\end{figure*}

\subsection{\label{sec:sublevel54}Injection of the center waveguide in an odd symmetric array}

Here, we analyze the properties of the solution in Sec. \ref{sec:level3} for the special case of an injection in the center waveguide of an odd symmetric array, which satisfies the condition $C_j=C_{N+1-j}$ with $j=1,\dots,N$---for example, a homogeneous or parabolic coupling profile. Note that no other specific requirement is asked for the coupling profile apart from the constraint stated above.
If only the waveguide $q$ is injected, the entries of $\tilde{P}$ simplify to $\tilde{P}_{nm}=S_{nq}S_{mq}$, see Eq. \eqref{Eq:14A_P_tilde}. In the following, we consider and odd symmetric array injected in the center waveguide with index $q=(N+1)/2$.

In a symmetric array, $S_{m,q}=(-1)^{m+1} S_{m,N+1-q}$ \cite{Efremidis2005}. Consequently, $S_{m,(N+1)/2}=0$ when $m$ is even. As a result, the even rows and columns of $\tilde{P}$ vanish, and hence those of $\tilde{K}(z)$. Thus, only odd supermodes participate in the evolution, as reported elsewhere \cite{Barral2020}. In fact, only the symmetric pairs fulfilling $\lambda_{n} + \lambda_{N+1-n}=0$ satisfy the phase-matching condition and build up along the propagation. These pairs present a degeneracy for the index $n=(N+1)/2$ corresponding to the zero supermode, that is only excited if this index is odd. These odd supermodes interfere to create the resulting pattern in the individual-mode basis. 

The arguments exposed above are illustrated in Fig. \ref{fig:matrices_inj_center}, which corresponds to the case of a homogeneous seven-waveguide array injected only in the center (waveguide 4). It can be seen in Figs. \ref{fig:matrices_inj_center}(a) and \ref{fig:matrices_inj_center}(i) that the even rows and columns of $\tilde{P}$ are null. Thus, the even supermodes do not participate in the evolution, as depicted in Figs. \ref{fig:matrices_inj_center}(c) and \ref{fig:matrices_inj_center}(k). Interestingly, the zero supermode is not excited here because it is associated with the index 4, which is even. The phases of the nonvanishing entries of $\tilde{P}$ present values of either $0$ or $\pi$, as depicted in Figs. \ref{fig:matrices_inj_center}(e) and \ref{fig:matrices_inj_center}(m). It is clearly observed that the antidiagonal nature of $\tilde{T}(z)$ increases with the propagation---see Figs. \ref{fig:matrices_inj_center}(b), and (j), which depict the absolute value of $\tilde{T}(z)$ at $C_0 z=1$, and $C_0 z=20$, respectively. This is consistent with the symmetric pairs of supermodes becoming dominant with the propagation due to the phase-matching condition. As described in Eq. \eqref{Eq:16_K_tilde_matrix}, $\tilde{K}(z)$ is proportional to an element-wise multiplication of  $\tilde{P}$ and  $\tilde{T}(z)$---see Figs. \ref{fig:matrices_inj_center}(a), \ref{fig:matrices_inj_center}(b), and \ref{fig:matrices_inj_center}(c) corresponding to the propagation distante $C_0 z =1$, and Figs. \ref{fig:matrices_inj_center}(i), \ref{fig:matrices_inj_center}(j), and \ref{fig:matrices_inj_center}(k) associated with $C_0 z =20$. 

The solution in the individual-mode basis is given by $K(z)$. In the beginning of the propagation, the probability of finding the photons is higher in the vicinity of the center waveguide---see Fig. \ref{fig:matrices_inj_center}(d). As the propagation increases, the intensity pattern moves away from the center, and concentrates on the eigenstates related to the diagonal and antidiagonal of $K(z)$---see Fig. \ref{fig:matrices_inj_center}(l). In general, the pattern of the phases of $\tilde{T}(z)$, $\tilde{K}(z)$, and $K(z)$ varies during the evolution---compare Figs. \ref{fig:matrices_inj_center}(f), \ref{fig:matrices_inj_center}(g), and \ref{fig:matrices_inj_center}(h) with Figs. \ref{fig:matrices_inj_center}(n), \ref{fig:matrices_inj_center}(o), and \ref{fig:matrices_inj_center}(p), respectively.

The analysis around Fig. \ref{fig:matrices_inj_center} shows that the definition of $\tilde{K}(z)$ as given by Eq. \eqref{Eq:16_K_tilde_matrix} is conceptually useful because it isolates two important aspects that determine the nature of the propagation. That is, the components of the pump in the supermode basis, and the phase-matching effects, which are described by $\tilde{P}$, and $\tilde{T}(z)$, respectively. We show in Appendix \ref{Appendix_B} examples of the entries of the matrices $P$, $\tilde{P}$, $\tilde{T}(z)$, $\tilde{K}(z)$, and $K(z)$ for other injection profiles, including a case considering a parabolic coupling profile.

\section{\label{sec:level6}Inverse problem\protect}

For a small array, it is easy to solve an inverse problem by means of a simple inspection of the analytic solution presented here. That is, to determine the initial conditions required to achieve a target state. Particularly, this is illustrated in Secs. \ref{sec:sublevel52} and \ref{sec:sublevel53}. However, the larger the array, the more intricate the analytic expression becomes.

A powerful advantage of our method is that it is scalable. Thus, it can be used to tackle inverse problems, even for large arrays.  In fact, for a general case with an arbitrary coupling profile, the most demanding task---in the calculation of the output state---would be to find the eigenvalues and eigenvectors of the matrix $\Omega$. However, these can be easily obtained numerically, and even analytically in some cases.

As a proof of concept, we present here an example of an optimization method, based on the analytic solution, to determine the injection profile required to produce an output state with certain desired characteristics. In fact, as it was mentioned in Secs. \ref{sec:sublevel52} and \ref{sec:sublevel53}, not every output state can be generated in an ANW. Indeed, one of the evident restrictions for $\tilde{K}(z)$ is that it must be symmetric due to the degeneracy of the photons, see Eqs. \eqref{Eq:14A_P_tilde}, \eqref{Eq:14B_T_tilde}, and \eqref{Eq:16_K_tilde_matrix}. Hence, in some cases, one can at most try to obtain an approximation of certain output states. Actually, it would be interesting to develope a theoretical understanding of which output states can (or cannot) be produced in an ANW. However, this is out of the scope of this work. 

We introduce the correlation matrix $\Gamma$, associated with an output state $|\Psi \rangle$, with entries defined as $\Gamma_{kq}=|\langle\dots,1_k,\dots,1_q,\dots|\Psi\rangle|^2/\langle\Psi|\Psi\rangle$ for $k \neq q$, and $\Gamma_{kk}=|\langle\dots,2_k,\dots|\Psi\rangle|^2/\langle\Psi|\Psi\rangle$ for the diagonal. In terms of the entries of the matrix $K(z)$, we can write the general expression
\begin{equation}   \Gamma_{kq}=|K_{kq}|^2/\sum_{i,j=1}^N 2^{\delta_{ij} - 1}|K_{ij}|^2. \label{Eq:35_Gamma_general}
\end{equation}
Particularly, for $k \neq q$, $\Gamma_{kq}$ corresponds to the probability of detecting exactly one photon at the waveguide $k$ and the other at the waveguide $q$. In addition, $\Gamma_{kk}$ gives the probability of detecting both photons at the waveguide $k$  \cite{Peruzzo2010}.

\begin{figure}
    \centering    \includegraphics[height=0.9\textheight]{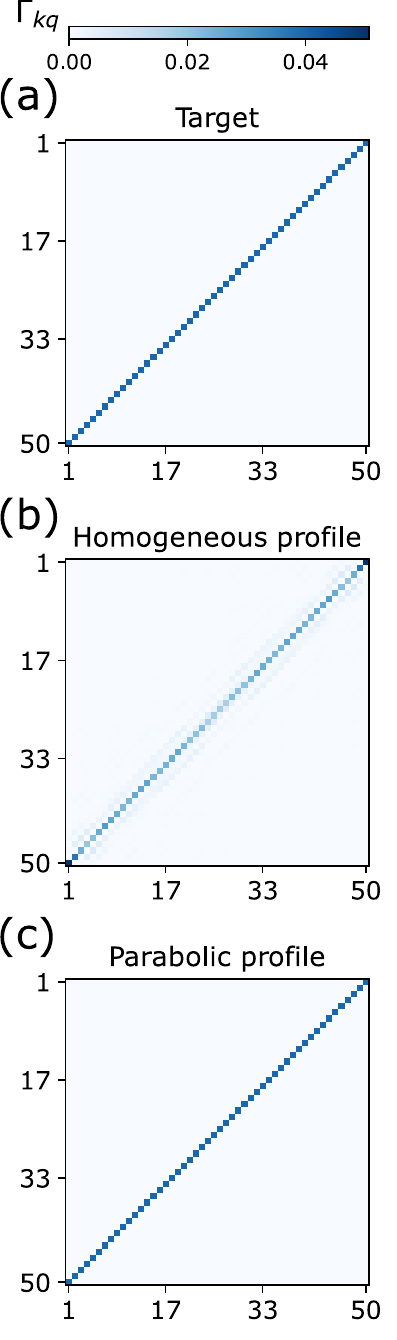}
    \caption{(a) Target correlation matrix for an array of $50$ waveguides. Panels (b), and (c) depict the obtained correlation matrix after optimization using a homogeneous, and a parabolic coupling profile, respectively. The color bar indicates the value of the correlation matrix entries $\Gamma_{kq}$ in all cases. The similarities in (b), and (c) are $\mathcal{S}=0.6320$, and $\mathcal{S}=0.9998$, respectively.}
    \label{fig:optimization}
\end{figure}

Regarding the example presented here, we chose an array of 50 waveguides and a target consisting of an antidiagonal correlation matrix with equal nonzero entries. This target correlation matrix is depicted in Fig. \ref{fig:optimization}(a), and its entries are given by
\begin{equation}
\Gamma'_{qk}=\frac{\delta_{k,N+1-q}}{\lceil N/2 \rceil}.\label{Eq:36_antidiagonal}
\end{equation}

Since our method is compatible with any coupling profile, we compared the results obtained for a homogeneous ($C_j=C_0$), and a parabolic ($C_j= \sqrt{j(N - j)} C_0/2$) coupling profile, and the corresponding results are shown in Figs.  \ref{fig:optimization}(b), and \ref{fig:optimization}(c), respectively. 

\begin{figure*}
    \centering
    \includegraphics[width=\textwidth]{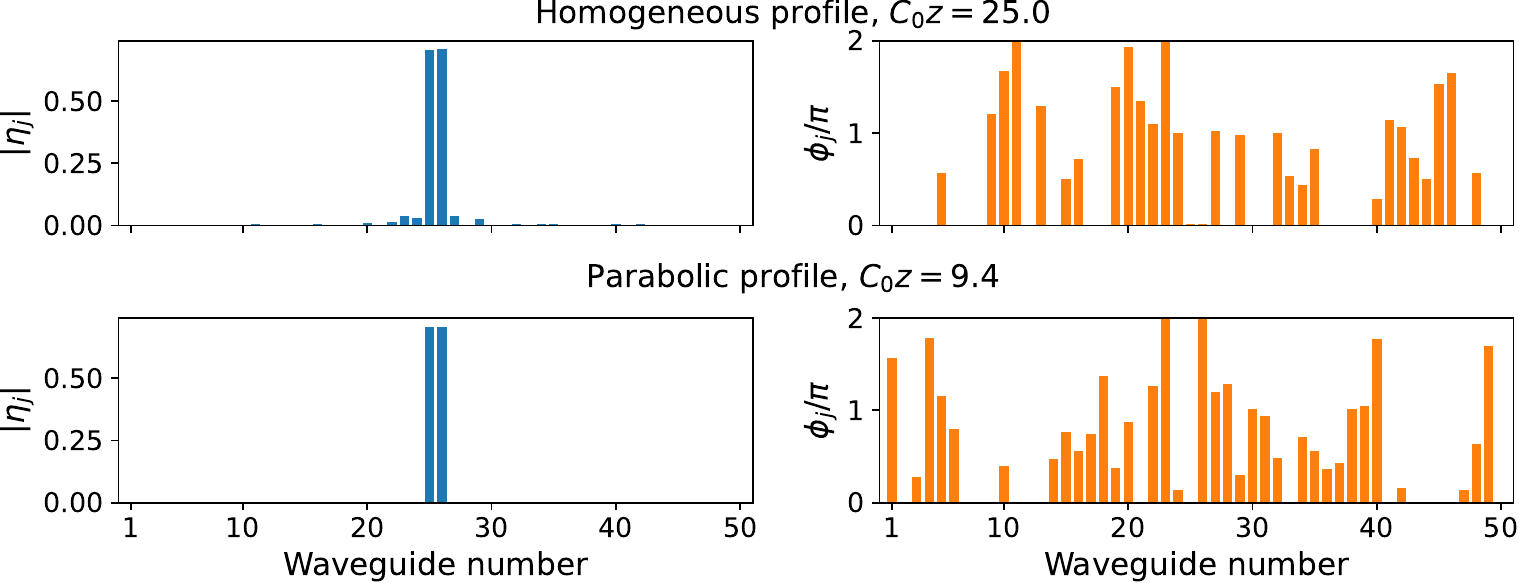}
    \caption{Amplitudes $|\eta_j|$ (left column) and phases $\phi_j$ (right column) of the pump profile that optimized the merit function $\mathrm{MF}$ using a homogeneous (top panel) and parabolic (bottom panel) coupling profile. For each case, the position that optimized the merit function is depicted in the respective panel. Those were $C_0z=25.0$, and $C_0z=9.4$ for the homogeneous, and parabolic coupling profiles, respectively.}
    \label{fig:optimized_pump}
\end{figure*}

In order to quantify the closeness of the optimized result to the target correlation matrix, we used the similarity $\mathcal{S}=(\sum_{i,j=1}^N\sqrt{\Gamma_{ij}\Gamma'_{ij}})^2/[(\sum_{i,j=1}^N\Gamma_{ij})(\sum_{i,j=1}^N\Gamma'_{ij})]$ between two probability distributions $\Gamma$ and $\Gamma'$ \cite{Peruzzo2010}. The similarity $\mathcal{S}$ takes values between $0$ and $1$, with $\mathcal{S}=1$ for a perfect overlap. In general, the closer $\mathcal{S}$ is to $1$ the more similar are $\Gamma$ and $\Gamma'$.

We defined the following merit function aiming to obtain the target correlation matrix described above
\begin{eqnarray}
    \text{MF} &=& \sum_{i,j=1}^N \left(\Gamma_{ij} - \Gamma'_{ij}\right)^2   \label{Eq:37_merit_function}
\end{eqnarray}
with $\mathrm{MF}$ a function of the position and the injection profile.
For each case, the position and injection profile that minimized the merit function defined in Eq. \eqref{Eq:37_merit_function} were determined by means of an optimization algorithm---see more details in Appendix \ref{Appendix_inverse_proc}. The optimal parameters for the analyzed coupling profiles are depicted in Fig. \ref{fig:optimized_pump}---these were the ones which gave the lowest value of the merit function. Interestingly, the injection profiles shown in Fig. \ref{fig:optimized_pump} are along the same line of what has been reported elsewhere regarding the conditions that produce a mostly antidiagonal correlation matrix in ANW \cite{Solntsev2012, Bai2016, Raymond2024}---that is, injecting two consecutive waveguides in the middle of the array with the same amplitudes and phases.

We obtained $\mathcal{S}=0.6320$, and $\mathcal{S}=0.9998$ for the homogeneous, and parabolic coupling profiles, respectively. This is consistent with the results depicted in Fig. \ref{fig:optimized_pump}, in which it is possible to see with the naked eye that the correlation matrix corresponding to the parabolic profile is more similar to the target than that of the homogeneous profile. We emphasize the case presented here is a proof of concept. In fact, in general, we do not discard that better similarity values could be achieved with a more sophisticated optimization algorithm. Note that in the example exposed here we were not only aiming for a pure antidiagonal correlation matrix, but also one with equal antidiagonal entries. The fact that we were able to obtain a set of input parameters that can satisfactorily approximate such a state shows our method is efficient and capable of tackling inversion problems, even for ANW with a large number of waveguides. We recall that, for the ANW with 50 waveguides studied here, the parabolic coupling profile case gives an excellent similarity value ($\mathcal{S}=0.9998$). In fact, we analyzed the same problem presented here for an array of 100 waveguides and obtained $\mathcal{S}=0.99991$ for the case of the parabolic coupling profile---see Appendix \ref{Appendix_C}, where we show this and other examples of optimization problems for some particular target correlation matrices. 

In this section, we looked for a state associated with a specific correlation matrix. A similar approach could be used to study the feasibility of achieving a target state with fixed values of $K_{ij}$ (the biphoton amplitudes of probability in the individual-mode basis), as well as to estimate the corresponding fidelity between the desired and obtained states. In Appendix \ref{Appendix_scalability}, we comment on the scalability of the inversion problem (see Fig. \ref{fig:benchmarking_inverse}).

\section{\label{sec:conclusions}Conclusions\protect}

We used a supermode approach to obtain an analytic expression for the output states in an ANW for the case of degenerate spontaneous parametric down-conversion, and assuming the undepleted pump approximation. The expression developed here is general for ANW of any number of waveguides and coupling profiles. An analytic solution gives physical insight about light propagation in an ANW. A significant advantage of the expression presented here is that it does not require computationally-expensive calculations. In fact, once the eigenvalues and eigenvectors of the coupling matrix are obtained, our method provides an explicit analytic expression, which makes it scalable and highly efficient for numerical implementations. In this regard, we showed a proof-of-concept example of how to use our method to study inversion problems in arrays of nonlinear waveguides. The approach presented in this work could be extended to non-degenerate biphoton states, allowing for the analysis of more general quantum states. In future work, the tools developed here could be used to explore interesting problems related to waveguide arrays, such as topological protection
\cite{BlancoRedondo2018,MedinaDuenas2021,Ren2022,Doyle2022}, propagation regimes and the effects of disorder in quantum walks \cite{Eichelkraut2013,Bai2016,Kokkinakis2024}, and the development of efficient optimization methods to invert arbitrary output states.

\section{\label{sec:data_availability}Data availability\protect}

The data that support the findings of this article are openly available \cite{Repository_ANW}.

\begin{acknowledgments}
 E.A.R.-G. and J.D.-Q. acknowledge support from the Vice-rectory for Research at the University of Costa Rica (project no. C2128). D.B. acknowledges the support from MICINN through the European Union NextGenerationEU recovery plan (PRTR-C17.I1), the Galician Regional Government through “Planes Complementarios de I+D+I con las Comunidades Autónomas” in Quantum Communication.
\end{acknowledgments}
 

\appendix

\section{Analytic Solution} \label{Appendix_A}
We begin by identifying that $\hat{M}_{\mathrm{L}}$, with the form given in Eq. \eqref{Eq:10_ML_b}, corresponds to the free-rotating part of the momentum operator $\hat{M}$ in the linear supermode basis. Hence, it is natural to change to the interaction picture, and obtain
\begin{equation}
|\Psi(z)\rangle_{\mathrm{I}} = \mathrm{exp} \left[\frac{i}{\hbar} \int^z_0 \hat{M}^{\mathrm{(I)}}_\mathrm{NL}(z') \mathrm{d}z'\right]  |0\rangle,\label{AEq:1_PhioutMlInt}
\end{equation}
with $|\Psi(z)\rangle_{\mathrm{I}}=\mathrm{exp}(-i\hat{M}_{\mathrm{L}} z/\hbar)|\Psi(z)\rangle$, and
\begin{eqnarray}
\hat{M}^{\mathrm{(I)}}_\mathrm{NL}(z) &=& \hbar g||\vec{\alpha}||\sum^N_{n,m=1}  \sum^N_{j=1} |\eta_j| e^{i\phi_j} S_{nj} S_{mj} \hat{B}^{\dagger}_n \hat{B}^{\dagger}_m  e^{-i(\lambda_n+\lambda_m)z} \nonumber\\
& &+ \mathrm{h.c.},\label{AEq:2_MLSint}
\end{eqnarray}
with $\hat{B}_k=\hat{b}^{\mathrm{(I)}}_k e^{-i\lambda_k z}$ the slowly-varying amplitude part of the supermodes in the interaction picture, $\hat{b}^{\mathrm{(I)}}_k$. As a result
\begin{align}
\int^z_0 \hat{M}^{\mathrm{(I)}}_\mathrm{NL}(z') \mathrm{d}z'=& \hbar \sum^N_{n,m=1} (-i)\tilde{Q}_{nm}(z) e^{-i(\lambda_n+\lambda_m)z} \hat{B}^{\dagger}_n \hat{B}^{\dagger}_m \nonumber\\
&+ \mathrm{h.c.},\label{AEq:3_Int_MLSint}
\end{align}
with $\tilde{Q}_{nm}$ the $nm$ entry of a symmetric matrix $\tilde{Q}(z)$, which can be expressed as
\begin{equation}
\tilde{Q}(z)=\delta(z)\tilde{P}\odot \tilde{T}(z),\label{AEq:4_K_matrix}
\end{equation}
with $\delta(z)=izg||\vec{\alpha}||$ a scalar factor, and $\odot$ denoting the Hadamard product (element-wise), and the pump $\tilde{P}$ and phase matching $\tilde{T}(z)$ matrices with entries
\begin{align} 
& \tilde{P}_{nm} = \sum^N_{j=1} |\eta_j| e^{i\phi_j} S_{nj} S_{mj}, 
\nonumber \\
& \tilde{T}_{nm}(z)=e^{i(\lambda_n+\lambda_m)z/2}\mathrm{sinc}\left[(\lambda_n+\lambda_m)z/2\right].\label{AEq:5_T_mn}
\end{align}
The result from Eq. \eqref{AEq:3_Int_MLSint} can be plugged into Eq. \eqref{AEq:1_PhioutMlInt}, which gives the following expression for the output state in the interaction picture 
\begin{eqnarray}
|\Psi(z)\rangle_{\mathrm{I}}  = \mathrm{exp} \Biggl\{ i &\Biggl[& \sum^N_{n,m=1} (-i)\tilde{Q}_{nm}(z) \hat{B}^{\dagger}_n \hat{B}^{\dagger}_m e^{-i(\lambda_n+\lambda_m)z}  \nonumber \\
& & + \mathrm{h.c.} \Biggr] \Biggr\}  |0\rangle.\label{AEq:6_sol_phi_z}
\end{eqnarray}
It is possible to return to $|\Psi(z)\rangle_\mathrm{b}$, in the basis $\{\hat{b}_k\}$, by means of the relation 
\begin{eqnarray}
|\Psi(z)\rangle_\mathrm{b}&=&\mathrm{exp}\left(i \hat{M}_{\mathrm{L}} \,z/\hbar\right)|\Psi(z)\rangle_{\mathrm{I}} \nonumber\\
&=&\mathrm{exp}\left[i z \sum^N_{j=1} \lambda_j \hat{b}^{\dagger}_j \hat{b}_j \right]|\Psi(z)\rangle_{\mathrm{I}}.\label{AEq:7_phi_from_phi_int}
\end{eqnarray}
Thus, $|\Psi(z)\rangle_\mathrm{b}$ can be expressed at first order in $\eta_j$ as follows
\begin{eqnarray}
|\Psi(z)\rangle_\mathrm{b}  &=&  \frac{1}{\sqrt{\mathcal{N}}} \left[1+\sum^N_{n,m=1} \tilde{Q}_{nm}(z) \hat{b}^{\dagger}_n \hat{b}^{\dagger}_m  \right]  |0\rangle_\mathrm{b} \label{AEq:8_sol_phi_z_1_order_b}  \\
 &=&\frac{1}{\sqrt{\mathcal{N}}} \Bigg\{  |0\rangle_\mathrm{b} +  \sum^N_{n=1} \tilde{K}_{nn}(z) |\dots,2_n,\dots\rangle_{\mathrm{b}} \nonumber \\
& &+ \sum^{N-1}_{n=1} \sum^N_{m=n+1} \tilde{K}_{nm}(z) |\dots,1_n,\dots,1_m,\dots\rangle_{\mathrm{b}} \Bigg\}, \nonumber\\ 
\label{AEq:9_sol_phi_z_1_order_an} 
\end{eqnarray}
with $\mathcal{N}$ an appropriate normalization constant and $\tilde{K}_{nm}(z)$ the entries of the matrix $\tilde{K}(z)$ defined as
\begin{align}
\tilde{K}(z)&= D\odot \tilde{Q}(z) \nonumber\\
&=\delta(z)D\odot\tilde{P}\odot \tilde{T}(z), \label{AEq:10_Ktilde_matrix}
\end{align}
with $D$ the degeneracy matrix with entries 
\begin{equation}
D_{nm}=2^{1-\delta_{nm}/2}. \label{AEq:11_Bunching_matrix}
\end{equation}
In this way, the amplitudes of the $|\dots,1_n,\dots,1_m,\dots\rangle_{\mathrm{b}} $, and $|\dots,2_n,\dots\rangle_{\mathrm{b}}$ biphoton eigenstates in the supermode basis are given by $\tilde{K}_{nm}(z)$ (with $n \neq m$), and $\tilde{K}_{nn}(z)$, respectively. It is worth mentioning we are interested in biphoton states. Hence, we assume to be working in a low-injection regime such that the approximation done in Eq. \eqref{AEq:8_sol_phi_z_1_order_b} is valid.

The corresponding output state in the individual-mode basis $|\Psi(z)\rangle$ can be expressed as
\begin{eqnarray}
|\Psi(z)\rangle  &=& \frac{1}{\sqrt{\mathcal{N}}}\Bigg\{|0\rangle +  \sum^N_{n=1} K_{qq}(z) |\dots,2_n,\dots\rangle \nonumber \\
& &+ \sum^{N-1}_{k=1} \sum^N_{q=k+1} K_{kq}(z) |\dots,1_k,\dots,1_q,\dots\rangle \Bigg\}, \nonumber\\ 
\label{AEq:12_sol_analytic_1_order_an} 
\end{eqnarray}
with $K_{kq}(z)$ the $kq$ entry of the matrix $K(z)$, which can be obtained by applying a change of basis to $\tilde{Q}(z)$ as follows 
\begin{equation}
K(z)  = D \odot Q(z),
\label{AEq:13_K_definition}
\end{equation}
with
\begin{equation}
Q(z)  = S^T \tilde{Q}(z) S.
\label{AEq:14_Q_trans_definition}
\end{equation}
Thus, the amplitudes of the biphoton eigenstates in the individual-mode basis are given by the entries of $K(z)$.


\section{Inversion Procedure}
\label{Appendix_inverse_proc}
Our analytic solution presented in Sec. \ref{sec:level3} allowed us to efficiently compute the output state and, consequently, the correlation matrix. We then used the \textit{minimize} function from the \textit{SciPy} library to optimize the merit function $\mathrm{MF}$, defined in Eq. (\ref{Eq:37_merit_function}), specifying the initial guess and bounds for the variables---namely, the position $z$ and the pump profile ${|\eta_j|,\phi_j}$---while using default values for the remaining parameters of the minimization routine. The bounds for the position and pump parameters were set as $0<z<20$, $0<|\eta_j|<1$, and $0<\phi_j<2\pi$ for all waveguides. This optimization was performed in the individual-mode basis, although it could also be accomplished in the supermode basis, as shown in Appendix E4.

To slightly improve the optimization results, we performed 10 iterations of the \textit{minimize} function, using a randomly generated initial guess each time. The reported results correspond to the iteration which gave the smallest $\mathrm{MF}$. Although the improvement from multiple iterations relies partly on chance, we deemed it suitable for our purposes, since our goal was merely to demonstrate a proof of concept, and a single iteration already yielded satisfactory results for most cases. A more sophisticated approach could be developed to obtain better results and a faster algorithm. The script used for all the calculations can be found in a GitHub repository \cite{Repository_ANW}.

\begin{figure}
    \centering
    \includegraphics[width=\columnwidth]{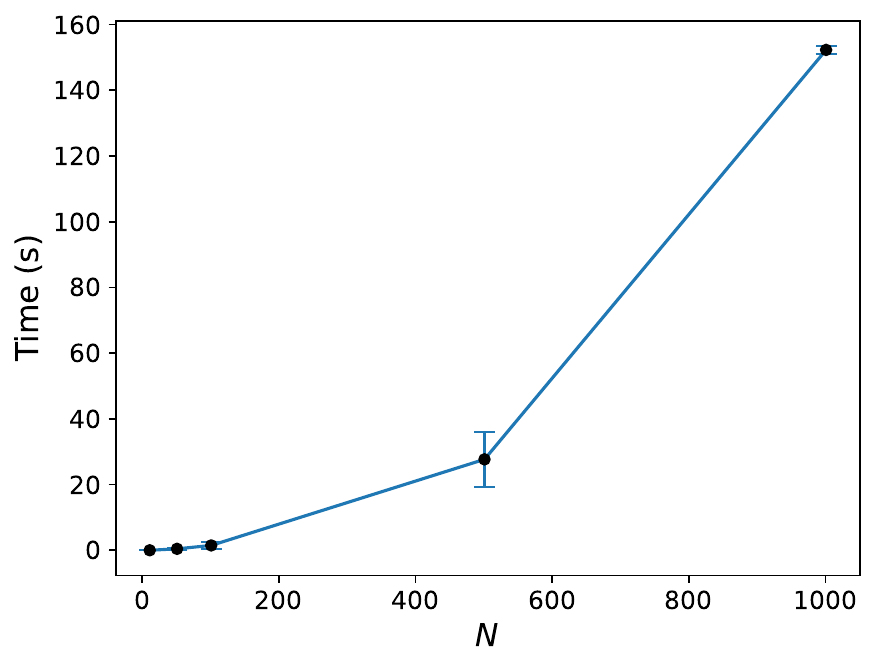}
    \caption{Calculation time of the output state as a function of the size $N$ of an odd homogeneous array for the case of an injection only in the center waveguide. The symbols, joined by straight lines, correspond to the average calculated values and the error bars denote one standard deviation around each average value. Note that, for some data points, the error bars are smaller than the symbols.}
    \label{fig:benchmarking_direct}
\end{figure}

\begin{figure}
    \centering
    \includegraphics[width=\columnwidth]{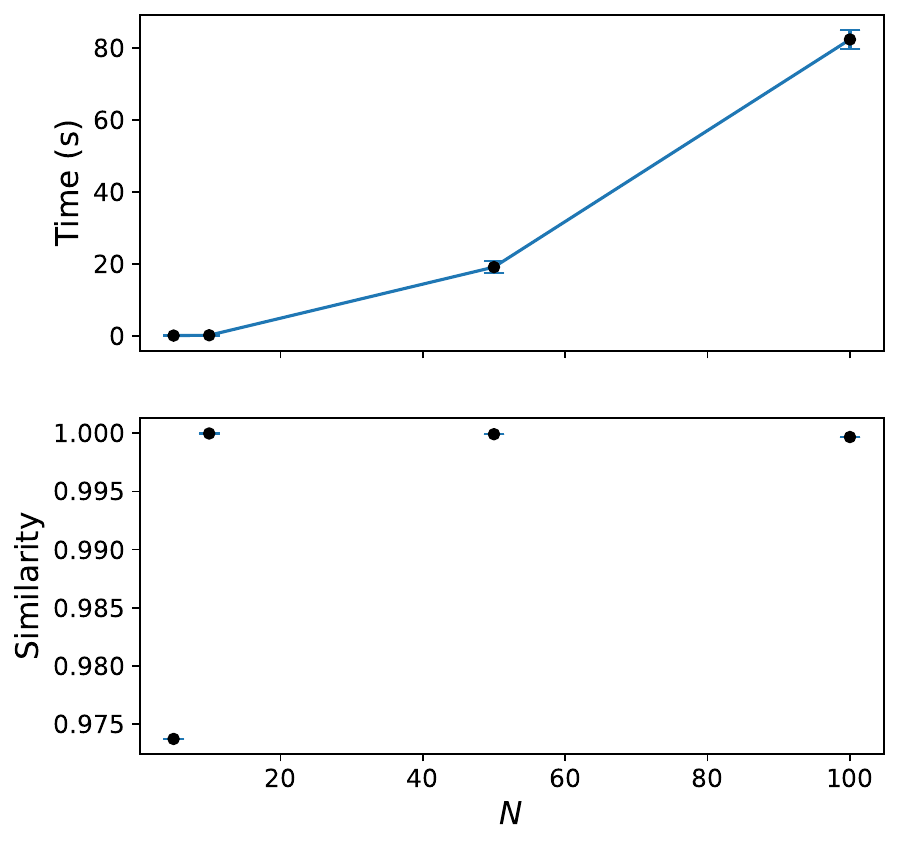}
    \caption{(Top panel) Calculation time of the input parameters as a function of the system size $N$ for the inverse problem corresponding to an array with a parabolic coupling profile and a target consisting of an antidiagonal correlation matrix with equal entries. (Bottom panel) Similarity values obtained for the inverse problems related to the top panel. The black circles denote the average calculated values and the error bars (in some cases smaller than the symbols) depict one standard deviation around the corresponding averages. In the top panel, the calculated data points are connected by straight lines.}
    \label{fig:benchmarking_inverse}
\end{figure}

\section{Scalability}
\label{Appendix_scalability}
To assess the scalability of our method as the system size increases, we measured the time required to compute the output state for the direct problem and that to determine the input parameters (position and pump profile) for the inverse problem. The execution times were measured in Google Colab \cite{Colab} with its free resources (without additional compute units, equipped with an Intel\textsuperscript{\tiny\textregistered} Xeon\textsuperscript{\tiny\textregistered} CPU @ 2.20\, GHz, 12.7\, GB of RAM, and Python version 3.11.12) using the \textit{\%timeit} function, which benchmarks the runtime of Python code snippets. We set the number of repeats to 10---that is, each execution time is the average of 10 repetitions. All the other parameters of this function were kept at their default values. For the inverse problem, the reported similarity also corresponds to the average of 10 repetitions.

In Fig. \ref{fig:benchmarking_direct}, we studied the direct propagation problem for the case of an injection only in the center waveguide of odd homogeneous arrays of different sizes $N$. It can be seen that our method only requires a few minutes, even for a number of waveguides as large as 1000. Similarly, in Fig. \ref{fig:benchmarking_inverse}, reasonable times were observed for large values of $N$ when solving the inverse problem for an array with a parabolic coupling profile and a target consisting of an antidiagonal correlation matrix with equal entries. Moreover, the similarity tends to be consistent and close to one as $N$ increases.

In this section, we chose to diagonalize the matrix $\Omega$ using a numerical method instead of employing the corresponding explicit analytic expressions available for the eigenvectors and eigenvalues for the homogeneous and parabolic coupling profiles. Thus, the times reported in Figs. \ref{fig:benchmarking_direct} and \ref{fig:benchmarking_inverse} include the step associated with the numerical diagonalization of $\Omega$. Additionally, we performed only one iteration of the \textit{minimize} function, always using as an initial guess the same injection for all waveguides to ensure a fair comparison across different values of $N$. The script used for obtaining the results presented in this section can be found in \cite{Repository_ANW}.

\onecolumngrid

\newpage

\section{Entries of relevant matrices for different cases}\label{Appendix_B}

\subsection{Flat pump (homogeneous coupling profile, $N=7$)}

\begin{figure}[b]
    \centering
    \includegraphics[width=0.9\linewidth]{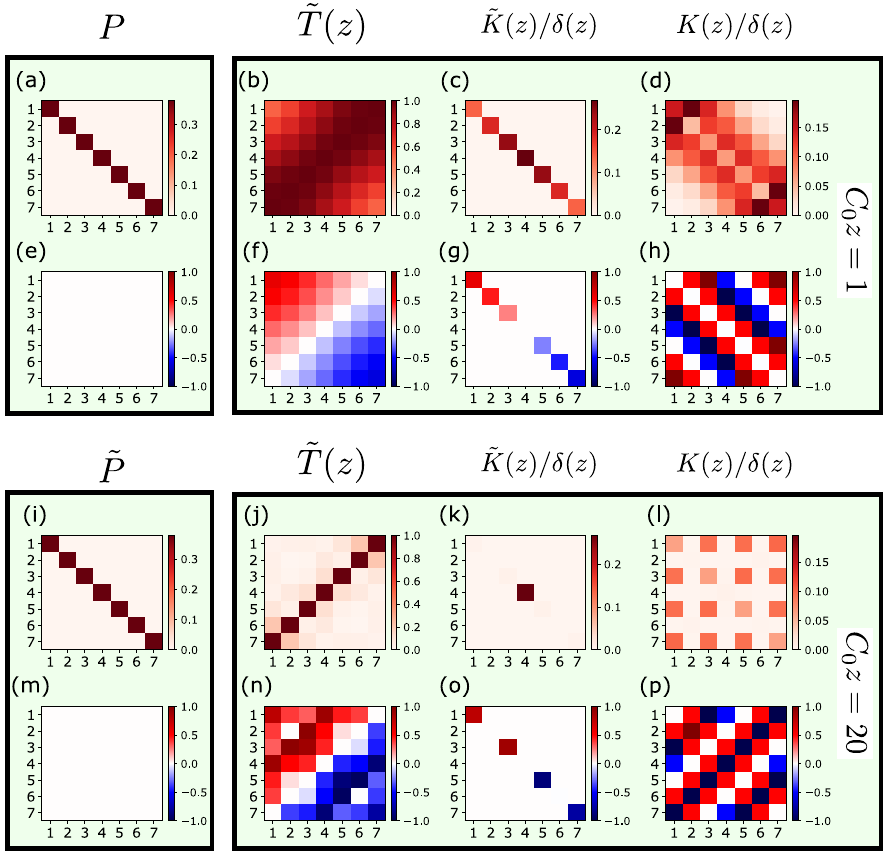}
    \caption{Absolute values (first and third rows) and phases (second and fourth rows) of the entries of the matrices $P=\mathrm{diag}(\eta_1,\dots,\eta_N)$ (upper left block), $\tilde{P}$ (bottom left block), $\tilde{T}(z)$ (second column), $\tilde{K}(z)/\delta(z)$ (third column), and $K(z)/\delta(z)$ (fourth column) for the case of a flat pump injection in a homogeneous seven-waveguide ANW (see the associated $P$ matrix). The results are shown for the propagation distances given by $C_0z=1$ (upper right block) and $C_0z=20$ (bottom right block). For convenience, the phase values are shown in units of $\pi$, and $\tilde{K}(z)$ and $K(z)$ are divided by the multiplicative scalar factor $\delta(z)=izg||\vec{\alpha}||$. The vertical, and horizontal axis in each plot correspond to the rows, and columns of the related matrix, respectively.}
    \label{fig:Hadamard1}
\end{figure}

Figure \ref{fig:Hadamard1} depicts the entries of the relevant matrices for the case of a homogeneous seven-waveguide ANW with a flat pump injection, in which all waveguides are excited with the same pump amplitude and phase. The data are presented for two propagation distances given by $C_0z=1$ and $C_0z=20$.

\newpage


\subsection{Flat pump amplitude with alternating $\pi$ phase (homogeneous coupling profile, $N=7$)}

\begin{figure}[b]
    \centering
    \includegraphics[width=0.9\linewidth]{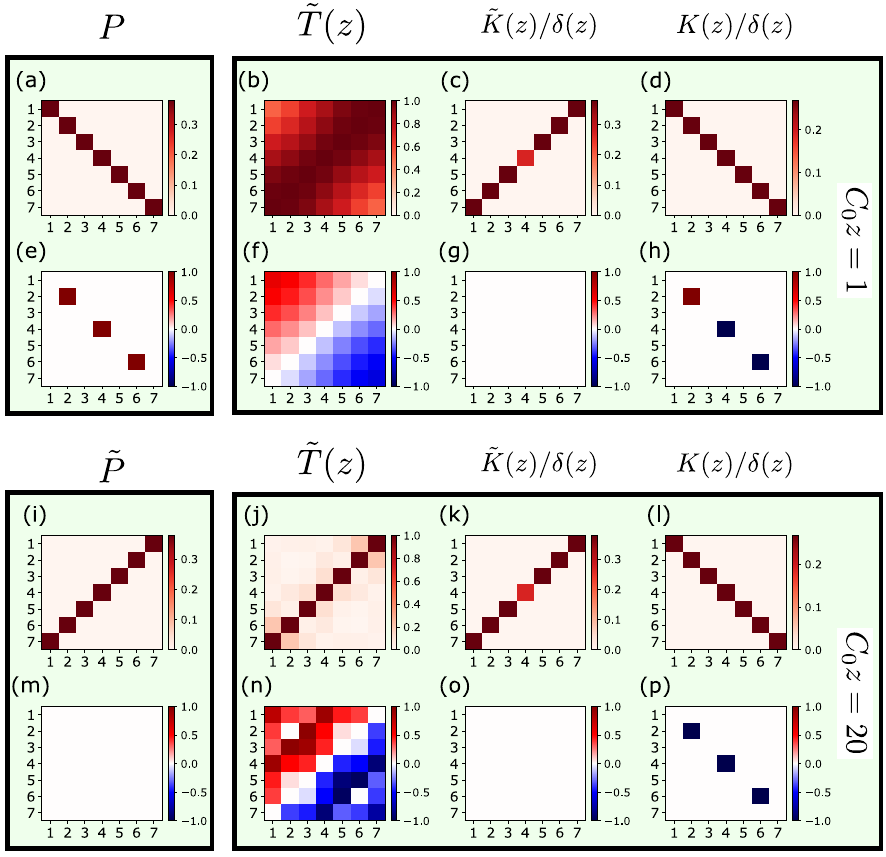}
    \caption{Absolute values (first and third rows) and phases (second and fourth rows) of the entries of the matrices $P=\mathrm{diag}(\eta_1,\dots,\eta_N)$ (upper left block), $\tilde{P}$ (bottom left block), $\tilde{T}(z)$ (second column), $\tilde{K}(z)/\delta(z)$ (third column), and $K(z)/\delta(z)$ (fourth column) for the case of an injection consisting of a flat pump amplitude with alternating $\pi$ phase in a homogeneous seven-waveguide ANW (see the associated $P$ matrix). The results are shown for the propagation distances given by $C_0z=1$ (upper right block) and $C_0z=20$ (bottom right block). For convenience, the phase values are shown in units of $\pi$, and $\tilde{K}(z)$ and $K(z)$ are divided by the multiplicative scalar factor $\delta(z)=izg||\vec{\alpha}||$. The vertical, and horizontal axis in each plot correspond to the rows, and columns of the related matrix, respectively.}
    \label{fig:Hadamard2}
\end{figure}

Figure \ref{fig:Hadamard2} depicts the entries of the relevant matrices for the case of a homogeneous seven-waveguide ANW with an injection consisting of a flat pump amplitude with alternating $\pi$ phase---in this case, the pump amplitudes are the same for all waveguides while the corresponding phases present an alternating $\pi$ profile. The data are presented for two propagation distances given by $C_0z=1$ and $C_0z=20$.

\newpage


\subsection{Two consecutive waveguides in the middle of the array injected with the same amplitude and phase (parabolic coupling profile, $N=8$)}

\begin{figure}[b]
    \centering
    \includegraphics[width=0.9\linewidth]{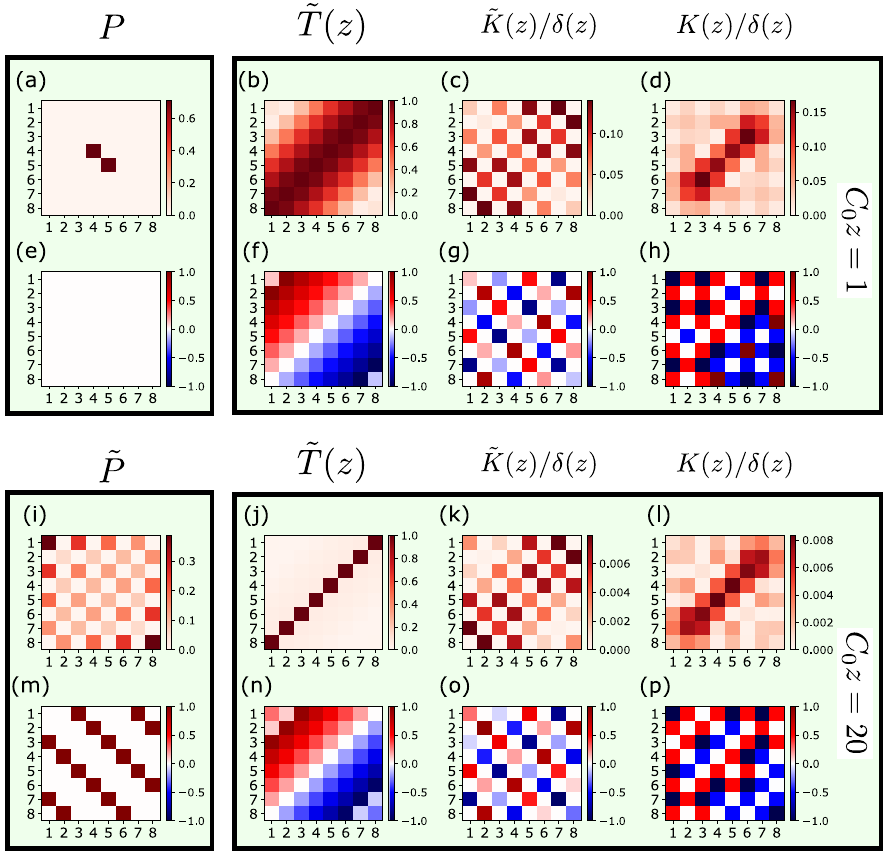}
    \caption{Absolute values (first and third rows) and phases (second and fourth rows) of the entries of the matrices $P=\mathrm{diag}(\eta_1,\dots,\eta_N)$ (upper left block), $\tilde{P}$ (bottom left block), $\tilde{T}(z)$ (second column), $\tilde{K}(z)/\delta(z)$ (third column), and $K(z)/\delta(z)$ (fourth column) for the case of an injection only in two consecutive waveguides with identical amplitudes and phases in the center of a seven-waveguide ANW with a parabolic coupling profile (see the associated $P$ matrix). The results are shown for the propagation distances given by $C_0z=1$ (upper right block) and $C_0z=20$ (bottom right block). For convenience, the phase values are shown in units of $\pi$, and $\tilde{K}(z)$ and $K(z)$ are divided by the multiplicative scalar factor $\delta(z)=izg||\vec{\alpha}||$. The vertical, and horizontal axis in each plot correspond to the rows, and columns of the related matrix, respectively.}  \label{fig:Hadamard3}
\end{figure}

Figure \ref{fig:Hadamard3} depicts the entries of the relevant matrices for the case of a seven-waveguide ANW with a parabolic coupling profile and injecting only two consecutive waveguides in the center of the array with the same amplitude and phase. The data are presented for two propagation distances given by $C_0z=1$ and $C_0z=20$.

\newpage


\section{Optimization examples}\label{Appendix_C}

\begin{figure}[h]
    \centering
    \includegraphics[width=0.6\linewidth]{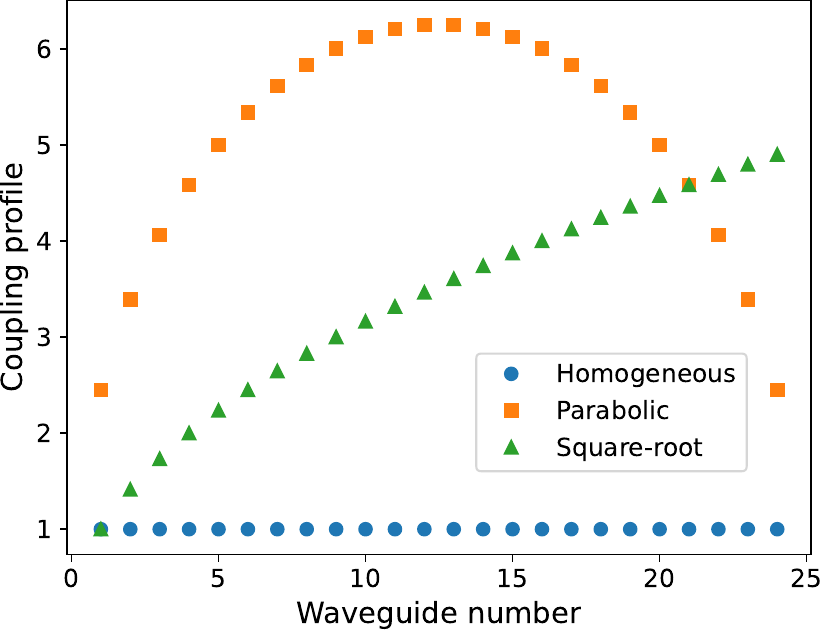}
    \caption{The homogeneous $C_j=C_0$ (blue circles), parabolic $C_j=\sqrt{j(N-j)} C_0/2$ (orange squares) and square-root $C_j=\sqrt{j}C_0$ (green triangles) coupling profiles were considered in the following optimization examples. Here, the data sets are shown in units of $C_0$.}
    \label{fig:couplings}
\end{figure}

Here, we present illustrative application examples of the optimization method. In each case, the optimized parameters were obtained for the homogeneous, parabolic, and square-root coupling profiles (depicted in Fig. \ref{fig:couplings}). 

\newpage

\subsection{Antidiagonal correlation matrix with equal nonvanishing entries ($N=100$)}

\begin{figure}[h]
    \centering
    \includegraphics[width=0.9\linewidth]{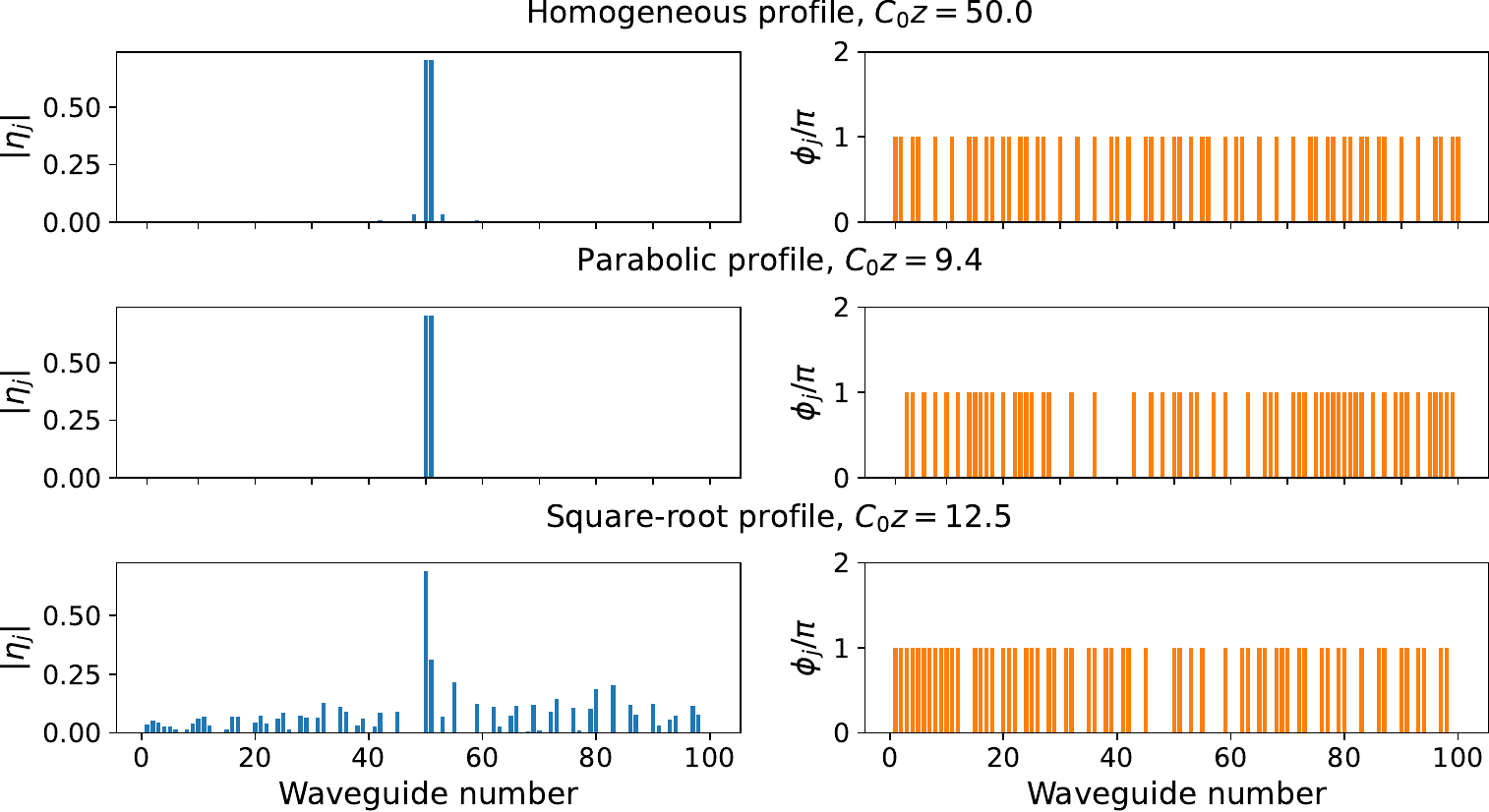}
    \caption{Amplitudes $|\eta_j|$ (left column) and phases $\phi_j$ (right column) of the pump profile that optimized the merit function $\mathrm{MF}$ using a homogeneous (first row), parabolic (second row), and square-root (third row) coupling profile. The target for an ANW of 100 waveguides consisted of an antidiagonal correlation matrix with equal entries in the antidiagonal. For each case, the position that optimized the merit function is depicted above the respective row. Those were $C_0z=50.000$, $C_0z=9.426$, and $C_0z=12.499$ for the homogeneous, parabolic, and square-root coupling profiles, respectively.}
    \label{fig:opt_pump0}
\end{figure}

\begin{figure}[h]
    \centering
    \includegraphics[width=0.75\linewidth]{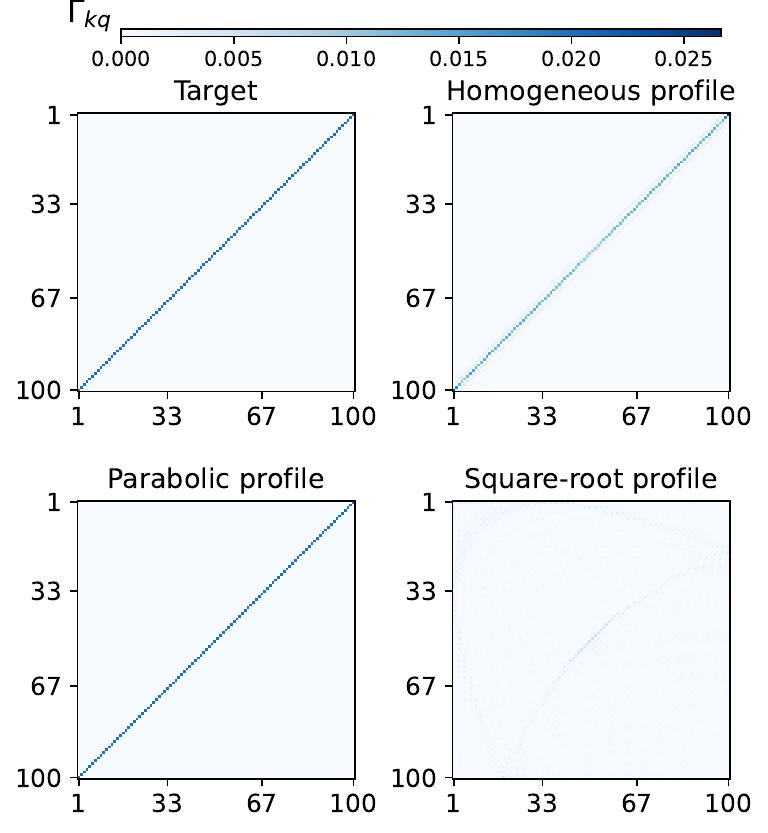}
    \caption{ Target (top left panel) for an ANW of 100 waveguides corresponding to an antidiagonal correlation matrix with equal entries in the antidiagonal. The obtained correlation matrices after optimization are depicted for the cases of the homogeneous (top right panel), parabolic (bottom left panel), and square-root (bottom right) coupling profiles. The color bar indicates the value of the correlation matrix entries $\Gamma_{kq}$ in all cases.}
    \label{fig:opt0}
\end{figure}

Here, we employed the optimization method with a target consisting of an antidiagonal correlation matrix with equal nonvanishing entries for an array of $100$ waveguides. The resulting similarity values for the different coupling profiles are presented in Table \ref{Table_fig:opt_pump0} and the corresponding optimized parameters are depicted in Fig. \ref{fig:opt_pump0}. In addition, the target and the obtained correlation matrices after the optimization are shown in Fig. \ref{fig:opt0}.

\begin{table}[h]
\caption{Similarity values obtained in the optimization problem for each analyzed coupling profile. The target for an ANW of 100 waveguides consisted of an antidiagonal correlation matrix with equal entries in the antidiagonal.}
\begin{center}
\begin{tabular}{ c  |  c }
 Coupling profile\,\,  &  \,\,Similarity $\mathcal{S}$ \\ 
 \hline
 Homogeneous & 0.6345 \\  
 Parabolic & 0.9999 \\
 Square-root & 0.02530 \\
\end{tabular}
\label{Table_fig:opt_pump0}
\end{center}
\end{table}

\newpage

\newpage

\subsection{Diagonal correlation matrix  with equal nonvanishing entries ($N=50$)}

\begin{figure}[h]
    \centering
    \includegraphics[width=0.9\linewidth]{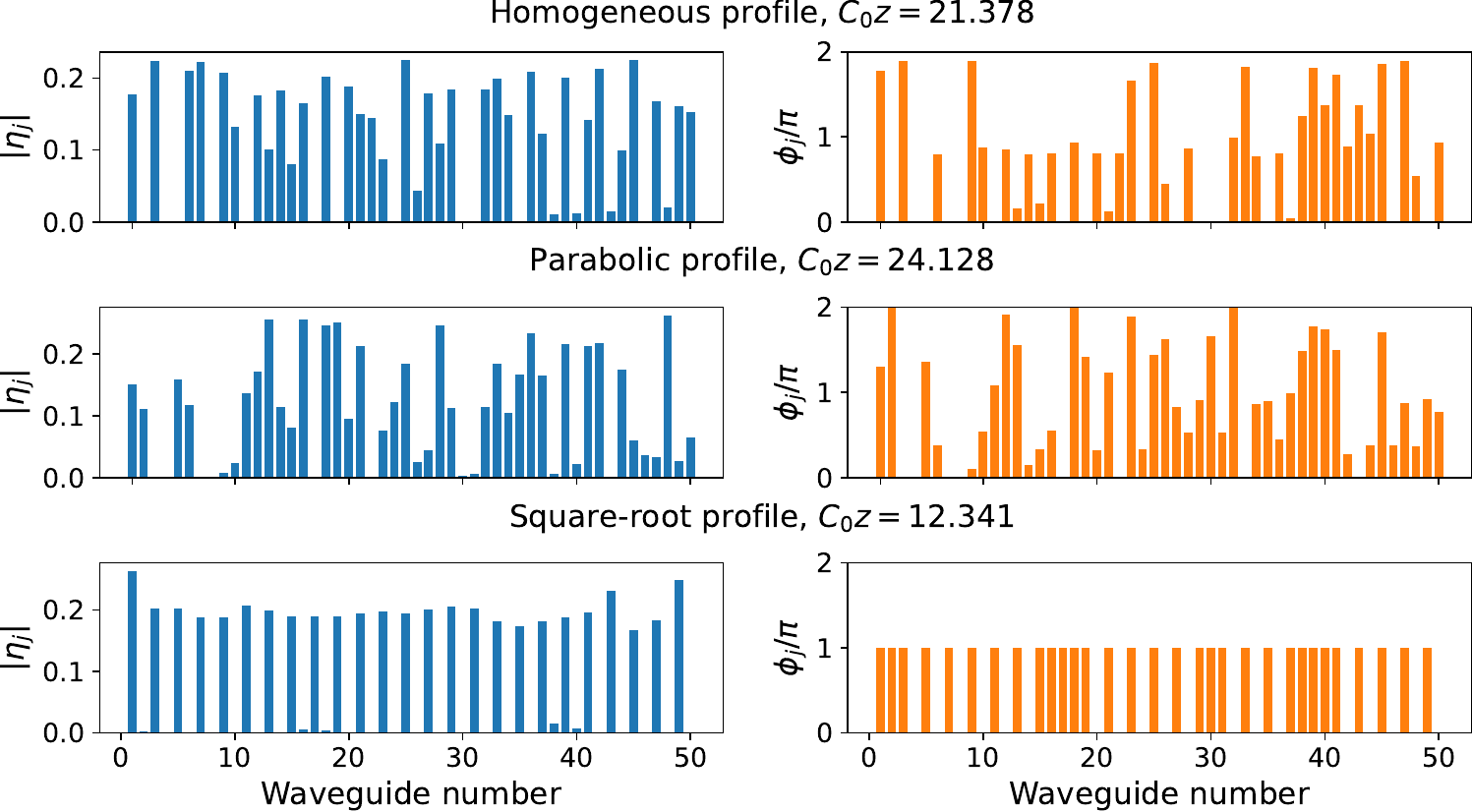}
    \caption{Amplitudes $|\eta_j|$ (left column) and phases $\phi_j$ (right column) of the pump profile that optimized the merit function $\mathrm{MF}$ using a homogeneous (first row), parabolic (second row), and square-root (third row) coupling profile. The target for an array of 50 waveguides consisted of a diagonal correlation matrix with equal entries in the diagonal. For each case, the position that optimized the merit function is depicted above the respective row. Those were $C_0z=21.378$, $C_0z=24.128$, and $C_0z=12.341$ for the homogeneous, parabolic, and square-root coupling profiles, respectively.}
    \label{fig:opt_pump1}
\end{figure}

\begin{figure}[h]
    \centering
    \includegraphics[width=0.75\linewidth]{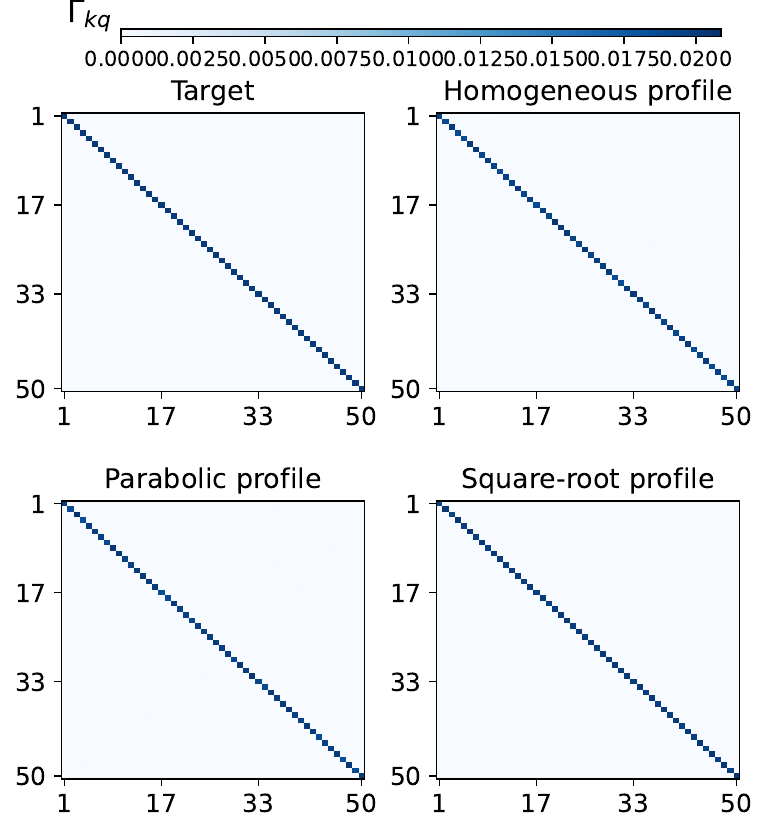}
    \caption{Target (top left panel) for an array of 50 waveguides corresponding to a diagonal correlation matrix with equal entries in the diagonal. The obtained correlation matrices after optimization are depicted for the cases of the homogeneous (top right panel), parabolic (bottom left panel), and square-root (bottom right) coupling profiles. The color bar indicates the value of the correlation matrix entries $\Gamma_{kq}$ in all cases.}
    \label{fig:opt1}
\end{figure}

Here, we employed the optimization method with a target consisting of an diagonal correlation matrix with equal nonvanishing entries for an array of $50$ waveguides. The resulting similarity values for the different coupling profiles are presented in Table \ref{Table_fig:opt_pump1} and the corresponding optimized parameters are depicted in Fig. \ref{fig:opt_pump1}. In addition, the target and the obtained correlation matrices after the optimization are shown in Fig. \ref{fig:opt1}.

\begin{table}[h]
\caption{Similarity values obtained in the optimization problem for each analyzed coupling profile. The target for an ANW of 50 waveguides consisted of a diagonal correlation matrix with equal entries in the diagonal.}
\begin{center}
\begin{tabular}{ c  |  c }
 Coupling profile\,\,  &  \,\,Similarity $\mathcal{S}$ \\ 
 \hline
 Homogeneous & 0.95481 \\  
 Parabolic & 0.96539 \\
 Square-root & 0.99366 \\
\end{tabular}
\label{Table_fig:opt_pump1}
\end{center}
\end{table}

\newpage

\newpage

\subsection{Correlation matrix with only odd individual modes and equal nonvanishing entries ($N=25$)}

\begin{figure}[h]
    \centering
    \includegraphics[width=0.77\linewidth]{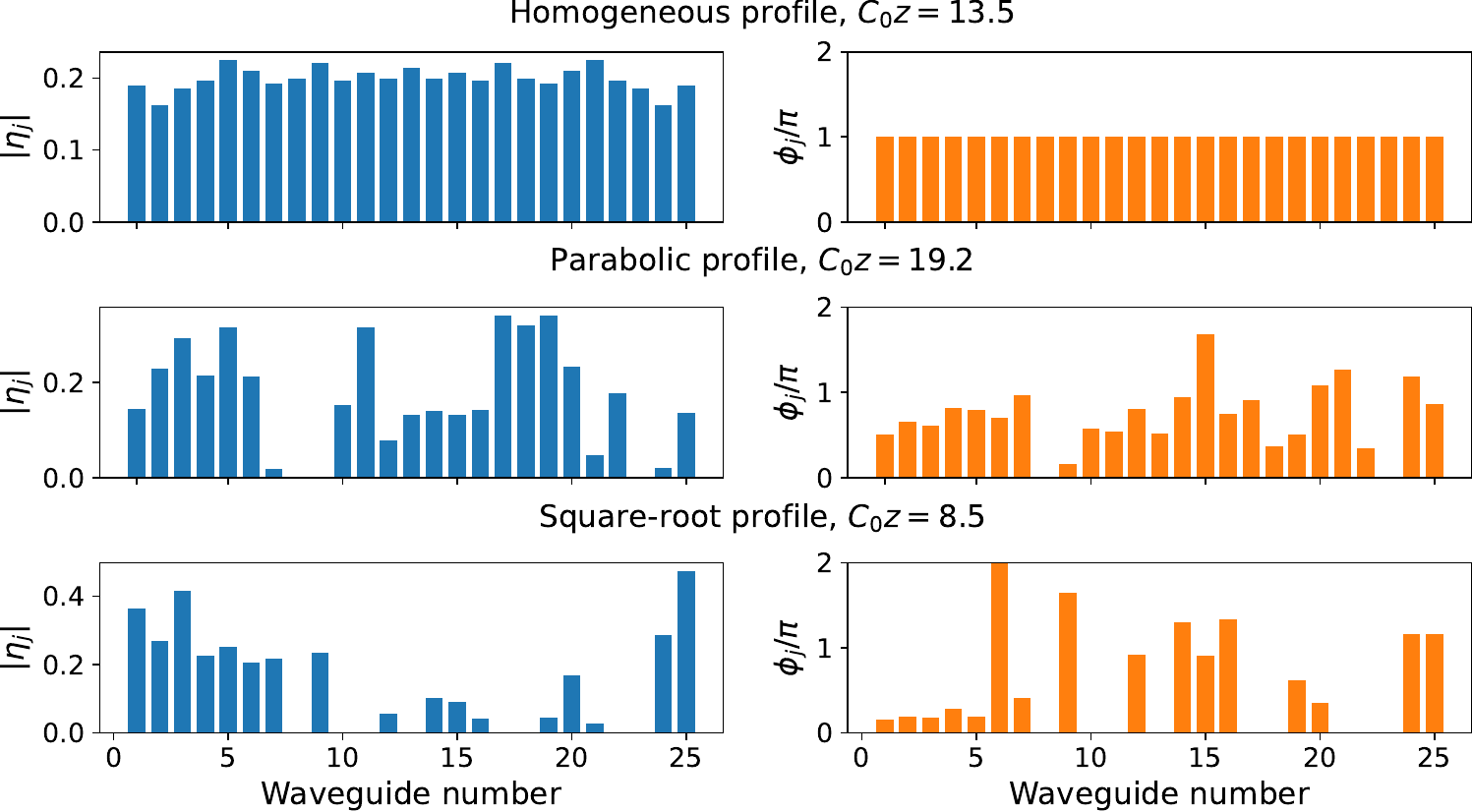}
    \caption{Amplitudes $|\eta_j|$ (left column) and phases $\phi_j$ (right column) of the pump profile that optimized the merit function $\mathrm{MF}$ using a homogeneous (first row), parabolic (second row), and square-root (third row) coupling profile. The target for an ANW of 25 waveguides consisted of a correlation matrix with null even rows and columns and with equal values for the remaining entries (the ones which are exclusively related to odd individual modes). For each case, the position that optimized the merit function is depicted above the respective row. Those were $C_0z=13.5$, $C_0z=19.2$, and $C_0z=8.541$ for the homogeneous, parabolic, and square-root coupling profiles, respectively.}
    \label{fig:opt_pump2}
\end{figure}

\begin{figure}[h]
    \centering
    \includegraphics[width=0.75\linewidth]{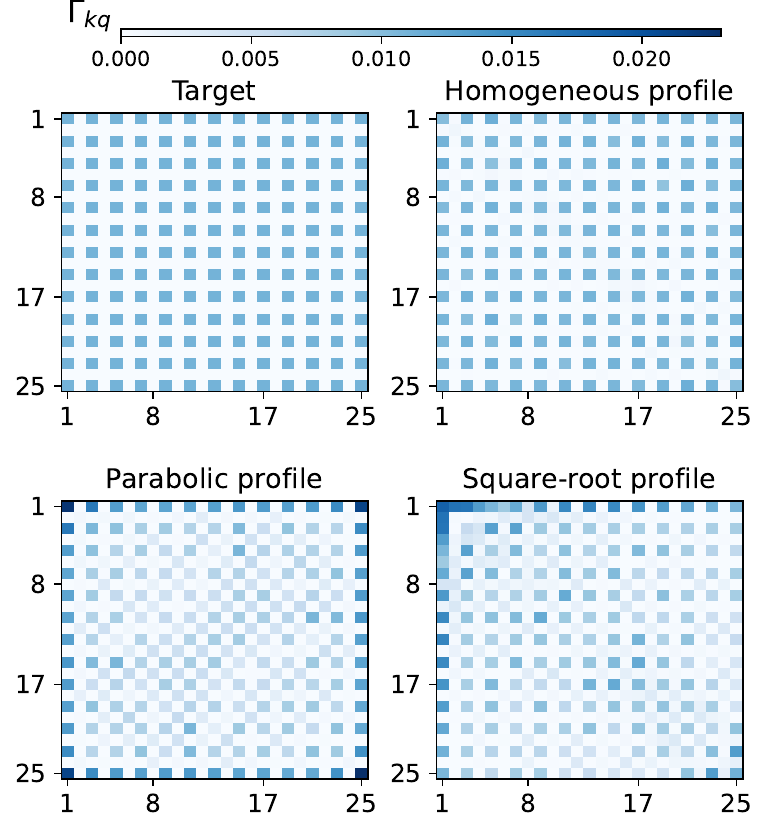}
    \caption{Target (top left panel) for an ANW of 25 waveguides corresponding to a correlation matrix with null even rows and columns and with equal values for the remaining entries (the ones which are exclusively related to odd individual modes). The obtained correlation matrices after optimization are depicted for the cases of the homogeneous (top right panel), parabolic (bottom left panel), and square-root (bottom right) coupling profiles. The color bar indicates the value of the correlation matrix entries $\Gamma_{kq}$ in all cases.}
    \label{fig:opt2}
\end{figure}

Here, we employed the optimization method with a target consisting of a correlation matrix with only odd individual modes and equal nonvanishing entries for an array of $25$ waveguides. The resulting similarity values for the different coupling profiles are presented in Table \ref{Table_fig:opt_pump2} and the corresponding optimized parameters are depicted in Fig. \ref{fig:opt_pump2}. In addition, the target and the obtained correlation matrices after the optimization are shown in Fig. \ref{fig:opt2}.

\begin{table}[h]
\caption{Similarity values obtained in the optimization problem for each analyzed coupling profile. The target for an ANW of 25 waveguides consisted of a correlation matrix with null even rows and columns and with equal values for the remaining entries (the ones which are exclusively related to odd individual modes).}
\begin{center}
\begin{tabular}{ c  |  c }
 Coupling profile\,\,  &  \,\,Similarity $\mathcal{S}$ \\ 
 \hline
 Homogeneous & 0.96914 \\  
 Parabolic & 0.78688 \\
 Square-root & 0.78094 \\
\end{tabular}
\label{Table_fig:opt_pump2}
\end{center}
\end{table}

\newpage

\newpage

\subsection{Supermode correlation matrix with only odd supermodes and equal nonvanishing entries  ($N=25$)}

\begin{figure}[h]
    \centering
    \includegraphics[width=0.9\linewidth]{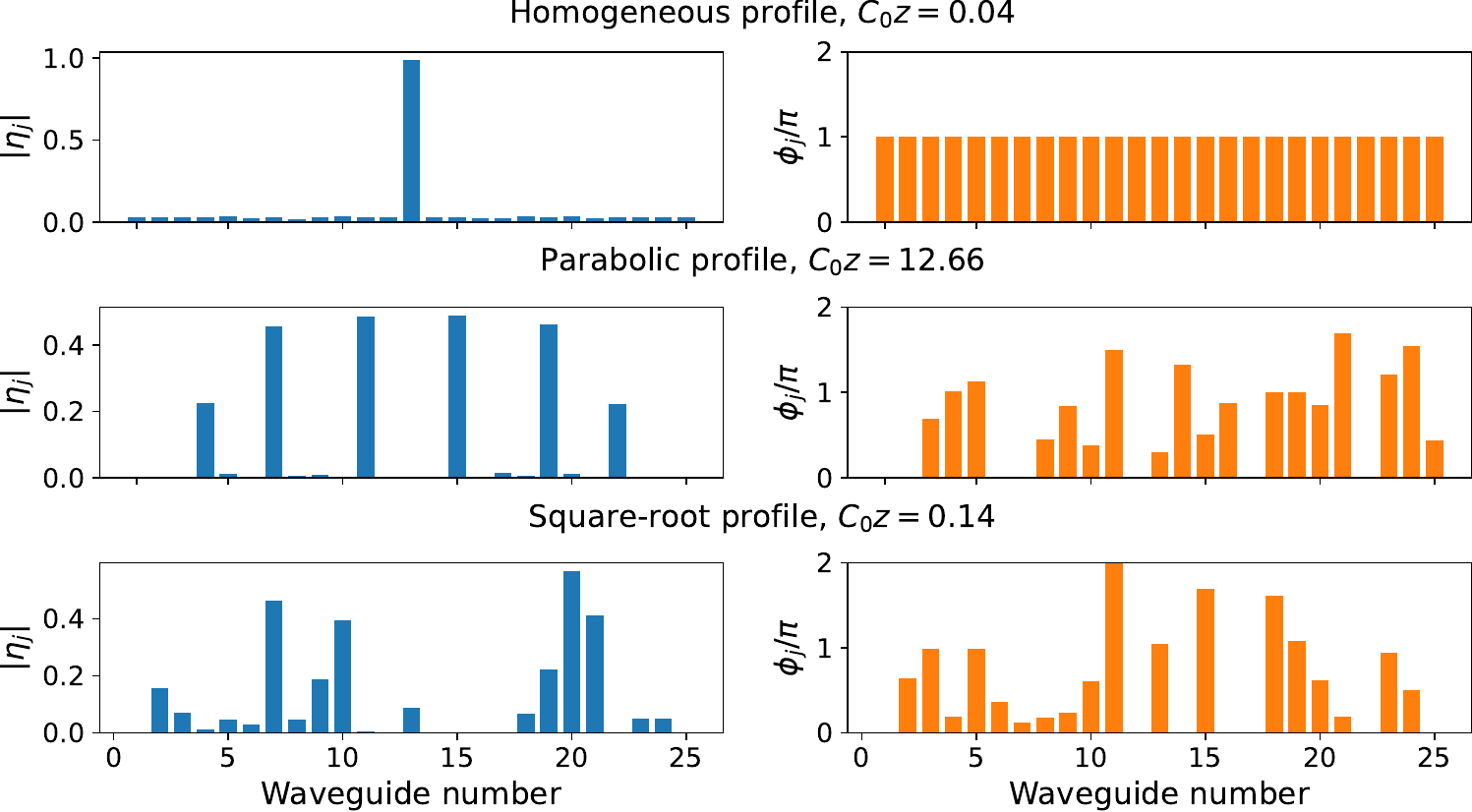}
    \caption{Amplitudes $|\eta_j|$ (left column) and phases $\phi_j$ (right column) of the pump profile that optimized the merit function $\mathrm{MF}$ using a homogeneous (first row), parabolic (second row), and square-root (third row) coupling profile. The target consisted of a supermode correlation matrix $\tilde{\Gamma}$ with null even rows and columns and with equal values for the remaining entries (the ones which are exclusively related to odd supermodes). For each case, the position that optimized the merit function is depicted above the respective row. Those were $C_0z=0.04$, $C_0z=12.66$, and $C_0z=0.14$ for the homogeneous, parabolic, and square-root coupling profiles, respectively.}
    \label{fig:opt_pump3}
\end{figure}

\begin{figure}[h]
    \centering
    \includegraphics[width=0.75\linewidth]{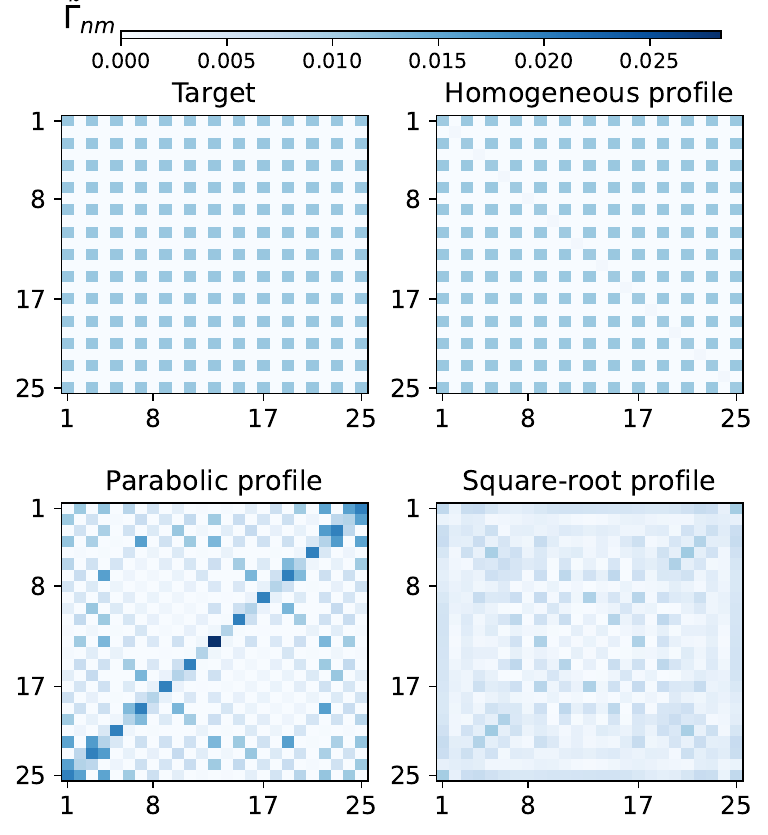}
    \caption{Target (top left panel) for an ANW of 25 waveguides corresponding to a supermode correlation matrix $\tilde{\Gamma}$ with null even rows and columns and with equal values for the remaining entries (the ones which are exclusively related to odd supermodes). The obtained correlation matrices after optimization are depicted for the cases of the homogeneous (top right panel), parabolic (bottom left panel), and square-root (bottom right) coupling profiles. The color bar indicates the value of the correlation matrix entries $\tilde{\Gamma}_{nm}$ in all cases.}
    \label{fig:opt3}
\end{figure}

Here, we employed the optimization method with a target consisting of a supermode correlation matrix with only odd supermodes and equal nonvanishing entries for an array of $25$ waveguides. The resulting similarity values for the different coupling profiles are presented in Table \ref{Table_fig:opt_pump3} and the corresponding optimized parameters are depicted in Fig. \ref{fig:opt_pump3}. In addition, the target and the obtained correlation matrices after the optimization are shown in Fig. \ref{fig:opt3}.

This is actually an optimization problem in the supermode basis. In fact, the target is given in terms of the supermode correlation matrix $\tilde{\Gamma}$, which is analogous to the individual-mode correlation matrix $\Gamma$.
The entries of $\tilde{\Gamma}$ are defined as $\tilde{\Gamma}_{nm}=|_\mathrm{b}\langle\dots,1_n,\dots,1_m,\dots|\Psi\rangle|^2/\langle\Psi|\Psi\rangle$ for $n \neq m$, and $\tilde{\Gamma}_{nn}=|_\mathrm{b}\langle\dots,2_n,\dots|\Psi\rangle|^2/\langle\Psi|\Psi\rangle$, with the subscript $\mathrm{b}$ denoting that the eigenstates are in the supermode basis. Similarly to the entries of $\Gamma$, $\tilde{\Gamma}_{nm}$ represents the probability of detecting one photon in the supermode $n$ and the other in the supermode $m$. Also, $\tilde{\Gamma}_{nn}$ corresponds to the probability of detecting two photons in the supermode $n$. In this case, the target is a supermode correlation matrix $\tilde{\Gamma}$ with null even rows and columns and with equal values for the remaining entries (the ones which are exclusively related to odd supermodes).

\begin{table}[h]
\caption{Similarity values obtained in the optimization problem for each analyzed coupling profile. The target for an ANW of 25 waveguides consisted of a supermode correlation matrix $\tilde{\Gamma}$ with null even rows and columns and with equal values for the remaining entries (the ones which are exclusively related to odd supermodes).}
\begin{center}
\begin{tabular}{ c  |  c }
 Coupling profile\,\,  &  \,\,Similarity $\mathcal{S}$ \\ 
 \hline
 Homogeneous & 0.99410 \\  
 Parabolic & 0.01197 \\
 Square-root & 0.41476 \\
\end{tabular}
\label{Table_fig:opt_pump3}
\end{center}
\end{table}

\clearpage

\twocolumngrid


\begin{thebibliography}{64}%
\makeatletter
\providecommand \@ifxundefined [1]{%
 \@ifx{#1\undefined}
}%
\providecommand \@ifnum [1]{%
 \ifnum #1\expandafter \@firstoftwo
 \else \expandafter \@secondoftwo
 \fi
}%
\providecommand \@ifx [1]{%
 \ifx #1\expandafter \@firstoftwo
 \else \expandafter \@secondoftwo
 \fi
}%
\providecommand \natexlab [1]{#1}%
\providecommand \enquote  [1]{``#1''}%
\providecommand \bibnamefont  [1]{#1}%
\providecommand \bibfnamefont [1]{#1}%
\providecommand \citenamefont [1]{#1}%
\providecommand \href@noop [0]{\@secondoftwo}%
\providecommand \href [0]{\begingroup \@sanitize@url \@href}%
\providecommand \@href[1]{\@@startlink{#1}\@@href}%
\providecommand \@@href[1]{\endgroup#1\@@endlink}%
\providecommand \@sanitize@url [0]{\catcode `\\12\catcode `\$12\catcode `\&12\catcode `\#12\catcode `\^12\catcode `\_12\catcode `\%12\relax}%
\providecommand \@@startlink[1]{}%
\providecommand \@@endlink[0]{}%
\providecommand \url  [0]{\begingroup\@sanitize@url \@url }%
\providecommand \@url [1]{\endgroup\@href {#1}{\urlprefix }}%
\providecommand \urlprefix  [0]{URL }%
\providecommand \Eprint [0]{\href }%
\providecommand \doibase [0]{https://doi.org/}%
\providecommand \selectlanguage [0]{\@gobble}%
\providecommand \bibinfo  [0]{\@secondoftwo}%
\providecommand \bibfield  [0]{\@secondoftwo}%
\providecommand \translation [1]{[#1]}%
\providecommand \BibitemOpen [0]{}%
\providecommand \bibitemStop [0]{}%
\providecommand \bibitemNoStop [0]{.\EOS\space}%
\providecommand \EOS [0]{\spacefactor3000\relax}%
\providecommand \BibitemShut  [1]{\csname bibitem#1\endcsname}%
\let\auto@bib@innerbib\@empty
\bibitem [{\citenamefont {Wang}\ \emph {et~al.}(2019)\citenamefont {Wang}, \citenamefont {Sciarrino}, \citenamefont {Laing},\ and\ \citenamefont {Thompson}}]{Wang2019}%
  \BibitemOpen
  \bibfield  {author} {\bibinfo {author} {\bibfnamefont {J.}~\bibnamefont {Wang}}, \bibinfo {author} {\bibfnamefont {F.}~\bibnamefont {Sciarrino}}, \bibinfo {author} {\bibfnamefont {A.}~\bibnamefont {Laing}},\ and\ \bibinfo {author} {\bibfnamefont {M.~G.}\ \bibnamefont {Thompson}},\ }\bibfield  {title} {\bibinfo {title} {Integrated photonic quantum technologies},\ }\href {https://doi.org/10.1038/s41566-019-0532-1} {\bibfield  {journal} {\bibinfo  {journal} {Nature Photonics}\ }\textbf {\bibinfo {volume} {14}},\ \bibinfo {pages} {273} (\bibinfo {year} {2019})}\BibitemShut {NoStop}%
\bibitem [{\citenamefont {Pelucchi}\ \emph {et~al.}(2021)\citenamefont {Pelucchi}, \citenamefont {Fagas}, \citenamefont {Aharonovich}, \citenamefont {Englund}, \citenamefont {Figueroa}, \citenamefont {Gong}, \citenamefont {Hannes}, \citenamefont {Liu}, \citenamefont {Lu}, \citenamefont {Matsuda}, \citenamefont {Pan}, \citenamefont {Schreck}, \citenamefont {Sciarrino}, \citenamefont {Silberhorn}, \citenamefont {Wang},\ and\ \citenamefont {Jöns}}]{Pelucchi2021}%
  \BibitemOpen
  \bibfield  {author} {\bibinfo {author} {\bibfnamefont {E.}~\bibnamefont {Pelucchi}}, \bibinfo {author} {\bibfnamefont {G.}~\bibnamefont {Fagas}}, \bibinfo {author} {\bibfnamefont {I.}~\bibnamefont {Aharonovich}}, \bibinfo {author} {\bibfnamefont {D.}~\bibnamefont {Englund}}, \bibinfo {author} {\bibfnamefont {E.}~\bibnamefont {Figueroa}}, \bibinfo {author} {\bibfnamefont {Q.}~\bibnamefont {Gong}}, \bibinfo {author} {\bibfnamefont {H.}~\bibnamefont {Hannes}}, \bibinfo {author} {\bibfnamefont {J.}~\bibnamefont {Liu}}, \bibinfo {author} {\bibfnamefont {C.-Y.}\ \bibnamefont {Lu}}, \bibinfo {author} {\bibfnamefont {N.}~\bibnamefont {Matsuda}}, \bibinfo {author} {\bibfnamefont {J.-W.}\ \bibnamefont {Pan}}, \bibinfo {author} {\bibfnamefont {F.}~\bibnamefont {Schreck}}, \bibinfo {author} {\bibfnamefont {F.}~\bibnamefont {Sciarrino}}, \bibinfo {author} {\bibfnamefont {C.}~\bibnamefont {Silberhorn}}, \bibinfo {author} {\bibfnamefont {J.}~\bibnamefont {Wang}},\ and\ \bibinfo {author} {\bibfnamefont {K.~D.}\
  \bibnamefont {Jöns}},\ }\bibfield  {title} {\bibinfo {title} {The potential and global outlook of integrated photonics for quantum technologies},\ }\href {https://doi.org/10.1038/s42254-021-00398-z} {\bibfield  {journal} {\bibinfo  {journal} {Nature Reviews Physics}\ }\textbf {\bibinfo {volume} {4}},\ \bibinfo {pages} {194} (\bibinfo {year} {2021})}\BibitemShut {NoStop}%
\bibitem [{\citenamefont {Ramakrishnan}\ \emph {et~al.}(2023)\citenamefont {Ramakrishnan}, \citenamefont {Ravichandran}, \citenamefont {Mishra}, \citenamefont {Kaushalram}, \citenamefont {Hegde}, \citenamefont {Talabattula},\ and\ \citenamefont {Rohde}}]{Ramakrishnan2023}%
  \BibitemOpen
  \bibfield  {author} {\bibinfo {author} {\bibfnamefont {R.~K.}\ \bibnamefont {Ramakrishnan}}, \bibinfo {author} {\bibfnamefont {A.~B.}\ \bibnamefont {Ravichandran}}, \bibinfo {author} {\bibfnamefont {A.}~\bibnamefont {Mishra}}, \bibinfo {author} {\bibfnamefont {A.}~\bibnamefont {Kaushalram}}, \bibinfo {author} {\bibfnamefont {G.}~\bibnamefont {Hegde}}, \bibinfo {author} {\bibfnamefont {S.}~\bibnamefont {Talabattula}},\ and\ \bibinfo {author} {\bibfnamefont {P.~P.}\ \bibnamefont {Rohde}},\ }\bibfield  {title} {\bibinfo {title} {Integrated photonic platforms for quantum technology: a review},\ }\href {https://doi.org/10.1007/s41683-023-00115-1} {\bibfield  {journal} {\bibinfo  {journal} {ISSS Journal of Micro and Smart Systems}\ }\textbf {\bibinfo {volume} {12}},\ \bibinfo {pages} {83} (\bibinfo {year} {2023})}\BibitemShut {NoStop}%
\bibitem [{\citenamefont {Baboux}\ \emph {et~al.}(2023)\citenamefont {Baboux}, \citenamefont {Moody},\ and\ \citenamefont {Ducci}}]{Baboux2023}%
  \BibitemOpen
  \bibfield  {author} {\bibinfo {author} {\bibfnamefont {F.}~\bibnamefont {Baboux}}, \bibinfo {author} {\bibfnamefont {G.}~\bibnamefont {Moody}},\ and\ \bibinfo {author} {\bibfnamefont {S.}~\bibnamefont {Ducci}},\ }\bibfield  {title} {\bibinfo {title} {Nonlinear integrated quantum photonics with AlGaAs},\ }\href {https://doi.org/10.1364/optica.481385} {\bibfield  {journal} {\bibinfo  {journal} {Optica}\ }\textbf {\bibinfo {volume} {10}},\ \bibinfo {pages} {917} (\bibinfo {year} {2023})}\BibitemShut {NoStop}%
\bibitem [{\citenamefont {Zhang}\ \emph {et~al.}(2023)\citenamefont {Zhang}, \citenamefont {Tran}, \citenamefont {Zhang}, \citenamefont {Dorche}, \citenamefont {Shen}, \citenamefont {Shen}, \citenamefont {Asawa}, \citenamefont {Kim}, \citenamefont {Kim}, \citenamefont {Levinson}, \citenamefont {Bowers},\ and\ \citenamefont {Komljenovic}}]{Zhang2023}%
  \BibitemOpen
  \bibfield  {author} {\bibinfo {author} {\bibfnamefont {C.}~\bibnamefont {Zhang}}, \bibinfo {author} {\bibfnamefont {M.~A.}\ \bibnamefont {Tran}}, \bibinfo {author} {\bibfnamefont {Z.}~\bibnamefont {Zhang}}, \bibinfo {author} {\bibfnamefont {A.~E.}\ \bibnamefont {Dorche}}, \bibinfo {author} {\bibfnamefont {Y.}~\bibnamefont {Shen}}, \bibinfo {author} {\bibfnamefont {B.}~\bibnamefont {Shen}}, \bibinfo {author} {\bibfnamefont {K.}~\bibnamefont {Asawa}}, \bibinfo {author} {\bibfnamefont {G.}~\bibnamefont {Kim}}, \bibinfo {author} {\bibfnamefont {N.}~\bibnamefont {Kim}}, \bibinfo {author} {\bibfnamefont {F.}~\bibnamefont {Levinson}}, \bibinfo {author} {\bibfnamefont {J.~E.}\ \bibnamefont {Bowers}},\ and\ \bibinfo {author} {\bibfnamefont {T.}~\bibnamefont {Komljenovic}},\ }\bibfield  {title} {\bibinfo {title} {Integrated photonics beyond communications},\ }\href {https://doi.org/10.1063/5.0184677} {\bibfield  {journal} {\bibinfo  {journal} {Applied Physics Letters}\ }\textbf {\bibinfo {volume} {123}},\ \bibinfo {pages} {230501}  (\bibinfo {year} {2023})}\BibitemShut
  {NoStop}%
\bibitem [{\citenamefont {Shekhar}\ \emph {et~al.}(2024)\citenamefont {Shekhar}, \citenamefont {Bogaerts}, \citenamefont {Chrostowski}, \citenamefont {Bowers}, \citenamefont {Hochberg}, \citenamefont {Soref},\ and\ \citenamefont {Shastri}}]{Shekhar2024}%
  \BibitemOpen
  \bibfield  {author} {\bibinfo {author} {\bibfnamefont {S.}~\bibnamefont {Shekhar}}, \bibinfo {author} {\bibfnamefont {W.}~\bibnamefont {Bogaerts}}, \bibinfo {author} {\bibfnamefont {L.}~\bibnamefont {Chrostowski}}, \bibinfo {author} {\bibfnamefont {J.~E.}\ \bibnamefont {Bowers}}, \bibinfo {author} {\bibfnamefont {M.}~\bibnamefont {Hochberg}}, \bibinfo {author} {\bibfnamefont {R.}~\bibnamefont {Soref}},\ and\ \bibinfo {author} {\bibfnamefont {B.~J.}\ \bibnamefont {Shastri}},\ }\bibfield  {title} {\bibinfo {title} {Roadmapping the next generation of silicon photonics},\ }\href {https://doi.org/10.1038/s41467-024-44750-0} {\bibfield  {journal} {\bibinfo  {journal} {Nature Communications}\ }\textbf {\bibinfo {volume} {15}},\ \bibinfo {pages} {751} (\bibinfo {year} {2024})}\BibitemShut {NoStop}%
\bibitem [{\citenamefont {Labont{\'e}}\ \emph {et~al.}(2024)\citenamefont {Labont{\'e}}, \citenamefont {Alibart}, \citenamefont {D’Auria}, \citenamefont {Doutre}, \citenamefont {Etesse}, \citenamefont {Sauder}, \citenamefont {Martin}, \citenamefont {Picholle},\ and\ \citenamefont {Tanzilli}}]{Labonte2024}%
  \BibitemOpen
  \bibfield  {author} {\bibinfo {author} {\bibfnamefont {L.}~\bibnamefont {Labont{\'e}}}, \bibinfo {author} {\bibfnamefont {O.}~\bibnamefont {Alibart}}, \bibinfo {author} {\bibfnamefont {V.}~\bibnamefont {D’Auria}}, \bibinfo {author} {\bibfnamefont {F.}~\bibnamefont {Doutre}}, \bibinfo {author} {\bibfnamefont {J.}~\bibnamefont {Etesse}}, \bibinfo {author} {\bibfnamefont {G.}~\bibnamefont {Sauder}}, \bibinfo {author} {\bibfnamefont {A.}~\bibnamefont {Martin}}, \bibinfo {author} {\bibfnamefont {{\'E}.}~\bibnamefont {Picholle}},\ and\ \bibinfo {author} {\bibfnamefont {S.}~\bibnamefont {Tanzilli}},\ }\bibfield  {title} {\bibinfo {title} {Integrated photonics for quantum communications and metrology},\ }\href {https://doi.org/10.1103/prxquantum.5.010101} {\bibfield  {journal} {\bibinfo  {journal} {PRX Quantum}\ }\textbf {\bibinfo {volume} {5}},\ \bibinfo {pages} {010101} (\bibinfo {year} {2024})}\BibitemShut {NoStop}%
\bibitem [{\citenamefont {Dutt}\ \emph {et~al.}(2024)\citenamefont {Dutt}, \citenamefont {Mohanty}, \citenamefont {Gaeta},\ and\ \citenamefont {Lipson}}]{Dutt2024}%
  \BibitemOpen
  \bibfield  {author} {\bibinfo {author} {\bibfnamefont {A.}~\bibnamefont {Dutt}}, \bibinfo {author} {\bibfnamefont {A.}~\bibnamefont {Mohanty}}, \bibinfo {author} {\bibfnamefont {A.~L.}\ \bibnamefont {Gaeta}},\ and\ \bibinfo {author} {\bibfnamefont {M.}~\bibnamefont {Lipson}},\ }\bibfield  {title} {\bibinfo {title} {Nonlinear and quantum photonics using integrated optical materials},\ }\href {https://doi.org/10.1038/s41578-024-00668-z} {\bibfield  {journal} {\bibinfo  {journal} {Nature Reviews Materials}\ }\textbf {\bibinfo {volume} {9}},\ \bibinfo {pages} {321} (\bibinfo {year} {2024})}\BibitemShut {NoStop}%
\bibitem [{\citenamefont {Reck}\ \emph {et~al.}(1994)\citenamefont {Reck}, \citenamefont {Zeilinger}, \citenamefont {Bernstein},\ and\ \citenamefont {Bertani}}]{Reck1994}%
  \BibitemOpen
  \bibfield  {author} {\bibinfo {author} {\bibfnamefont {M.}~\bibnamefont {Reck}}, \bibinfo {author} {\bibfnamefont {A.}~\bibnamefont {Zeilinger}}, \bibinfo {author} {\bibfnamefont {H.~J.}\ \bibnamefont {Bernstein}},\ and\ \bibinfo {author} {\bibfnamefont {P.}~\bibnamefont {Bertani}},\ }\bibfield  {title} {\bibinfo {title} {Experimental realization of any discrete unitary operator},\ }\href {https://doi.org/10.1103/physrevlett.73.58} {\bibfield  {journal} {\bibinfo  {journal} {Physical Review Letters}\ }\textbf {\bibinfo {volume} {73}},\ \bibinfo {pages} {58} (\bibinfo {year} {1994})}\BibitemShut {NoStop}%
\bibitem [{\citenamefont {Clements}\ \emph {et~al.}(2016)\citenamefont {Clements}, \citenamefont {Humphreys}, \citenamefont {Metcalf}, \citenamefont {Kolthammer},\ and\ \citenamefont {Walsmley}}]{Clements2016}%
  \BibitemOpen
  \bibfield  {author} {\bibinfo {author} {\bibfnamefont {W.~R.}\ \bibnamefont {Clements}}, \bibinfo {author} {\bibfnamefont {P.~C.}\ \bibnamefont {Humphreys}}, \bibinfo {author} {\bibfnamefont {B.~J.}\ \bibnamefont {Metcalf}}, \bibinfo {author} {\bibfnamefont {W.~S.}\ \bibnamefont {Kolthammer}},\ and\ \bibinfo {author} {\bibfnamefont {I.~A.}\ \bibnamefont {Walsmley}},\ }\bibfield  {title} {\bibinfo {title} {Optimal design for universal multiport interferometers},\ }\href {https://doi.org/10.1364/optica.3.001460} {\bibfield  {journal} {\bibinfo  {journal} {Optica}\ }\textbf {\bibinfo {volume} {3}},\ \bibinfo {pages} {1460} (\bibinfo {year} {2016})}\BibitemShut {NoStop}%
\bibitem [{\citenamefont {Taballione}\ \emph {et~al.}(2023)\citenamefont {Taballione}, \citenamefont {Anguita}, \citenamefont {de~Goede}, \citenamefont {Venderbosch}, \citenamefont {Kassenberg}, \citenamefont {Snijders}, \citenamefont {Kannan}, \citenamefont {Vleeshouwers}, \citenamefont {Smith}, \citenamefont {Epping}, \citenamefont {van~der Meer}, \citenamefont {Pinkse}, \citenamefont {van~den Vlekkert},\ and\ \citenamefont {Renema}}]{Taballione2023}%
  \BibitemOpen
  \bibfield  {author} {\bibinfo {author} {\bibfnamefont {C.}~\bibnamefont {Taballione}}, \bibinfo {author} {\bibfnamefont {M.~C.}\ \bibnamefont {Anguita}}, \bibinfo {author} {\bibfnamefont {M.}~\bibnamefont {de~Goede}}, \bibinfo {author} {\bibfnamefont {P.}~\bibnamefont {Venderbosch}}, \bibinfo {author} {\bibfnamefont {B.}~\bibnamefont {Kassenberg}}, \bibinfo {author} {\bibfnamefont {H.}~\bibnamefont {Snijders}}, \bibinfo {author} {\bibfnamefont {N.}~\bibnamefont {Kannan}}, \bibinfo {author} {\bibfnamefont {W.~L.}\ \bibnamefont {Vleeshouwers}}, \bibinfo {author} {\bibfnamefont {D.}~\bibnamefont {Smith}}, \bibinfo {author} {\bibfnamefont {J.~P.}\ \bibnamefont {Epping}}, \bibinfo {author} {\bibfnamefont {R.}~\bibnamefont {van~der Meer}}, \bibinfo {author} {\bibfnamefont {P.~W.~H.}\ \bibnamefont {Pinkse}}, \bibinfo {author} {\bibfnamefont {H.}~\bibnamefont {van~den Vlekkert}},\ and\ \bibinfo {author} {\bibfnamefont {J.~J.}\ \bibnamefont {Renema}},\ }\bibfield  {title} {\bibinfo {title} {20-{M}ode {U}niversal
  {Q}uantum {P}hotonic {P}rocessor},\ }\href {https://doi.org/10.22331/q-2023-08-01-1071} {\bibfield  {journal} {\bibinfo  {journal} {Quantum}\ }\textbf {\bibinfo {volume} {7}},\ \bibinfo {pages} {1071} (\bibinfo {year} {2023})}\BibitemShut {NoStop}%
\bibitem [{\citenamefont {Carolan}\ \emph {et~al.}(2015)\citenamefont {Carolan}, \citenamefont {Harrold}, \citenamefont {Sparrow}, \citenamefont {Martín-López}, \citenamefont {Russell}, \citenamefont {Silverstone}, \citenamefont {Shadbolt}, \citenamefont {Matsuda}, \citenamefont {Oguma}, \citenamefont {Itoh}, \citenamefont {Marshall}, \citenamefont {Thompson}, \citenamefont {Matthews}, \citenamefont {Hashimoto}, \citenamefont {O’Brien},\ and\ \citenamefont {Laing}}]{Carolan2015}%
  \BibitemOpen
  \bibfield  {author} {\bibinfo {author} {\bibfnamefont {J.}~\bibnamefont {Carolan}}, \bibinfo {author} {\bibfnamefont {C.}~\bibnamefont {Harrold}}, \bibinfo {author} {\bibfnamefont {C.}~\bibnamefont {Sparrow}}, \bibinfo {author} {\bibfnamefont {E.}~\bibnamefont {Martín-López}}, \bibinfo {author} {\bibfnamefont {N.~J.}\ \bibnamefont {Russell}}, \bibinfo {author} {\bibfnamefont {J.~W.}\ \bibnamefont {Silverstone}}, \bibinfo {author} {\bibfnamefont {P.~J.}\ \bibnamefont {Shadbolt}}, \bibinfo {author} {\bibfnamefont {N.}~\bibnamefont {Matsuda}}, \bibinfo {author} {\bibfnamefont {M.}~\bibnamefont {Oguma}}, \bibinfo {author} {\bibfnamefont {M.}~\bibnamefont {Itoh}}, \bibinfo {author} {\bibfnamefont {G.~D.}\ \bibnamefont {Marshall}}, \bibinfo {author} {\bibfnamefont {M.~G.}\ \bibnamefont {Thompson}}, \bibinfo {author} {\bibfnamefont {J.~C.~F.}\ \bibnamefont {Matthews}}, \bibinfo {author} {\bibfnamefont {T.}~\bibnamefont {Hashimoto}}, \bibinfo {author} {\bibfnamefont {J.~L.}\ \bibnamefont {O’Brien}},\ and\
  \bibinfo {author} {\bibfnamefont {A.}~\bibnamefont {Laing}},\ }\bibfield  {title} {\bibinfo {title} {Universal linear optics},\ }\href {https://doi.org/10.1126/science.aab3642} {\bibfield  {journal} {\bibinfo  {journal} {Science}\ }\textbf {\bibinfo {volume} {349}},\ \bibinfo {pages} {711} (\bibinfo {year} {2015})}\BibitemShut {NoStop}%
\bibitem [{\citenamefont {Maring}\ \emph {et~al.}(2024)\citenamefont {Maring}, \citenamefont {Fyrillas}, \citenamefont {Pont}, \citenamefont {Ivanov}, \citenamefont {Stepanov}, \citenamefont {Margaria}, \citenamefont {Hease}, \citenamefont {Pishchagin}, \citenamefont {Lemaître}, \citenamefont {Sagnes}, \citenamefont {Au}, \citenamefont {Boissier}, \citenamefont {Bertasi}, \citenamefont {Baert}, \citenamefont {Valdivia}, \citenamefont {Billard}, \citenamefont {Acar}, \citenamefont {Brieussel}, \citenamefont {Mezher}, \citenamefont {Wein}, \citenamefont {Salavrakos}, \citenamefont {Sinnott}, \citenamefont {Fioretto}, \citenamefont {Emeriau}, \citenamefont {Belabas}, \citenamefont {Mansfield}, \citenamefont {Senellart}, \citenamefont {Senellart},\ and\ \citenamefont {Somaschi}}]{Maring2024}%
  \BibitemOpen
  \bibfield  {author} {\bibinfo {author} {\bibfnamefont {N.}~\bibnamefont {Maring}}, \bibinfo {author} {\bibfnamefont {A.}~\bibnamefont {Fyrillas}}, \bibinfo {author} {\bibfnamefont {M.}~\bibnamefont {Pont}}, \bibinfo {author} {\bibfnamefont {E.}~\bibnamefont {Ivanov}}, \bibinfo {author} {\bibfnamefont {P.}~\bibnamefont {Stepanov}}, \bibinfo {author} {\bibfnamefont {N.}~\bibnamefont {Margaria}}, \bibinfo {author} {\bibfnamefont {W.}~\bibnamefont {Hease}}, \bibinfo {author} {\bibfnamefont {A.}~\bibnamefont {Pishchagin}}, \bibinfo {author} {\bibfnamefont {A.}~\bibnamefont {Lemaître}}, \bibinfo {author} {\bibfnamefont {I.}~\bibnamefont {Sagnes}}, \bibinfo {author} {\bibfnamefont {T.~H.}\ \bibnamefont {Au}}, \bibinfo {author} {\bibfnamefont {S.}~\bibnamefont {Boissier}}, \bibinfo {author} {\bibfnamefont {E.}~\bibnamefont {Bertasi}}, \bibinfo {author} {\bibfnamefont {A.}~\bibnamefont {Baert}}, \bibinfo {author} {\bibfnamefont {M.}~\bibnamefont {Valdivia}}, \bibinfo {author} {\bibfnamefont {M.}~\bibnamefont
  {Billard}}, \bibinfo {author} {\bibfnamefont {O.}~\bibnamefont {Acar}}, \bibinfo {author} {\bibfnamefont {A.}~\bibnamefont {Brieussel}}, \bibinfo {author} {\bibfnamefont {R.}~\bibnamefont {Mezher}}, \bibinfo {author} {\bibfnamefont {S.~C.}\ \bibnamefont {Wein}}, \bibinfo {author} {\bibfnamefont {A.}~\bibnamefont {Salavrakos}}, \bibinfo {author} {\bibfnamefont {P.}~\bibnamefont {Sinnott}}, \bibinfo {author} {\bibfnamefont {D.~A.}\ \bibnamefont {Fioretto}}, \bibinfo {author} {\bibfnamefont {P.-E.}\ \bibnamefont {Emeriau}}, \bibinfo {author} {\bibfnamefont {N.}~\bibnamefont {Belabas}}, \bibinfo {author} {\bibfnamefont {S.}~\bibnamefont {Mansfield}}, \bibinfo {author} {\bibfnamefont {P.}~\bibnamefont {Senellart}}, \bibinfo {author} {\bibfnamefont {J.}~\bibnamefont {Senellart}},\ and\ \bibinfo {author} {\bibfnamefont {N.}~\bibnamefont {Somaschi}},\ }\bibfield  {title} {\bibinfo {title} {A versatile single-photon-based quantum computing platform},\ }\href {https://doi.org/10.1038/s41566-024-01403-4} {\bibfield
  {journal} {\bibinfo  {journal} {Nature Photonics}\ }\textbf {\bibinfo {volume} {18}},\ \bibinfo {pages} {603} (\bibinfo {year} {2024})}\BibitemShut {NoStop}%
\bibitem [{\citenamefont {Peruzzo}\ \emph {et~al.}(2010)\citenamefont {Peruzzo}, \citenamefont {Lobino}, \citenamefont {Matthews}, \citenamefont {Matsuda}, \citenamefont {Politi}, \citenamefont {Poulios}, \citenamefont {Zhou}, \citenamefont {Lahini}, \citenamefont {Ismail}, \citenamefont {Wörhoff}, \citenamefont {Bromberg}, \citenamefont {Silberberg}, \citenamefont {Thompson},\ and\ \citenamefont {OBrien}}]{Peruzzo2010}%
  \BibitemOpen
  \bibfield  {author} {\bibinfo {author} {\bibfnamefont {A.}~\bibnamefont {Peruzzo}}, \bibinfo {author} {\bibfnamefont {M.}~\bibnamefont {Lobino}}, \bibinfo {author} {\bibfnamefont {J.~C.~F.}\ \bibnamefont {Matthews}}, \bibinfo {author} {\bibfnamefont {N.}~\bibnamefont {Matsuda}}, \bibinfo {author} {\bibfnamefont {A.}~\bibnamefont {Politi}}, \bibinfo {author} {\bibfnamefont {K.}~\bibnamefont {Poulios}}, \bibinfo {author} {\bibfnamefont {X.-Q.}\ \bibnamefont {Zhou}}, \bibinfo {author} {\bibfnamefont {Y.}~\bibnamefont {Lahini}}, \bibinfo {author} {\bibfnamefont {N.}~\bibnamefont {Ismail}}, \bibinfo {author} {\bibfnamefont {K.}~\bibnamefont {Wörhoff}}, \bibinfo {author} {\bibfnamefont {Y.}~\bibnamefont {Bromberg}}, \bibinfo {author} {\bibfnamefont {Y.}~\bibnamefont {Silberberg}}, \bibinfo {author} {\bibfnamefont {M.~G.}\ \bibnamefont {Thompson}},\ and\ \bibinfo {author} {\bibfnamefont {J.~L.}\ \bibnamefont {OBrien}},\ }\bibfield  {title} {\bibinfo {title} {Quantum walks of correlated photons},\ }\href
  {https://doi.org/10.1126/science.1193515} {\bibfield  {journal} {\bibinfo  {journal} {Science}\ }\textbf {\bibinfo {volume} {329}},\ \bibinfo {pages} {1500} (\bibinfo {year} {2010})}\BibitemShut {NoStop}%
\bibitem [{\citenamefont {Weimann}\ \emph {et~al.}(2016)\citenamefont {Weimann}, \citenamefont {Perez-Leija}, \citenamefont {Lebugle}, \citenamefont {Keil}, \citenamefont {Tichy}, \citenamefont {Gräfe}, \citenamefont {Heilmann}, \citenamefont {Nolte}, \citenamefont {Moya-Cessa}, \citenamefont {Weihs}, \citenamefont {Christodoulides},\ and\ \citenamefont {Szameit}}]{Weimann2016}%
  \BibitemOpen
  \bibfield  {author} {\bibinfo {author} {\bibfnamefont {S.}~\bibnamefont {Weimann}}, \bibinfo {author} {\bibfnamefont {A.}~\bibnamefont {Perez-Leija}}, \bibinfo {author} {\bibfnamefont {M.}~\bibnamefont {Lebugle}}, \bibinfo {author} {\bibfnamefont {R.}~\bibnamefont {Keil}}, \bibinfo {author} {\bibfnamefont {M.}~\bibnamefont {Tichy}}, \bibinfo {author} {\bibfnamefont {M.}~\bibnamefont {Gräfe}}, \bibinfo {author} {\bibfnamefont {R.}~\bibnamefont {Heilmann}}, \bibinfo {author} {\bibfnamefont {S.}~\bibnamefont {Nolte}}, \bibinfo {author} {\bibfnamefont {H.}~\bibnamefont {Moya-Cessa}}, \bibinfo {author} {\bibfnamefont {G.}~\bibnamefont {Weihs}}, \bibinfo {author} {\bibfnamefont {D.~N.}\ \bibnamefont {Christodoulides}},\ and\ \bibinfo {author} {\bibfnamefont {A.}~\bibnamefont {Szameit}},\ }\bibfield  {title} {\bibinfo {title} {Implementation of quantum and classical discrete fractional fourier transforms},\ }\href {https://doi.org/10.1038/ncomms11027} {\bibfield  {journal} {\bibinfo  {journal} {Nature Communications}\ }\textbf {\bibinfo
  {volume} {7}},\ \bibinfo {pages} {11027}  (\bibinfo {year} {2016})}\BibitemShut {NoStop}%
\bibitem [{\citenamefont {Solntsev}\ \emph {et~al.}(2014)\citenamefont {Solntsev}, \citenamefont {Setzpfandt}, \citenamefont {Clark}, \citenamefont {Wu}, \citenamefont {Collins}, \citenamefont {Xiong}, \citenamefont {Schreiber}, \citenamefont {Katzschmann}, \citenamefont {Eilenberger}, \citenamefont {Schiek}, \citenamefont {Sohler}, \citenamefont {Mitchell}, \citenamefont {Silberhorn}, \citenamefont {Eggleton}, \citenamefont {Pertsch}, \citenamefont {Sukhorukov}, \citenamefont {Neshev},\ and\ \citenamefont {Kivshar}}]{Solntsev2014}%
  \BibitemOpen
  \bibfield  {author} {\bibinfo {author} {\bibfnamefont {A.~S.}\ \bibnamefont {Solntsev}}, \bibinfo {author} {\bibfnamefont {F.}~\bibnamefont {Setzpfandt}}, \bibinfo {author} {\bibfnamefont {A.~S.}\ \bibnamefont {Clark}}, \bibinfo {author} {\bibfnamefont {C.~W.}\ \bibnamefont {Wu}}, \bibinfo {author} {\bibfnamefont {M.~J.}\ \bibnamefont {Collins}}, \bibinfo {author} {\bibfnamefont {C.}~\bibnamefont {Xiong}}, \bibinfo {author} {\bibfnamefont {A.}~\bibnamefont {Schreiber}}, \bibinfo {author} {\bibfnamefont {F.}~\bibnamefont {Katzschmann}}, \bibinfo {author} {\bibfnamefont {F.}~\bibnamefont {Eilenberger}}, \bibinfo {author} {\bibfnamefont {R.}~\bibnamefont {Schiek}}, \bibinfo {author} {\bibfnamefont {W.}~\bibnamefont {Sohler}}, \bibinfo {author} {\bibfnamefont {A.}~\bibnamefont {Mitchell}}, \bibinfo {author} {\bibfnamefont {C.}~\bibnamefont {Silberhorn}}, \bibinfo {author} {\bibfnamefont {B.~J.}\ \bibnamefont {Eggleton}}, \bibinfo {author} {\bibfnamefont {T.}~\bibnamefont {Pertsch}}, \bibinfo {author}
  {\bibfnamefont {A.~A.}\ \bibnamefont {Sukhorukov}}, \bibinfo {author} {\bibfnamefont {D.~N.}\ \bibnamefont {Neshev}},\ and\ \bibinfo {author} {\bibfnamefont {Y.~S.}\ \bibnamefont {Kivshar}},\ }\bibfield  {title} {\bibinfo {title} {Generation of nonclassical biphoton states through cascaded quantum walks on a nonlinear chip},\ }\href {https://doi.org/10.1103/physrevx.4.031007} {\bibfield  {journal} {\bibinfo  {journal} {Physical Review X}\ }\textbf {\bibinfo {volume} {4}},\ \bibinfo {pages} {031007} (\bibinfo {year} {2014})}\BibitemShut {NoStop}%
\bibitem [{\citenamefont {Christodoulides}\ \emph {et~al.}(2003)\citenamefont {Christodoulides}, \citenamefont {Lederer},\ and\ \citenamefont {Silberberg}}]{Christodoulides2003}%
  \BibitemOpen
  \bibfield  {author} {\bibinfo {author} {\bibfnamefont {D.~N.}\ \bibnamefont {Christodoulides}}, \bibinfo {author} {\bibfnamefont {F.}~\bibnamefont {Lederer}},\ and\ \bibinfo {author} {\bibfnamefont {Y.}~\bibnamefont {Silberberg}},\ }\bibfield  {title} {\bibinfo {title} {Discretizing light behaviour in linear and nonlinear waveguide lattices},\ }\href {https://doi.org/10.1038/nature01936} {\bibfield  {journal} {\bibinfo  {journal} {Nature}\ }\textbf {\bibinfo {volume} {424}},\ \bibinfo {pages} {817} (\bibinfo {year} {2003})}\BibitemShut {NoStop}%
\bibitem [{\citenamefont {Barral}\ \emph {et~al.}(2020{\natexlab{a}})\citenamefont {Barral}, \citenamefont {Walschaers}, \citenamefont {Bencheikh}, \citenamefont {Parigi}, \citenamefont {Levenson}, \citenamefont {Treps},\ and\ \citenamefont {Belabas}}]{Barral2020a}%
  \BibitemOpen
  \bibfield  {author} {\bibinfo {author} {\bibfnamefont {D.}~\bibnamefont {Barral}}, \bibinfo {author} {\bibfnamefont {M.}~\bibnamefont {Walschaers}}, \bibinfo {author} {\bibfnamefont {K.}~\bibnamefont {Bencheikh}}, \bibinfo {author} {\bibfnamefont {V.}~\bibnamefont {Parigi}}, \bibinfo {author} {\bibfnamefont {J.~A.}\ \bibnamefont {Levenson}}, \bibinfo {author} {\bibfnamefont {N.}~\bibnamefont {Treps}},\ and\ \bibinfo {author} {\bibfnamefont {N.}~\bibnamefont {Belabas}},\ }\bibfield  {title} {\bibinfo {title} {Versatile photonic entanglement synthesizer in the spatial domain},\ }\href {https://doi.org/10.1103/physrevapplied.14.044025} {\bibfield  {journal} {\bibinfo  {journal} {Physical Review Applied}\ }\textbf {\bibinfo {volume} {14}},\ \bibinfo {pages} {044025} (\bibinfo {year} {2020}{\natexlab{a}})}\BibitemShut {NoStop}%
\bibitem [{\citenamefont {Kruse}\ \emph {et~al.}(2013)\citenamefont {Kruse}, \citenamefont {Katzschmann}, \citenamefont {Christ}, \citenamefont {Schreiber}, \citenamefont {Wilhelm}, \citenamefont {Laiho}, \citenamefont {Gábris}, \citenamefont {Hamilton}, \citenamefont {Jex},\ and\ \citenamefont {Silberhorn}}]{Kruse2013}%
  \BibitemOpen
  \bibfield  {author} {\bibinfo {author} {\bibfnamefont {R.}~\bibnamefont {Kruse}}, \bibinfo {author} {\bibfnamefont {F.}~\bibnamefont {Katzschmann}}, \bibinfo {author} {\bibfnamefont {A.}~\bibnamefont {Christ}}, \bibinfo {author} {\bibfnamefont {A.}~\bibnamefont {Schreiber}}, \bibinfo {author} {\bibfnamefont {S.}~\bibnamefont {Wilhelm}}, \bibinfo {author} {\bibfnamefont {K.}~\bibnamefont {Laiho}}, \bibinfo {author} {\bibfnamefont {A.}~\bibnamefont {Gábris}}, \bibinfo {author} {\bibfnamefont {C.~S.}\ \bibnamefont {Hamilton}}, \bibinfo {author} {\bibfnamefont {I.}~\bibnamefont {Jex}},\ and\ \bibinfo {author} {\bibfnamefont {C.}~\bibnamefont {Silberhorn}},\ }\bibfield  {title} {\bibinfo {title} {Spatio-spectral characteristics of parametric down-conversion in waveguide arrays},\ }\href {https://doi.org/10.1088/1367-2630/15/8/083046} {\bibfield  {journal} {\bibinfo  {journal} {New Journal of Physics}\ }\textbf {\bibinfo {volume} {15}},\ \bibinfo {pages} {083046} (\bibinfo {year} {2013})}\BibitemShut {NoStop}%
\bibitem [{\citenamefont {Raymond}\ \emph {et~al.}(2024)\citenamefont {Raymond}, \citenamefont {Zecchetto}, \citenamefont {Palomo}, \citenamefont {Morassi}, \citenamefont {Lemaître}, \citenamefont {Raineri}, \citenamefont {Amanti}, \citenamefont {Ducci},\ and\ \citenamefont {Baboux}}]{Raymond2024}%
  \BibitemOpen
  \bibfield  {author} {\bibinfo {author} {\bibfnamefont {A.}~\bibnamefont {Raymond}}, \bibinfo {author} {\bibfnamefont {A.}~\bibnamefont {Zecchetto}}, \bibinfo {author} {\bibfnamefont {J.}~\bibnamefont {Palomo}}, \bibinfo {author} {\bibfnamefont {M.}~\bibnamefont {Morassi}}, \bibinfo {author} {\bibfnamefont {A.}~\bibnamefont {Lemaître}}, \bibinfo {author} {\bibfnamefont {F.}~\bibnamefont {Raineri}}, \bibinfo {author} {\bibfnamefont {M.}~\bibnamefont {Amanti}}, \bibinfo {author} {\bibfnamefont {S.}~\bibnamefont {Ducci}},\ and\ \bibinfo {author} {\bibfnamefont {F.}~\bibnamefont {Baboux}},\ }\bibfield  {title} {\bibinfo {title} {Tunable generation of spatial entanglement in nonlinear waveguide arrays},\ }\href {https://doi.org/10.1103/physrevlett.133.233602} {\bibfield  {journal} {\bibinfo  {journal} {Physical Review Letters}\ }\textbf {\bibinfo {volume} {133}},\ \bibinfo {pages} {233602} (\bibinfo {year} {2024})}\BibitemShut {NoStop}%
\bibitem [{\citenamefont {Herec}\ \emph {et~al.}(2003)\citenamefont {Herec}, \citenamefont {Fiurá\v{s}ek},\ and\ \citenamefont {Mi\v{s}ta}}]{Herec2003}%
  \BibitemOpen
  \bibfield  {author} {\bibinfo {author} {\bibfnamefont {J.}~\bibnamefont {Herec}}, \bibinfo {author} {\bibfnamefont {J.}~\bibnamefont {Fiurá\v{s}ek}},\ and\ \bibinfo {author} {\bibfnamefont {L.}~\bibnamefont {Mi\v{s}ta}},\ }\bibfield  {title} {\bibinfo {title} {Entanglement generation in continuously coupled parametric generators},\ }\href {https://doi.org/10.1088/1464-4266/5/5/008} {\bibfield  {journal} {\bibinfo  {journal} {Journal of Optics B: Quantum and Semiclassical Optics}\ }\textbf {\bibinfo {volume} {5}},\ \bibinfo {pages} {419} (\bibinfo {year} {2003})}\BibitemShut {NoStop}%
\bibitem [{\citenamefont {Rai}\ and\ \citenamefont {Angelakis}(2012)}]{Rai2012}%
  \BibitemOpen
  \bibfield  {author} {\bibinfo {author} {\bibfnamefont {A.}~\bibnamefont {Rai}}\ and\ \bibinfo {author} {\bibfnamefont {D.~G.}\ \bibnamefont {Angelakis}},\ }\bibfield  {title} {\bibinfo {title} {Dynamics of nonclassical light in integrated nonlinear waveguide arrays and generation of robust continuous-variable entanglement},\ }\href {https://doi.org/10.1103/physreva.85.052330} {\bibfield  {journal} {\bibinfo  {journal} {Physical Review A}\ }\textbf {\bibinfo {volume} {85}},\ \bibinfo {pages} {052330} (\bibinfo {year} {2012})}\BibitemShut {NoStop}%
\bibitem [{\citenamefont {Barral}\ \emph {et~al.}(2020{\natexlab{b}})\citenamefont {Barral}, \citenamefont {Walschaers}, \citenamefont {Bencheikh}, \citenamefont {Parigi}, \citenamefont {Levenson}, \citenamefont {Treps},\ and\ \citenamefont {Belabas}}]{Barral2020}%
  \BibitemOpen
  \bibfield  {author} {\bibinfo {author} {\bibfnamefont {D.}~\bibnamefont {Barral}}, \bibinfo {author} {\bibfnamefont {M.}~\bibnamefont {Walschaers}}, \bibinfo {author} {\bibfnamefont {K.}~\bibnamefont {Bencheikh}}, \bibinfo {author} {\bibfnamefont {V.}~\bibnamefont {Parigi}}, \bibinfo {author} {\bibfnamefont {J.~A.}\ \bibnamefont {Levenson}}, \bibinfo {author} {\bibfnamefont {N.}~\bibnamefont {Treps}},\ and\ \bibinfo {author} {\bibfnamefont {N.}~\bibnamefont {Belabas}},\ }\bibfield  {title} {\bibinfo {title} {Quantum state engineering in arrays of nonlinear waveguides},\ }\href {https://doi.org/10.1103/physreva.102.043706} {\bibfield  {journal} {\bibinfo  {journal} {Physical Review A}\ }\textbf {\bibinfo {volume} {102}},\ \bibinfo {pages} {043706} (\bibinfo {year} {2020}{\natexlab{b}})}\BibitemShut {NoStop}%
\bibitem [{\citenamefont {Bacaoco}\ \emph {et~al.}(2024)\citenamefont {Bacaoco}, \citenamefont {Galettis}, \citenamefont {Huang}, \citenamefont {Ilin},\ and\ \citenamefont {Solntsev}}]{Bacaoco2024}%
  \BibitemOpen
  \bibfield  {author} {\bibinfo {author} {\bibfnamefont {M.~Y.}\ \bibnamefont {Bacaoco}}, \bibinfo {author} {\bibfnamefont {M.~J.}\ \bibnamefont {Galettis}}, \bibinfo {author} {\bibfnamefont {J.}~\bibnamefont {Huang}}, \bibinfo {author} {\bibfnamefont {D.}~\bibnamefont {Ilin}},\ and\ \bibinfo {author} {\bibfnamefont {A.~S.}\ \bibnamefont {Solntsev}},\ }\bibfield  {title} {\bibinfo {title} {{G}eneration of {T}unable {T}hree‐{P}hoton {E}ntanglement in {C}ubic {N}onlinear {C}oupled {W}aveguides},\ }\href {https://doi.org/10.1002/qute.202400409} {\bibfield  {journal} {\bibinfo  {journal} {Advanced Quantum Technologies}\ }\textbf {\bibinfo {volume} {8}},\ \bibinfo {pages} {2400409}  (\bibinfo {year} {2025})}\BibitemShut {NoStop}%
\bibitem [{\citenamefont {Hamilton}\ \emph {et~al.}(2014)\citenamefont {Hamilton}, \citenamefont {Kruse}, \citenamefont {Sansoni}, \citenamefont {Silberhorn},\ and\ \citenamefont {Jex}}]{Hamilton2014}%
  \BibitemOpen
  \bibfield  {author} {\bibinfo {author} {\bibfnamefont {C.~S.}\ \bibnamefont {Hamilton}}, \bibinfo {author} {\bibfnamefont {R.}~\bibnamefont {Kruse}}, \bibinfo {author} {\bibfnamefont {L.}~\bibnamefont {Sansoni}}, \bibinfo {author} {\bibfnamefont {C.}~\bibnamefont {Silberhorn}},\ and\ \bibinfo {author} {\bibfnamefont {I.}~\bibnamefont {Jex}},\ }\bibfield  {title} {\bibinfo {title} {Driven quantum walks},\ }\href {https://doi.org/10.1103/physrevlett.113.083602} {\bibfield  {journal} {\bibinfo  {journal} {Physical Review Letters}\ }\textbf {\bibinfo {volume} {113}},\ \bibinfo {pages} {083602} (\bibinfo {year} {2014})}\BibitemShut {NoStop}%
\bibitem [{\citenamefont {Leykam}\ \emph {et~al.}(2015)\citenamefont {Leykam}, \citenamefont {Solntsev}, \citenamefont {Sukhorukov},\ and\ \citenamefont {Desyatnikov}}]{Leykam2015}%
  \BibitemOpen
  \bibfield  {author} {\bibinfo {author} {\bibfnamefont {D.}~\bibnamefont {Leykam}}, \bibinfo {author} {\bibfnamefont {A.~S.}\ \bibnamefont {Solntsev}}, \bibinfo {author} {\bibfnamefont {A.~A.}\ \bibnamefont {Sukhorukov}},\ and\ \bibinfo {author} {\bibfnamefont {A.~S.}\ \bibnamefont {Desyatnikov}},\ }\bibfield  {title} {\bibinfo {title} {Lattice topology and spontaneous parametric down-conversion in quadratic nonlinear waveguide arrays},\ }\href {https://doi.org/10.1103/physreva.92.033815} {\bibfield  {journal} {\bibinfo  {journal} {Physical Review A}\ }\textbf {\bibinfo {volume} {92}},\ \bibinfo {pages} {033815} (\bibinfo {year} {2015})}\BibitemShut {NoStop}%
\bibitem [{\citenamefont {Doyle}\ \emph {et~al.}(2022)\citenamefont {Doyle}, \citenamefont {Zhang}, \citenamefont {Wang}, \citenamefont {Bell}, \citenamefont {Bartlett},\ and\ \citenamefont {Blanco-Redondo}}]{Doyle2022}%
  \BibitemOpen
  \bibfield  {author} {\bibinfo {author} {\bibfnamefont {C.}~\bibnamefont {Doyle}}, \bibinfo {author} {\bibfnamefont {W.-W.}\ \bibnamefont {Zhang}}, \bibinfo {author} {\bibfnamefont {M.}~\bibnamefont {Wang}}, \bibinfo {author} {\bibfnamefont {B.~A.}\ \bibnamefont {Bell}}, \bibinfo {author} {\bibfnamefont {S.~D.}\ \bibnamefont {Bartlett}},\ and\ \bibinfo {author} {\bibfnamefont {A.}~\bibnamefont {Blanco-Redondo}},\ }\bibfield  {title} {\bibinfo {title} {Biphoton entanglement of topologically distinct modes},\ }\href {https://doi.org/10.1103/physreva.105.023513} {\bibfield  {journal} {\bibinfo  {journal} {Physical Review A}\ }\textbf {\bibinfo {volume} {105}},\ \bibinfo {pages} {023513} (\bibinfo {year} {2022})}\BibitemShut {NoStop}%
\bibitem [{\citenamefont {Bai}\ \emph {et~al.}(2016)\citenamefont {Bai}, \citenamefont {Xu}, \citenamefont {Lu}, \citenamefont {Zhong},\ and\ \citenamefont {Zhu}}]{Bai2016}%
  \BibitemOpen
  \bibfield  {author} {\bibinfo {author} {\bibfnamefont {Y.~F.}\ \bibnamefont {Bai}}, \bibinfo {author} {\bibfnamefont {P.}~\bibnamefont {Xu}}, \bibinfo {author} {\bibfnamefont {L.~L.}\ \bibnamefont {Lu}}, \bibinfo {author} {\bibfnamefont {M.~L.}\ \bibnamefont {Zhong}},\ and\ \bibinfo {author} {\bibfnamefont {S.~N.}\ \bibnamefont {Zhu}},\ }\bibfield  {title} {\bibinfo {title} {Two-photon anderson localization in a disordered quadratic waveguide array},\ }\href {https://doi.org/10.1088/2040-8978/18/5/055201} {\bibfield  {journal} {\bibinfo  {journal} {Journal of Optics}\ }\textbf {\bibinfo {volume} {18}},\ \bibinfo {pages} {055201} (\bibinfo {year} {2016})}\BibitemShut {NoStop}%
\bibitem [{\citenamefont {Yang}\ \emph {et~al.}(2014)\citenamefont {Yang}, \citenamefont {Xu}, \citenamefont {Lu},\ and\ \citenamefont {Zhu}}]{Yang2014}%
  \BibitemOpen
  \bibfield  {author} {\bibinfo {author} {\bibfnamefont {Y.}~\bibnamefont {Yang}}, \bibinfo {author} {\bibfnamefont {P.}~\bibnamefont {Xu}}, \bibinfo {author} {\bibfnamefont {L.~L.}\ \bibnamefont {Lu}},\ and\ \bibinfo {author} {\bibfnamefont {S.~N.}\ \bibnamefont {Zhu}},\ }\bibfield  {title} {\bibinfo {title} {Manipulation of a two-photon state in a $\chi^{(2)}$-modulated nonlinear waveguide array},\ }\href {https://doi.org/10.1103/physreva.90.043842} {\bibfield  {journal} {\bibinfo  {journal} {Physical Review A}\ }\textbf {\bibinfo {volume} {90}},\ \bibinfo {pages} {043842} (\bibinfo {year} {2014})}\BibitemShut {NoStop}%
\bibitem [{\citenamefont {Hamilton}\ \emph {et~al.}(2022)\citenamefont {Hamilton}, \citenamefont {Christ}, \citenamefont {Barkhofen}, \citenamefont {Barnett}, \citenamefont {Jex},\ and\ \citenamefont {Silberhorn}}]{Hamilton2022}%
  \BibitemOpen
  \bibfield  {author} {\bibinfo {author} {\bibfnamefont {C.~S.}\ \bibnamefont {Hamilton}}, \bibinfo {author} {\bibfnamefont {R.}~\bibnamefont {Christ}}, \bibinfo {author} {\bibfnamefont {S.}~\bibnamefont {Barkhofen}}, \bibinfo {author} {\bibfnamefont {S.~M.}\ \bibnamefont {Barnett}}, \bibinfo {author} {\bibfnamefont {I.}~\bibnamefont {Jex}},\ and\ \bibinfo {author} {\bibfnamefont {C.}~\bibnamefont {Silberhorn}},\ }\bibfield  {title} {\bibinfo {title} {Quantum-state creation in nonlinear-waveguide arrays},\ }\href {https://doi.org/10.1103/physreva.105.042622} {\bibfield  {journal} {\bibinfo  {journal} {Physical Review A}\ }\textbf {\bibinfo {volume} {105}},\ \bibinfo {pages} {042622} (\bibinfo {year} {2022})}\BibitemShut {NoStop}%
\bibitem [{\citenamefont {Gisin}\ and\ \citenamefont {Thew}(2007)}]{Gisin2007}%
  \BibitemOpen
  \bibfield  {author} {\bibinfo {author} {\bibfnamefont {N.}~\bibnamefont {Gisin}}\ and\ \bibinfo {author} {\bibfnamefont {R.}~\bibnamefont {Thew}},\ }\bibfield  {title} {\bibinfo {title} {Quantum communication},\ }\href {https://doi.org/10.1038/nphoton.2007.22} {\bibfield  {journal} {\bibinfo  {journal} {Nature Photonics}\ }\textbf {\bibinfo {volume} {1}},\ \bibinfo {pages} {165} (\bibinfo {year} {2007})}\BibitemShut {NoStop}%
\bibitem [{\citenamefont {Couteau}\ \emph {et~al.}(2023)\citenamefont {Couteau}, \citenamefont {Barz}, \citenamefont {Durt}, \citenamefont {Gerrits}, \citenamefont {Huwer}, \citenamefont {Prevedel}, \citenamefont {Rarity}, \citenamefont {Shields},\ and\ \citenamefont {Weihs}}]{Couteau2023}%
  \BibitemOpen
  \bibfield  {author} {\bibinfo {author} {\bibfnamefont {C.}~\bibnamefont {Couteau}}, \bibinfo {author} {\bibfnamefont {S.}~\bibnamefont {Barz}}, \bibinfo {author} {\bibfnamefont {T.}~\bibnamefont {Durt}}, \bibinfo {author} {\bibfnamefont {T.}~\bibnamefont {Gerrits}}, \bibinfo {author} {\bibfnamefont {J.}~\bibnamefont {Huwer}}, \bibinfo {author} {\bibfnamefont {R.}~\bibnamefont {Prevedel}}, \bibinfo {author} {\bibfnamefont {J.}~\bibnamefont {Rarity}}, \bibinfo {author} {\bibfnamefont {A.}~\bibnamefont {Shields}},\ and\ \bibinfo {author} {\bibfnamefont {G.}~\bibnamefont {Weihs}},\ }\bibfield  {title} {\bibinfo {title} {Applications of single photons to quantum communication and computing},\ }\href {https://doi.org/10.1038/s42254-023-00583-2} {\bibfield  {journal} {\bibinfo  {journal} {Nature Reviews Physics}\ }\textbf {\bibinfo {volume} {5}},\ \bibinfo {pages} {326} (\bibinfo {year} {2023})}\BibitemShut {NoStop}%
\bibitem [{\citenamefont {Luo}\ \emph {et~al.}(2023)\citenamefont {Luo}, \citenamefont {Cao}, \citenamefont {Shi}, \citenamefont {Wan}, \citenamefont {Zhang}, \citenamefont {Li}, \citenamefont {Chen}, \citenamefont {Li}, \citenamefont {Li}, \citenamefont {Wang}, \citenamefont {Sun}, \citenamefont {Karim}, \citenamefont {Cai}, \citenamefont {Kwek},\ and\ \citenamefont {Liu}}]{Luo2023}%
  \BibitemOpen
  \bibfield  {author} {\bibinfo {author} {\bibfnamefont {W.}~\bibnamefont {Luo}}, \bibinfo {author} {\bibfnamefont {L.}~\bibnamefont {Cao}}, \bibinfo {author} {\bibfnamefont {Y.}~\bibnamefont {Shi}}, \bibinfo {author} {\bibfnamefont {L.}~\bibnamefont {Wan}}, \bibinfo {author} {\bibfnamefont {H.}~\bibnamefont {Zhang}}, \bibinfo {author} {\bibfnamefont {S.}~\bibnamefont {Li}}, \bibinfo {author} {\bibfnamefont {G.}~\bibnamefont {Chen}}, \bibinfo {author} {\bibfnamefont {Y.}~\bibnamefont {Li}}, \bibinfo {author} {\bibfnamefont {S.}~\bibnamefont {Li}}, \bibinfo {author} {\bibfnamefont {Y.}~\bibnamefont {Wang}}, \bibinfo {author} {\bibfnamefont {S.}~\bibnamefont {Sun}}, \bibinfo {author} {\bibfnamefont {M.~F.}\ \bibnamefont {Karim}}, \bibinfo {author} {\bibfnamefont {H.}~\bibnamefont {Cai}}, \bibinfo {author} {\bibfnamefont {L.~C.}\ \bibnamefont {Kwek}},\ and\ \bibinfo {author} {\bibfnamefont {A.~Q.}\ \bibnamefont {Liu}},\ }\bibfield  {title} {\bibinfo {title} {Recent progress in quantum photonic chips for quantum communication and internet},\ }\href {https://doi.org/10.1038/s41377-023-01173-8} {\bibfield  {journal} {\bibinfo  {journal} {Light: Science \& Applications}\ }\textbf {\bibinfo {volume} {12}},\ \bibinfo {pages} {175} (\bibinfo {year} {2023})}\BibitemShut {NoStop}%
\bibitem [{\citenamefont {Kok}\ \emph {et~al.}(2007)\citenamefont {Kok}, \citenamefont {Munro}, \citenamefont {Nemoto}, \citenamefont {Ralph}, \citenamefont {Dowling},\ and\ \citenamefont {Milburn}}]{Kok2007}%
  \BibitemOpen
  \bibfield  {author} {\bibinfo {author} {\bibfnamefont {P.}~\bibnamefont {Kok}}, \bibinfo {author} {\bibfnamefont {W.~J.}\ \bibnamefont {Munro}}, \bibinfo {author} {\bibfnamefont {K.}~\bibnamefont {Nemoto}}, \bibinfo {author} {\bibfnamefont {T.~C.}\ \bibnamefont {Ralph}}, \bibinfo {author} {\bibfnamefont {J.~P.}\ \bibnamefont {Dowling}},\ and\ \bibinfo {author} {\bibfnamefont {G.~J.}\ \bibnamefont {Milburn}},\ }\bibfield  {title} {\bibinfo {title} {Linear optical quantum computing with photonic qubits},\ }\href {https://doi.org/10.1103/revmodphys.79.135} {\bibfield  {journal} {\bibinfo  {journal} {Reviews of Modern Physics}\ }\textbf {\bibinfo {volume} {79}},\ \bibinfo {pages} {135} (\bibinfo {year} {2007})}\BibitemShut {NoStop}%
\bibitem [{\citenamefont {Kim}\ \emph {et~al.}(2024)\citenamefont {Kim}, \citenamefont {Hong}, \citenamefont {Kim}, \citenamefont {Kim}, \citenamefont {Lee}, \citenamefont {Pooser}, \citenamefont {Oh}, \citenamefont {Lee}, \citenamefont {Lee},\ and\ \citenamefont {Lim}}]{Kim2024}%
  \BibitemOpen
  \bibfield  {author} {\bibinfo {author} {\bibfnamefont {D.-H.}\ \bibnamefont {Kim}}, \bibinfo {author} {\bibfnamefont {S.}~\bibnamefont {Hong}}, \bibinfo {author} {\bibfnamefont {Y.-S.}\ \bibnamefont {Kim}}, \bibinfo {author} {\bibfnamefont {Y.}~\bibnamefont {Kim}}, \bibinfo {author} {\bibfnamefont {S.-W.}\ \bibnamefont {Lee}}, \bibinfo {author} {\bibfnamefont {R.~C.}\ \bibnamefont {Pooser}}, \bibinfo {author} {\bibfnamefont {K.}~\bibnamefont {Oh}}, \bibinfo {author} {\bibfnamefont {S.-Y.}\ \bibnamefont {Lee}}, \bibinfo {author} {\bibfnamefont {C.}~\bibnamefont {Lee}},\ and\ \bibinfo {author} {\bibfnamefont {H.-T.}\ \bibnamefont {Lim}},\ }\bibfield  {title} {\bibinfo {title} {Distributed quantum sensing of multiple phases with fewer photons},\ }\href {https://doi.org/10.1038/s41467-023-44204-z} {\bibfield  {journal} {\bibinfo  {journal} {Nature Communications}\ }\textbf {\bibinfo {volume} {15}},\ \bibinfo {pages} {266} (\bibinfo {year} {2024})}\BibitemShut {NoStop}%
\bibitem [{\citenamefont {Solntsev}\ \emph {et~al.}(2012{\natexlab{a}})\citenamefont {Solntsev}, \citenamefont {Sukhorukov}, \citenamefont {Neshev},\ and\ \citenamefont {Kivshar}}]{Solntsev2012}%
  \BibitemOpen
  \bibfield  {author} {\bibinfo {author} {\bibfnamefont {A.~S.}\ \bibnamefont {Solntsev}}, \bibinfo {author} {\bibfnamefont {A.~A.}\ \bibnamefont {Sukhorukov}}, \bibinfo {author} {\bibfnamefont {D.~N.}\ \bibnamefont {Neshev}},\ and\ \bibinfo {author} {\bibfnamefont {Y.~S.}\ \bibnamefont {Kivshar}},\ }\bibfield  {title} {\bibinfo {title} {Spontaneous parametric down-conversion and quantum walks in arrays of quadratic nonlinear waveguides},\ }\href {https://doi.org/10.1103/physrevlett.108.023601} {\bibfield  {journal} {\bibinfo  {journal} {Physical Review Letters}\ }\textbf {\bibinfo {volume} {108}},\ \bibinfo {pages} {023601} (\bibinfo {year} {2012}{\natexlab{a}})}\BibitemShut {NoStop}%
\bibitem [{\citenamefont {Gräfe}\ \emph {et~al.}(2012)\citenamefont {Gräfe}, \citenamefont {Solntsev}, \citenamefont {Keil}, \citenamefont {Sukhorukov}, \citenamefont {Heinrich}, \citenamefont {Tünnermann}, \citenamefont {Nolte}, \citenamefont {Szameit},\ and\ \citenamefont {Kivshar}}]{Graefe2012}%
  \BibitemOpen
  \bibfield  {author} {\bibinfo {author} {\bibfnamefont {M.}~\bibnamefont {Gräfe}}, \bibinfo {author} {\bibfnamefont {A.~S.}\ \bibnamefont {Solntsev}}, \bibinfo {author} {\bibfnamefont {R.}~\bibnamefont {Keil}}, \bibinfo {author} {\bibfnamefont {A.~A.}\ \bibnamefont {Sukhorukov}}, \bibinfo {author} {\bibfnamefont {M.}~\bibnamefont {Heinrich}}, \bibinfo {author} {\bibfnamefont {A.}~\bibnamefont {Tünnermann}}, \bibinfo {author} {\bibfnamefont {S.}~\bibnamefont {Nolte}}, \bibinfo {author} {\bibfnamefont {A.}~\bibnamefont {Szameit}},\ and\ \bibinfo {author} {\bibfnamefont {Y.~S.}\ \bibnamefont {Kivshar}},\ }\bibfield  {title} {\bibinfo {title} {Biphoton generation in quadratic waveguide arrays: A classical optical simulation},\ }\href {https://doi.org/10.1038/srep00562} {\bibfield  {journal} {\bibinfo  {journal} {Scientific Reports}\ }\textbf {\bibinfo {volume} {2}},\ \bibinfo {pages} {562} (\bibinfo {year} {2012})}\BibitemShut {NoStop}%
\bibitem [{\citenamefont {Solntsev}\ \emph {et~al.}(2012{\natexlab{b}})\citenamefont {Solntsev}, \citenamefont {Sukhorukov}, \citenamefont {Neshev},\ and\ \citenamefont {Kivshar}}]{Solntsev2012a}%
  \BibitemOpen
  \bibfield  {author} {\bibinfo {author} {\bibfnamefont {A.~S.}\ \bibnamefont {Solntsev}}, \bibinfo {author} {\bibfnamefont {A.~A.}\ \bibnamefont {Sukhorukov}}, \bibinfo {author} {\bibfnamefont {D.~N.}\ \bibnamefont {Neshev}},\ and\ \bibinfo {author} {\bibfnamefont {Y.~S.}\ \bibnamefont {Kivshar}},\ }\bibfield  {title} {\bibinfo {title} {Photon-pair generation in arrays of cubic nonlinear waveguides},\ }\href {https://doi.org/10.1364/oe.20.027441} {\bibfield  {journal} {\bibinfo  {journal} {Optics Express}\ }\textbf {\bibinfo {volume} {20}},\ \bibinfo {pages} {27441} (\bibinfo {year} {2012}{\natexlab{b}})}\BibitemShut {NoStop}%
\bibitem [{\citenamefont {Belsley}\ \emph {et~al.}(2020)\citenamefont {Belsley}, \citenamefont {Pertsch},\ and\ \citenamefont {Setzpfandt}}]{Belsley2020}%
  \BibitemOpen
  \bibfield  {author} {\bibinfo {author} {\bibfnamefont {A.}~\bibnamefont {Belsley}}, \bibinfo {author} {\bibfnamefont {T.}~\bibnamefont {Pertsch}},\ and\ \bibinfo {author} {\bibfnamefont {F.}~\bibnamefont {Setzpfandt}},\ }\bibfield  {title} {\bibinfo {title} {Generating path entangled states in waveguide systems with second-order nonlinearity},\ }\href {https://doi.org/10.1364/oe.401303} {\bibfield  {journal} {\bibinfo  {journal} {Optics Express}\ }\textbf {\bibinfo {volume} {28}},\ \bibinfo {pages} {28792} (\bibinfo {year} {2020})}\BibitemShut {NoStop}%
\bibitem [{\citenamefont {He}\ \emph {et~al.}(2024)\citenamefont {He}, \citenamefont {Xia}, \citenamefont {Leykam},\ and\ \citenamefont {Chen}}]{He2024}%
  \BibitemOpen
  \bibfield  {author} {\bibinfo {author} {\bibfnamefont {Y.}~\bibnamefont {He}}, \bibinfo {author} {\bibfnamefont {S.}~\bibnamefont {Xia}}, \bibinfo {author} {\bibfnamefont {D.}~\bibnamefont {Leykam}},\ and\ \bibinfo {author} {\bibfnamefont {Z.}~\bibnamefont {Chen}},\ }\bibfield  {title} {\bibinfo {title} {Optimizing biphoton generation via reconfigurable nonlinear waveguide arrays based on scattering tensor},\ }\href {https://doi.org/10.1364/oe.533728} {\bibfield  {journal} {\bibinfo  {journal} {Optics Express}\ }\textbf {\bibinfo {volume} {32}},\ \bibinfo {pages} {32244} (\bibinfo {year} {2024})}\BibitemShut {NoStop}%
\bibitem [{\citenamefont {Meng}\ \emph {et~al.}(2004)\citenamefont {Meng}, \citenamefont {Guo}, \citenamefont {Tan},\ and\ \citenamefont {Huang}}]{Meng2004}%
  \BibitemOpen
  \bibfield  {author} {\bibinfo {author} {\bibfnamefont {Y.-C.}\ \bibnamefont {Meng}}, \bibinfo {author} {\bibfnamefont {Q.-Z.}\ \bibnamefont {Guo}}, \bibinfo {author} {\bibfnamefont {W.-H.}\ \bibnamefont {Tan}},\ and\ \bibinfo {author} {\bibfnamefont {Z.-M.}\ \bibnamefont {Huang}},\ }\bibfield  {title} {\bibinfo {title} {Analytical solutions of coupled-mode equations for multiwaveguide systems, obtained by use of {C}hebyshev and generalized {C}hebyshev polynomials},\ }\href {https://doi.org/10.1364/josaa.21.001518} {\bibfield  {journal} {\bibinfo  {journal} {Journal of the Optical Society of America A}\ }\textbf {\bibinfo {volume} {21}},\ \bibinfo {pages} {1518} (\bibinfo {year} {2004})}\BibitemShut {NoStop}%
\bibitem [{\citenamefont {Bossé}\ and\ \citenamefont {Vinet}(2017)}]{Bosse2017}%
  \BibitemOpen
  \bibfield  {author} {\bibinfo {author} {\bibfnamefont { .-O.}\ \bibnamefont {Bossé}}\ and\ \bibinfo {author} {\bibfnamefont {L.}~\bibnamefont {Vinet}},\ }\bibfield  {title} {\bibinfo {title} {Coherent transport in photonic lattices: A survey of recent analytic results},\ }\href {https://doi.org/10.3842/sigma.2017.074} {\bibfield  {journal} {\bibinfo  {journal} {Symmetry, Integrability and Geometry: Methods and Applications}\ } \textbf {\bibinfo {volume} {13}},\ \bibinfo {pages} {074} (\bibinfo {year} {2017})}\BibitemShut {NoStop}%
\bibitem [{\citenamefont {Perez-Leija}\ \emph {et~al.}(2013)\citenamefont {Perez-Leija}, \citenamefont {Keil}, \citenamefont {Moya-Cessa}, \citenamefont {Szameit},\ and\ \citenamefont {Christodoulides}}]{PerezLeija2013}%
  \BibitemOpen
  \bibfield  {author} {\bibinfo {author} {\bibfnamefont {A.}~\bibnamefont {Perez-Leija}}, \bibinfo {author} {\bibfnamefont {R.}~\bibnamefont {Keil}}, \bibinfo {author} {\bibfnamefont {H.}~\bibnamefont {Moya-Cessa}}, \bibinfo {author} {\bibfnamefont {A.}~\bibnamefont {Szameit}},\ and\ \bibinfo {author} {\bibfnamefont {D.~N.}\ \bibnamefont {Christodoulides}},\ }\bibfield  {title} {\bibinfo {title} {Perfect transfer of path-entangled photons in ${J}_x$ photonic lattices},\ }\href {https://doi.org/10.1103/physreva.87.022303} {\bibfield  {journal} {\bibinfo  {journal} {Physical Review A}\ }\textbf {\bibinfo {volume} {87}},\ \bibinfo {pages} {022303} (\bibinfo {year} {2013})}\BibitemShut {NoStop}%
\bibitem [{\citenamefont {Perez-Leija}\ \emph {et~al.}(2010)\citenamefont {Perez-Leija}, \citenamefont {Moya-Cessa}, \citenamefont {Szameit},\ and\ \citenamefont {Christodoulides}}]{PerezLeija2010}%
  \BibitemOpen
  \bibfield  {author} {\bibinfo {author} {\bibfnamefont {A.}~\bibnamefont {Perez-Leija}}, \bibinfo {author} {\bibfnamefont {H.}~\bibnamefont {Moya-Cessa}}, \bibinfo {author} {\bibfnamefont {A.}~\bibnamefont {Szameit}},\ and\ \bibinfo {author} {\bibfnamefont {D.~N.}\ \bibnamefont {Christodoulides}},\ }\bibfield  {title} {\bibinfo {title} {{G}lauber–{F}ock photonic lattices},\ }\href {https://doi.org/10.1364/ol.35.002409} {\bibfield  {journal} {\bibinfo  {journal} {Optics Letters}\ }\textbf {\bibinfo {volume} {35}},\ \bibinfo {pages} {2409} (\bibinfo {year} {2010})}\BibitemShut {NoStop}%
\bibitem [{\citenamefont {Rodríguez-Lara}(2011)}]{RodriguezLara2011}%
  \BibitemOpen
  \bibfield  {author} {\bibinfo {author} {\bibfnamefont {B.~M.}\ \bibnamefont {Rodríguez-Lara}},\ }\bibfield  {title} {\bibinfo {title} {Exact dynamics of finite {G}lauber-{F}ock photonic lattices},\ }\href {https://doi.org/10.1103/physreva.84.053845} {\bibfield  {journal} {\bibinfo  {journal} {Physical Review A}\ }\textbf {\bibinfo {volume} {84}},\ \bibinfo {pages} {053845} (\bibinfo {year} {2011})}\BibitemShut {NoStop}%
\bibitem [{\citenamefont {Grice}\ and\ \citenamefont {Walmsley}(1997)}]{Grice1997}%
  \BibitemOpen
  \bibfield  {author} {\bibinfo {author} {\bibfnamefont {W.~P.}\ \bibnamefont {Grice}}\ and\ \bibinfo {author} {\bibfnamefont {I.~A.}\ \bibnamefont {Walmsley}},\ }\bibfield  {title} {\bibinfo {title} {Spectral information and distinguishability in type-{II} down-conversion with a broadband pump},\ }\href {https://doi.org/10.1103/physreva.56.1627} {\bibfield  {journal} {\bibinfo  {journal} {Physical Review A}\ }\textbf {\bibinfo {volume} {56}},\ \bibinfo {pages} {1627} (\bibinfo {year} {1997})}\BibitemShut {NoStop}%
\bibitem [{\citenamefont {Noda}\ \emph {et~al.}(1981)\citenamefont {Noda}, \citenamefont {Fukuma},\ and\ \citenamefont {Mikami}}]{Noda1981}%
  \BibitemOpen
  \bibfield  {author} {\bibinfo {author} {\bibfnamefont {J.}~\bibnamefont {Noda}}, \bibinfo {author} {\bibfnamefont {M.}~\bibnamefont {Fukuma}},\ and\ \bibinfo {author} {\bibfnamefont {O.}~\bibnamefont {Mikami}},\ }\bibfield  {title} {\bibinfo {title} {Design calculations for directional couplers fabricated by {Ti}-diffused $\mathrm{LiNbO}_3$ waveguides},\ }\href {https://doi.org/10.1364/ao.20.002284} {\bibfield  {journal} {\bibinfo  {journal} {Applied Optics}\ }\textbf {\bibinfo {volume} {20}},\ \bibinfo {pages} {2284} (\bibinfo {year} {1981})}\BibitemShut {NoStop}%
\bibitem [{\citenamefont {Toren}\ and\ \citenamefont {Ben-Aryeh}(1994)}]{Toren1994}%
  \BibitemOpen
  \bibfield  {author} {\bibinfo {author} {\bibfnamefont {M.}~\bibnamefont {Toren}}\ and\ \bibinfo {author} {\bibfnamefont {Y.}~\bibnamefont {Ben-Aryeh}},\ }\bibfield  {title} {\bibinfo {title} {The problem of propagation in quantum optics, with applications to amplification, coupling of em modes and distributed feedback lasers},\ }\href {https://doi.org/10.1088/0954-8998/6/5/006} {\bibfield  {journal} {\bibinfo  {journal} {Quantum Optics: Journal of the European Optical Society Part B}\ }\textbf {\bibinfo {volume} {6}},\ \bibinfo {pages} {425} (\bibinfo {year} {1994})}\BibitemShut {NoStop}%
\bibitem [{\citenamefont {Horoshko}(2022)}]{Horoshko2022}%
  \BibitemOpen
  \bibfield  {author} {\bibinfo {author} {\bibfnamefont {D.~B.}\ \bibnamefont {Horoshko}},\ }\bibfield  {title} {\bibinfo {title} {Generator of spatial evolution of the electromagnetic field},\ }\href {https://doi.org/10.1103/physreva.105.013708} {\bibfield  {journal} {\bibinfo  {journal} {Physical Review A}\ }\textbf {\bibinfo {volume} {105}},\ \bibinfo {pages} {013708} (\bibinfo {year} {2022})}\BibitemShut {NoStop}%
\bibitem [{\citenamefont {Li{\~n}ares}\ \emph {et~al.}(2008)\citenamefont {Li{\~n}ares}, \citenamefont {Nistal},\ and\ \citenamefont {Barral}}]{Linares2008}%
  \BibitemOpen
  \bibfield  {author} {\bibinfo {author} {\bibfnamefont {J.}~\bibnamefont {Li{\~n}ares}}, \bibinfo {author} {\bibfnamefont {M.~C.}\ \bibnamefont {Nistal}},\ and\ \bibinfo {author} {\bibfnamefont {D.}~\bibnamefont {Barral}},\ }\bibfield  {title} {\bibinfo {title} {Quantization of coupled 1d vector modes in integrated photonic waveguides},\ }\href {https://doi.org/10.1088/1367-2630/10/6/063023} {\bibfield  {journal} {\bibinfo  {journal} {New Journal of Physics}\ }\textbf {\bibinfo {volume} {10}},\ \bibinfo {pages} {063023} (\bibinfo {year} {2008})}\BibitemShut {NoStop}%
\bibitem [{\citenamefont {Ben-Aryeh}\ and\ \citenamefont {Serulnik}(1991)}]{BenAryeh1991}%
  \BibitemOpen
  \bibfield  {author} {\bibinfo {author} {\bibfnamefont {Y.}~\bibnamefont {Ben-Aryeh}}\ and\ \bibinfo {author} {\bibfnamefont {S.}~\bibnamefont {Serulnik}},\ }\bibfield  {title} {\bibinfo {title} {The quantum treatment of propagation in non-linear optical media by the use of temporal modes},\ }\href {https://doi.org/10.1016/0375-9601(91)90650-w} {\bibfield  {journal} {\bibinfo  {journal} {Physics Letters A}\ }\textbf {\bibinfo {volume} {155}},\ \bibinfo {pages} {473} (\bibinfo {year} {1991})}\BibitemShut {NoStop}%
\bibitem [{\citenamefont {Yang}\ \emph {et~al.}(2024)\citenamefont {Yang}, \citenamefont {Chapman}, \citenamefont {Haylock}, \citenamefont {Lenzini}, \citenamefont {Joglekar}, \citenamefont {Lobino},\ and\ \citenamefont {Peruzzo}}]{Yang2024}%
  \BibitemOpen
  \bibfield  {author} {\bibinfo {author} {\bibfnamefont {Y.}~\bibnamefont {Yang}}, \bibinfo {author} {\bibfnamefont {R.~J.}\ \bibnamefont {Chapman}}, \bibinfo {author} {\bibfnamefont {B.}~\bibnamefont {Haylock}}, \bibinfo {author} {\bibfnamefont {F.}~\bibnamefont {Lenzini}}, \bibinfo {author} {\bibfnamefont {Y.~N.}\ \bibnamefont {Joglekar}}, \bibinfo {author} {\bibfnamefont {M.}~\bibnamefont {Lobino}},\ and\ \bibinfo {author} {\bibfnamefont {A.}~\bibnamefont {Peruzzo}},\ }\bibfield  {title} {\bibinfo {title} {Programmable high-dimensional {H}amiltonian in a photonic waveguide array},\ }\href {https://doi.org/10.1038/s41467-023-44185-z} {\bibfield  {journal} {\bibinfo  {journal} {Nature Communications}\ }\textbf {\bibinfo {volume} {15}},\ \bibinfo {pages} {50} (\bibinfo {year} {2024})}\BibitemShut {NoStop}%
\bibitem [{\citenamefont {Onodera}\ \emph {et~al.}(2024)\citenamefont {Onodera}, \citenamefont {Stein}, \citenamefont {Ash}, \citenamefont {Sohoni}, \citenamefont {Bosch}, \citenamefont {Yanagimoto}, \citenamefont {Jankowski}, \citenamefont {McKenna}, \citenamefont {Wang}, \citenamefont {Shvets}, \citenamefont {Shcherbakov}, \citenamefont {Wright},\ and\ \citenamefont {McMahon}}]{Onodera2024}%
  \BibitemOpen
  \bibfield  {author} {\bibinfo {author} {\bibfnamefont {T.}~\bibnamefont {Onodera}}, \bibinfo {author} {\bibfnamefont {M.~M.}\ \bibnamefont {Stein}}, \bibinfo {author} {\bibfnamefont {B.~A.}\ \bibnamefont {Ash}}, \bibinfo {author} {\bibfnamefont {M.~M.}\ \bibnamefont {Sohoni}}, \bibinfo {author} {\bibfnamefont {M.}~\bibnamefont {Bosch}}, \bibinfo {author} {\bibfnamefont {R.}~\bibnamefont {Yanagimoto}}, \bibinfo {author} {\bibfnamefont {M.}~\bibnamefont {Jankowski}}, \bibinfo {author} {\bibfnamefont {T.~P.}\ \bibnamefont {McKenna}}, \bibinfo {author} {\bibfnamefont {T.}~\bibnamefont {Wang}}, \bibinfo {author} {\bibfnamefont {G.}~\bibnamefont {Shvets}}, \bibinfo {author} {\bibfnamefont {M.~R.}\ \bibnamefont {Shcherbakov}}, \bibinfo {author} {\bibfnamefont {L.~G.}\ \bibnamefont {Wright}},\ and\ \bibinfo {author} {\bibfnamefont {P.~L.}\ \bibnamefont {McMahon}},\ }\bibfield  {title} {\bibinfo {title} {Scaling on-chip photonic neural processors using arbitrarily programmable wave propagation},\ }\href@noop {}
  {\bibfield  {journal} {\bibinfo  {journal} {arXiv:2402.17750 [physics.optics]}\ } (\bibinfo {year} {2024})}\BibitemShut {NoStop}%
\bibitem [{\citenamefont {Quesada}\ and\ \citenamefont {Sipe}(2014)}]{Quesada2014}%
  \BibitemOpen
  \bibfield  {author} {\bibinfo {author} {\bibfnamefont {N.}~\bibnamefont {Quesada}}\ and\ \bibinfo {author} {\bibfnamefont {J.~E.}\ \bibnamefont {Sipe}},\ }\bibfield  {title} {\bibinfo {title} {Effects of time ordering in quantum nonlinear optics},\ }\href {https://doi.org/10.1103/physreva.90.063840} {\bibfield  {journal} {\bibinfo  {journal} {Physical Review A}\ }\textbf {\bibinfo {volume} {90}},\ \bibinfo {pages} {063840} (\bibinfo {year} {2014})}\BibitemShut {NoStop}%
\bibitem [{\citenamefont {Kapon}\ \emph {et~al.}(1984)\citenamefont {Kapon}, \citenamefont {Katz},\ and\ \citenamefont {Yariv}}]{Kapon1984}%
  \BibitemOpen
  \bibfield  {author} {\bibinfo {author} {\bibfnamefont {E.}~\bibnamefont {Kapon}}, \bibinfo {author} {\bibfnamefont {J.}~\bibnamefont {Katz}},\ and\ \bibinfo {author} {\bibfnamefont {A.}~\bibnamefont {Yariv}},\ }\bibfield  {title} {\bibinfo {title} {Supermode analysis of phase-locked arrays of semiconductor lasers: erratum},\ }\href {https://doi.org/10.1364/ol.9.000318} {\bibfield  {journal} {\bibinfo  {journal} {Optics Letters}\ }\textbf {\bibinfo {volume} {9}},\ \bibinfo {pages} {318} (\bibinfo {year} {1984})}\BibitemShut {NoStop}%
\bibitem [{\citenamefont {Efremidis}\ and\ \citenamefont {Christodoulides}(2005)}]{Efremidis2005}%
  \BibitemOpen
  \bibfield  {author} {\bibinfo {author} {\bibfnamefont {N.~K.}\ \bibnamefont {Efremidis}}\ and\ \bibinfo {author} {\bibfnamefont {D.~N.}\ \bibnamefont {Christodoulides}},\ }\bibfield  {title} {\bibinfo {title} {Revivals in engineered waveguide arrays},\ }\href {https://doi.org/10.1016/j.optcom.2004.11.009} {\bibfield  {journal} {\bibinfo  {journal} {Optics Communications}\ }\textbf {\bibinfo {volume} {246}},\ \bibinfo {pages} {345} (\bibinfo {year} {2005})}\BibitemShut {NoStop}%
\bibitem [{\citenamefont {Kruse}\ \emph {et~al.}(2015)\citenamefont {Kruse}, \citenamefont {Sansoni}, \citenamefont {Brauner}, \citenamefont {Ricken}, \citenamefont {Hamilton}, \citenamefont {Jex},\ and\ \citenamefont {Silberhorn}}]{Kruse2015}%
  \BibitemOpen
  \bibfield  {author} {\bibinfo {author} {\bibfnamefont {R.}~\bibnamefont {Kruse}}, \bibinfo {author} {\bibfnamefont {L.}~\bibnamefont {Sansoni}}, \bibinfo {author} {\bibfnamefont {S.}~\bibnamefont {Brauner}}, \bibinfo {author} {\bibfnamefont {R.}~\bibnamefont {Ricken}}, \bibinfo {author} {\bibfnamefont {C.~S.}\ \bibnamefont {Hamilton}}, \bibinfo {author} {\bibfnamefont {I.}~\bibnamefont {Jex}},\ and\ \bibinfo {author} {\bibfnamefont {C.}~\bibnamefont {Silberhorn}},\ }\bibfield  {title} {\bibinfo {title} {Dual-path source engineering in integrated quantum optics},\ }\href {https://doi.org/10.1103/physreva.92.053841} {\bibfield  {journal} {\bibinfo  {journal} {Physical Review A}\ }\textbf {\bibinfo {volume} {92}},\ \bibinfo {pages} {053841} (\bibinfo {year} {2015})}\BibitemShut {NoStop}%
\bibitem [{\citenamefont {Blanco-Redondo}\ \emph {et~al.}(2018)\citenamefont {Blanco-Redondo}, \citenamefont {Bell}, \citenamefont {Oren}, \citenamefont {Eggleton},\ and\ \citenamefont {Segev}}]{BlancoRedondo2018}%
  \BibitemOpen
  \bibfield  {author} {\bibinfo {author} {\bibfnamefont {A.}~\bibnamefont {Blanco-Redondo}}, \bibinfo {author} {\bibfnamefont {B.}~\bibnamefont {Bell}}, \bibinfo {author} {\bibfnamefont {D.}~\bibnamefont {Oren}}, \bibinfo {author} {\bibfnamefont {B.~J.}\ \bibnamefont {Eggleton}},\ and\ \bibinfo {author} {\bibfnamefont {M.}~\bibnamefont {Segev}},\ }\bibfield  {title} {\bibinfo {title} {Topological protection of biphoton states},\ }\href {https://doi.org/10.1126/science.aau4296} {\bibfield  {journal} {\bibinfo  {journal} {Science}\ }\textbf {\bibinfo {volume} {362}},\ \bibinfo {pages} {568} (\bibinfo {year} {2018})}\BibitemShut {NoStop}%
\bibitem [{\citenamefont {Medina~Dueñas}\ \emph {et~al.}(2021)\citenamefont {Medina~Dueñas}, \citenamefont {O’Ryan~Pérez}, \citenamefont {Hermann-Avigliano},\ and\ \citenamefont {Foa~Torres}}]{MedinaDuenas2021}%
  \BibitemOpen
  \bibfield  {author} {\bibinfo {author} {\bibfnamefont {J.}~\bibnamefont {Medina~Dueñas}}, \bibinfo {author} {\bibfnamefont {G.}~\bibnamefont {O’Ryan~Pérez}}, \bibinfo {author} {\bibfnamefont {C.}~\bibnamefont {Hermann-Avigliano}},\ and\ \bibinfo {author} {\bibfnamefont {L.~E.~F.}\ \bibnamefont {Foa~Torres}},\ }\bibfield  {title} {\bibinfo {title} {Quadrature protection of squeezed states in a one-dimensional photonic topological insulator},\ }\href {https://doi.org/10.22331/q-2021-08-17-526} {\bibfield  {journal} {\bibinfo  {journal} {Quantum}\ }\textbf {\bibinfo {volume} {5}},\ \bibinfo {pages} {526} (\bibinfo {year} {2021})}\BibitemShut {NoStop}%
\bibitem [{\citenamefont {Ren}\ \emph {et~al.}(2022)\citenamefont {Ren}, \citenamefont {Lu}, \citenamefont {Jiang}, \citenamefont {Gao}, \citenamefont {Zhou}, \citenamefont {Wang}, \citenamefont {Jiao}, \citenamefont {Wang}, \citenamefont {Solntsev},\ and\ \citenamefont {Jin}}]{Ren2022}%
  \BibitemOpen
  \bibfield  {author} {\bibinfo {author} {\bibfnamefont {R.-J.}\ \bibnamefont {Ren}}, \bibinfo {author} {\bibfnamefont {Y.-H.}\ \bibnamefont {Lu}}, \bibinfo {author} {\bibfnamefont {Z.-K.}\ \bibnamefont {Jiang}}, \bibinfo {author} {\bibfnamefont {J.}~\bibnamefont {Gao}}, \bibinfo {author} {\bibfnamefont {W.-H.}\ \bibnamefont {Zhou}}, \bibinfo {author} {\bibfnamefont {Y.}~\bibnamefont {Wang}}, \bibinfo {author} {\bibfnamefont {Z.-Q.}\ \bibnamefont {Jiao}}, \bibinfo {author} {\bibfnamefont {X.-W.}\ \bibnamefont {Wang}}, \bibinfo {author} {\bibfnamefont {A.~S.}\ \bibnamefont {Solntsev}},\ and\ \bibinfo {author} {\bibfnamefont {X.-M.}\ \bibnamefont {Jin}},\ }\bibfield  {title} {\bibinfo {title} {Topologically protecting squeezed light on a photonic chip},\ }\href {https://doi.org/10.1364/prj.445728} {\bibfield  {journal} {\bibinfo  {journal} {Photonics Research}\ }\textbf {\bibinfo {volume} {10}},\ \bibinfo {pages} {456} (\bibinfo {year} {2022})}\BibitemShut {NoStop}%
\bibitem [{\citenamefont {Eichelkraut}\ \emph {et~al.}(2013)\citenamefont {Eichelkraut}, \citenamefont {Heilmann}, \citenamefont {Weimann}, \citenamefont {Stützer}, \citenamefont {Dreisow}, \citenamefont {Christodoulides}, \citenamefont {Nolte},\ and\ \citenamefont {Szameit}}]{Eichelkraut2013}%
  \BibitemOpen
  \bibfield  {author} {\bibinfo {author} {\bibfnamefont {T.}~\bibnamefont {Eichelkraut}}, \bibinfo {author} {\bibfnamefont {R.}~\bibnamefont {Heilmann}}, \bibinfo {author} {\bibfnamefont {S.}~\bibnamefont {Weimann}}, \bibinfo {author} {\bibfnamefont {S.}~\bibnamefont {Stützer}}, \bibinfo {author} {\bibfnamefont {F.}~\bibnamefont {Dreisow}}, \bibinfo {author} {\bibfnamefont {D.~N.}\ \bibnamefont {Christodoulides}}, \bibinfo {author} {\bibfnamefont {S.}~\bibnamefont {Nolte}},\ and\ \bibinfo {author} {\bibfnamefont {A.}~\bibnamefont {Szameit}},\ }\bibfield  {title} {\bibinfo {title} {Mobility transition from ballistic to diffusive transport in non-hermitian lattices},\ }\href {https://doi.org/10.1038/ncomms3533} {\bibfield  {journal} {\bibinfo  {journal} {Nature Communications}\ }\textbf {\bibinfo {volume} {4}},\ \bibinfo {pages} {2533} (\bibinfo {year} {2013})}\BibitemShut {NoStop}%
\bibitem [{\citenamefont {Kokkinakis}\ \emph {et~al.}(2024)\citenamefont {Kokkinakis}, \citenamefont {Makris},\ and\ \citenamefont {Economou}}]{Kokkinakis2024}%
  \BibitemOpen
  \bibfield  {author} {\bibinfo {author} {\bibfnamefont {E.~T.}\ \bibnamefont {Kokkinakis}}, \bibinfo {author} {\bibfnamefont {K.~G.}\ \bibnamefont {Makris}},\ and\ \bibinfo {author} {\bibfnamefont {E.~N.}\ \bibnamefont {Economou}},\ }\bibfield  {title} {\bibinfo {title} {Anderson localization versus hopping asymmetry in a disordered lattice},\ }\href {https://doi.org/10.1103/physreva.110.053517} {\bibfield  {journal} {\bibinfo  {journal} {Physical Review A}\ }\textbf {\bibinfo {volume} {110}},\ \bibinfo {pages} {053517} (\bibinfo {year} {2024})}\BibitemShut {NoStop}%
\bibitem [{Rep()}]{Repository_ANW}%
  \BibitemOpen
  \href@noop {} {}\bibinfo {note} {ANW{\textunderscore}sol (2025), https://github.com/{\linebreak[0]}Jefferson-Delgado/{\linebreak[0]}ANW{\textunderscore}sol.}\BibitemShut {Stop}%
\bibitem [{\citenamefont {Google}()}]{Colab}%
  \BibitemOpen
  \bibfield  {author} {\bibinfo {author} {\bibnamefont {Google}},\ }\href {https://colab.research.google.com} {\bibinfo {title} {{W}elcome {T}o {C}olab}},\ \bibinfo {note} {available online: https://colab.research.google.com (accessed on 8 May 2025)}\BibitemShut {NoStop}%
\end{thebibliography}

%

\end{document}